%% file: paper.tex
\documentclass[conference,compsoc,11pt]{IEEEtran}

\usepackage[english]{babel}
\usepackage[numbers,sort&compress]{natbib}
\usepackage{etex}
\usepackage{amsmath,amssymb,amsfonts}
\usepackage{algorithmic}
\usepackage{graphicx}
\usepackage{textcomp}
\usepackage{xcolor}
\usepackage{epsfig,endnotes}
\usepackage[noreview]{opera}
\usepackage{color}
\usepackage{hyperref}
\usepackage{url}

\usepackage{pgfplots, pgfplotstable}
\usepgfplotslibrary{groupplots}
\usetikzlibrary{patterns}
\usepackage[htt]{hyphenat}
\usepackage{pifont}
\usepackage[linesnumbered,lined,vlined,ruled,commentsnumbered,noend]{algorithm2e}
\SetAlCapNameFnt{\scriptsize}
\SetAlCapFnt{\scriptsize}
\usepackage{tabulary}
\usepackage{ragged2e}
\usepackage{tabularx}
\usepackage{booktabs}
\usepackage{bbm}
\usepackage{tikz}
\usetikzlibrary{matrix,positioning}
\usepackage{multirow}
\usepackage{multicol}
\usepackage{placeins}
\usepackage{comment}
\usepackage{mdframed}
\usepackage{soul}
\usepackage{bbding}
\usepackage{colortbl}
\usepackage{hhline}
\usepackage{subfigure}
\usepackage[normalem]{ulem}
\usepackage[moderate,tracking=normal]{savetrees}
\usepackage{flushend}

\newcolumntype{L}[1]{>{\hsize=#1\hsize\RaggedRight} X}

\newcommand*\circled[1]{{\raisebox{.5pt}{\textcircled{\raisebox{-.9pt} {#1}}}}}

\newcommand{\xmark}{\ding{55}}%

\usepackage{tikz}
\usetikzlibrary{backgrounds}
\usetikzlibrary{arrows,shapes}
\usetikzlibrary{tikzmark}
\usetikzlibrary{calc}

\colorlet{mhpurple}{red!80}

\begin{document}
\newtheorem{mydefinition}{Definition}
\newcommand{\esp}{esp.\@\xspace}
\newcommand{\camflow}{\texttt{CamFlow}\xspace}
\newcommand{\system}{\textsc{Kairos}\xspace}
\renewcommand{\figureautorefname}{Fig.\xspace}
\renewcommand{\tableautorefname}{Table}
\renewcommand{\algorithmautorefname}{Alg.}
\def\Snospace~{\S{}}
\renewcommand{\sectionautorefname}{\Snospace}
\renewcommand{\subsectionautorefname}{\Snospace}
\renewcommand{\subsubsectionautorefname}{\Snospace}
\providecommand*{\lstlistingautorefname}{Listing}
\newcommand{\subfigureautorefname}{\figureautorefname}

\newcommand{\refappendix}[1]{\hyperref[#1]{Appendix\xspace\ref*{#1}}}
           
\date{}

\newmdtheoremenv[
hidealllines=true,
leftline=true,
innertopmargin=0pt,
innerbottommargin=0pt,
linewidth=2pt,
linecolor=gray!40,
innerrightmargin=0pt,
]{definitionii}{Definition}

\title{\system: Practical Intrusion Detection and Investigation
\\ using Whole-system Provenance}

\author{\IEEEauthorblockN{Zijun Cheng\IEEEauthorrefmark{1}\IEEEauthorrefmark{2},
		Qiujian Lv\IEEEauthorrefmark{1},
		Jinyuan Liang\IEEEauthorrefmark{3}, 
		Yan Wang\IEEEauthorrefmark{1},
		Degang Sun\IEEEauthorrefmark{1},
		Thomas Pasquier\IEEEauthorrefmark{3} and
		Xueyuan Han\IEEEauthorrefmark{4}}
	\IEEEauthorblockA{\IEEEauthorrefmark{1}Institute of Information Engineering, Chinese Academy of Sciences, China}
	\IEEEauthorblockA{\IEEEauthorrefmark{2}School of Cyber Security, University of Chinese Academy of Sciences, China}
	\IEEEauthorblockA{\IEEEauthorrefmark{3}University of British Columbia, British Columbia, Canada}
	\IEEEauthorblockA{\IEEEauthorrefmark{4}Wake Forest University, North Carolina, United States}}

\maketitle

\begin{abstract}
\input{abstract}
\end{abstract}

\section{Introduction}
\label{sec:introduction}
\input{introduction}

\section{Background \& Motivation}
\label{sec:motivation}
\input{motivation}

\section{Threat Model}
\label{sec:threat}
\input{threat}

\section{\system Framework}
\label{sec:framework}
\input{framework}

\section{Evaluation}
\label{sec:evaluation}
\input{evaluation}

\section{Discussion}
\label{sec:discussion}
\input{discussion}

\section{Related Work}
\label{sec:rw}
\input{rw}

\section{Conclusion}
\label{sec:conclusion}
\input{conclusion}

\section*{Acknowledgments}
\input{ack}

\bibliographystyle{IEEEtran}
\bibliography{biblio}

\begin{appendices}

\vspace{-20pt}
\section{DARPA Dataset Details}
\label{sec:appendix:darpa}
\input{appendix_darpa_data}

\section{Hyperparameter Impact on Performance}
\label{sec:appendix:hyper}
\input{appendix_hyper}

\input{graph_examples}

\end{appendices}

\end{document}

%% file: abstract.tex
Provenance graphs are structured audit logs
that describe the history of a system's execution. 
Recent studies have explored a variety of techniques
to analyze provenance graphs
for automated host intrusion detection,
focusing particularly on advanced persistent threats.
Sifting through their design documents,
we identify four common dimensions
that drive the development of 
\emph{provenance-based intrusion detection systems} (PIDSes):
\emph{scope} (can PIDSes detect modern attacks that infiltrate across application boundaries?),
\emph{attack agnosticity} (can PIDSes detect novel attacks without a priori knowledge of attack characteristics?),
\emph{timeliness} (can PIDSes \change{efficiently} 
monitor host systems %
\change{as they run}?), and
\change{\emph{attack reconstruction}}
(can \change{PIDSes distill attack activity
from large provenance graphs so that
sysadmins can easily understand 
and quickly respond to system intrusion?}). %
We present \system,
the first PIDS that simultaneously satisfies
the desiderata in all four dimensions,
whereas existing approaches
sacrifice at least one
\emph{and} struggle to achieve 
comparable detection performance.

\system leverages a novel graph neural network based
encoder-decoder architecture
that learns the \emph{temporal} evolution of 
a provenance graph's \emph{structural} changes
to quantify
the degree of \emph{anomalousness}
for each system event.
Then,
based on this fine-grained information,
\system %
reconstructs
attack footprints,
generating compact \emph{summary graphs}
that accurately describe malicious activity
over a \emph{stream} of system audit logs.
Using state-of-the-art benchmark datasets,
we demonstrate
that \system outperforms previous approaches. %

\vspace{5mm}

\noindgras{Note:} This is a preprint version of the paper accepted at
the 45th IEEE Symposium on Security and Privacy (S\&P'24)~\cite{cheng2024kairos}.

%% file: introduction.tex
Recent work on intrusion detection~\cite{unicorn:han:2020, holmes:milajerdi:2019, mining:barre:2019,winnower:hassan:2018, pidas:xie:2016,zengy2022shadewatcher, wang2022threatrace}
uses kernel-level causal dependency graphs, 
or \emph{provenance graphs},
to combat today's increasingly sophisticated system intrusions,
such as \emph{advanced persistent threats} (APTs)~\cite{apt:chen:2014}.
These graphs, 
constructed from system-level logs,
describe interactions (represented by edges) 
between kernel objects (represented by nodes), 
such as processes, files, and sockets,
to structurally represent the history of a system's execution.

Various aspects 
govern the design 
of prior \emph{provenance-based intrusion detection systems} (PIDSes).
In particular,
we identify four key dimensions of PIDSes
emerging from a large body of work
in this line of research:

	\noindgras{Scope}: System provenance tracks an entire system's activity,
	as well as cross-host interactions through sockets~\cite{camflow:pasquier:2017}.
	Leveraging provenance's system-wide visibility,
	PIDSes that scale to a network of systems are better equipped
	to detect intrusions 
	that span multiple applications and hosts~\cite{holmes:milajerdi:2019}.
	
	\noindgras{Attack Agnosticity}: As zero-day exploits
	(\ie malware or vulnerabilities 
	that are \emph{not} known by security analysts)
	become increasingly common~\cite{zero-day},
	PIDSes can better generalize to detect new attacks
	if they do not rely on any attack signatures 
	\change{or signals} known a priori.
	\change{Security practitioners
	have repeatedly discovered new attacks
	that easily bypass signal-based detectors
	deployed in the wild~\cite{falco-bypass, falco-bypass2}.}
	\change{In contrast,}
	\emph{anomaly-based} PIDSes~\cite{pidas:xie:2016, unicorn:han:2020,sigl:han:2021} extract distinguishing features 
	from graphs of known \emph{benign} system execution
	and use these features to determine whether a system is under attack.
	\change{These PIDSes 
	not only outperform non-provenance-based approaches
	(\eg log analysis~\cite{DBLP:conf/ccs/LiuWZJXM19, DBLP:conf/ccs/Du0ZS17}),
	but more importantly,
	demonstrate great detection performance
	in the face of unknown attacks}.
	This is because system provenance provides rich contextual information
	(both spatial and temporal)
	through its dynamic graph topology. 
	Such contexts separate a benign system event
	from a malicious event,
	\change{even if they look almost identical
    in isolation~\cite{unicorn:han:2020}.}
	For example,
	repeated connections to a system 
	(represented \emph{spatially} in a provenance graph
	as a large number of edges connected to socket nodes)
	in a short period of time 
	(rapid \emph{temporal} changes in the graph)
	could suggest a DoS attack,
	which might differ significantly from a graph
	describing legitimate socket connections.

	\noindgras{Timeliness}: A provenance graph evolves
	to record system activity as the system runs.
	PIDSes that analyze the graph %
	in a \emph{streaming} fashion~\cite{unicorn:han:2020}
	\change{as it evolves}
	provide more timely protection than
	offline systems~\cite{poirot:milajerdi:2019} 
	that introduce delays 
	between provenance capture and threat detection.
	
	\noindgras{\change{Attack Reconstruction}}: 
	System provenance
	is instrumental in \emph{understanding intrusions}%
	\remove{because it structurally represents causal relationships 
	between kernel objects}~\cite{backtracking:king:2003, nodoze:hassan:2019}.
	\remove{As such, }%
	We can reason about chains of events
	that could have led to an intrusion
	and the potential damage inflicted on the system by the intrusion
	by navigating back and forth along the edges in the graph.
	However, 
	it is impractical to
	\emph{manually} investigate the entire graph,
	given the large size of a typical provenance graph
	and the fast growth rate of the graph over time~\cite{camflow:pasquier:2017}.
	Instead, more practical PIDSes %
	provide minimum graph data
	that \emph{reconstructs attack scenarios}%
	\remove{(\eg from multiple points of detection)}
	through the dependencies between kernel objects.
	Such PIDSes can greatly reduce the manual effort,
    enabling sysadmins
	to quickly understand an intrusion 
	and devise a timely response.
	For example,
	\change{Holmes~\cite{holmes:milajerdi:2019} correlates edges
	that match the behavior of known attacks 
	to identify %
	APTs
	and uses the subgraphs of the correlated edges}
	to facilitate attack comprehension.

Unfortunately, 
\emph{no existing PIDSes achieve the desiderata simultaneously in all four dimensions.}
Solutions meeting the first three properties
~\cite{frappuccino:han:2017, Manzoor2016FastMA, unicorn:han:2020, zengy2022shadewatcher, wang2022threatrace}
\change{provide little information
to help sysadmins understand their decisions
and reconstruct the attack,}
while systems satisfying the last %
either consider single applications~\cite{winnower:hassan:2018,sigl:han:2021},
detect only %
known attacks~\cite{holmes:milajerdi:2019, rapsheet:hassan:2020, poirot:milajerdi:2019},
or require offline analysis~\cite{poirot:milajerdi:2019}.

We introduce \system,
the first PIDS %
that fulfills~all~four desiderata while achieving high detection performance.
It leverages
fine-grained,
temporal-spatial graph learning
that scales to
provenance graphs of a network of systems
to monitor \change{run-time} %
system behavior.
Specifically,
\system quantifies
the degree of anomalousness
for \emph{individual} edges (\ie system events)
as they appear
in the streaming graph,
based on
how much the \emph{historical behavioral patterns}
of their corresponding nodes (\ie system entities)
deviate from the patterns
learned from known benign executions 
in the past.
\system' graph analysis
is highly contextualized,
taking into account
\emph{dynamic} changes (\ie temporality) 
of the edges
surrounding a node
and the node's neighborhood structure (\ie spatiality).
Edge-level anomalousness provides
the basis for a graph-level causality analysis
that \system performs periodically 
at run time.
This analysis correlates only
highly anomalous edges
based on information flow
and constructs compact but meaningful summary graphs
from original provenance graphs
to fully \change{and concisely} describe attack scenarios 
(like the one shown in~\autoref{fig:motivation:theia-gt}),
without any \emph{a priori}
knowledge of attack characteristics.

We evaluate \system
on recent, publicly available
benchmark datasets
from DARPA
that simulate
APT campaigns,
as well as datasets
that allow us
to fairly compare \system
with state-of-the-art
open-source PIDSes, 
Unicorn~\cite{unicorn:han:2020}
\change{and ThreaTrace~\cite{wang2022threatrace}}.
Our results show that
\system achieves
high detection accuracy,
outperforming \change{both systems}
while incurring low computational
and memory overhead.
More importantly,
\system
accurately reconstructs
attack footprints,
significantly reducing the number of
graph edges that
sysadmins must inspect
to understand an attack.
\system is available online at
\url{https://github.com/ProvenanceAnalytics/kairos}.

%% file: motivation.tex
\begin{figure*}
	\centering
	\includegraphics[width=\linewidth]{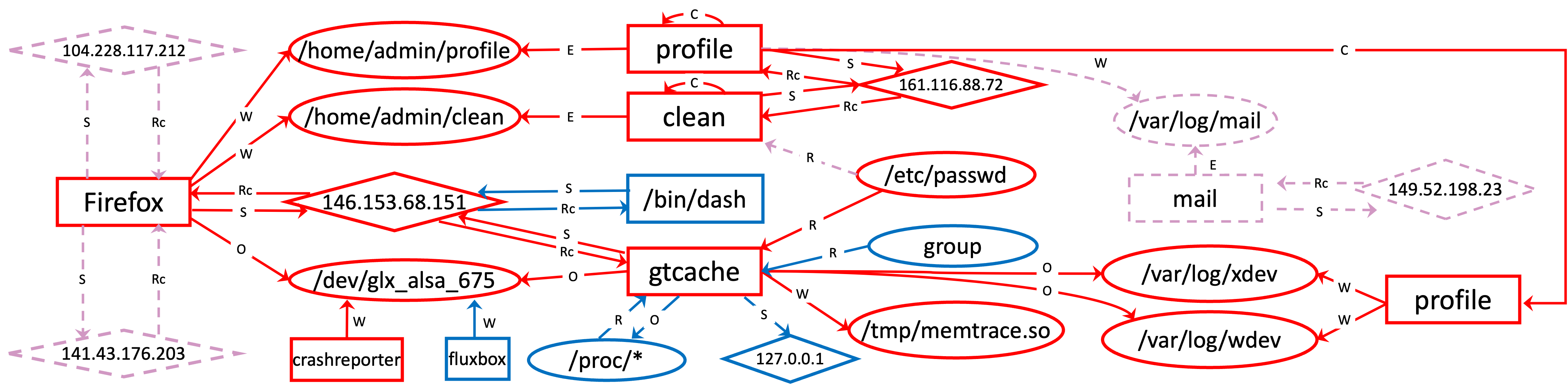}
	\caption{A provenance summary graph from DARPA E3-THEIA that describes attack activity in the motivating example 
		(\autoref{sec:motivation:example}), as automatically generated by \system.
		Rectangles, ovals, and diamonds represent processes, files, and sockets, respectively.
		R=Read, W=Write, O=Open, S=Send, Rc=Receive, C=Clone, and E=Execute.
		We add colors and dashed elements for clarity 
		to highlight the output that \system generates.
		Solid nodes and edges are extracted by \system
		from the original provenance graph to \change{reconstruct the attack}.
		Dashed pink nodes and edges are attack-related activities missed by \system,
		according to the attack ground truth.
		Blue nodes and edges are activities not explicitly mentioned in the ground truth
		but included by \system.
	}
	\label{fig:motivation:theia-gt}
	\vspace{-12pt}
\end{figure*}

We introduce 
the concept 
of system-level data provenance
in~\autoref{sec:motivation:provenance}.
In~\autoref{sec:motivation:example},
we use a real attack scenario
from DARPA
to motivate our design.

\vspace{-2mm}
\subsection{System-level Data Provenance}
\label{sec:motivation:provenance}
\vspace{-2mm}

System-level data provenance
\remove{describes the history of system execution,}%
records data flows between kernel-level objects,
\eg processes, files, and sockets.
Data provenance can be represented
as a directed graph,
called a \emph{provenance graph},
where nodes represent kernel-level objects
and edges represent various types of interactions
(\ie dependency relationships)
between these objects.
These interactions are typically the results of
system calls\remove{, \eg
\texttt{read}, \texttt{write},  and \texttt{open}}.
\autoref{fig:motivation:theia-gt} shows an example of a provenance graph.

We can capture system-level data provenance
using kernel event logging mechanisms,
such as Windows ETW and Linux Audit,
that are natively supported by different operating systems.
Alternatively,
specialized in-kernel reference monitors,
such as Hi-Fi~\cite{pohly2012hi} and CamFlow~\cite{camflow:pasquier:2017},
track fine-grained data flows
between low-level kernel abstractions (\eg inodes and processes)
to %
capture data~provenance.

\system is agnostic
to the underlying provenance capture mechanism,
analyzing \emph{all}\remove{system activities}%
\remove{and} kernel interactions of an \emph{entire network of systems}.
This \emph{whole-system}
(rather than \emph{application-specific})
visibility
is particularly crucial
in detecting modern sophisticated intrusions 
such as APTs,
because APTs often infect multiple applications
on a single host
and migrate from one infected host to another.
\remove{Unfortunately,
it is challenging
to effectively analyze whole-system data provenance
at run time
due to the rapid rate 
at which a provenance graph can grow~\cite{DBLP:conf/ccs/PasquierHMBHEBS18}.}

\vspace{-1.8mm}
\subsection{A Motivating Example}
\label{sec:motivation:example}
\vspace{-2.2mm}

We use a large-scale APT campaign 
simulated by DARPA~\cite{darpa-e3}
to illustrate the challenges 
faced by existing PIDSes
along the four desiderata 
\remove{(\ie scope, attack agnosticity, timeliness, 
and attack reconstruction)
discussed in~}(\autoref{sec:introduction}).
For comparison,
we briefly describe 
\system' output %
in this scenario
at the end\remove{of this section}.
\remove{Later,}%
In \autoref{sec:framework}, 
we provide an in-depth discussion 
of \system' design,
and in \autoref{sec:evaluation},
we give details 
of this experiment (among others).

\vspace{-1.5mm}
\subsubsection{Scenario (\autoref{fig:motivation:theia-gt})}

The attacker leverages a \texttt{Firefox} vulnerability
to establish a foothold on a victim machine,
which enables the attacker 
to write a malicious payload called \texttt{clean}
to \texttt{/home/admin/} on disk. %
The attacker then executes the payload
with escalated privileges.
This new attack process (with root privileges) communicates
with the attacker's command-and-control (C\&C) server
at \texttt{161.116.88.72} %
to download and execute
another %
malicious payload
called \texttt{profile}, again with root privileges.
\texttt{profile}, in turn, 
fetches a third payload
called \texttt{xdev} from the C\&C server
and stores the payload in \texttt{/var/log/}. %
\texttt{profile} and \texttt{xdev} lurk in the victim host
to prepare for %
subsequent attacks. 
A few days later, 
the attacker uses \texttt{profile} 
to %
inject malicious code in the \texttt{mail} process
and executes \texttt{mail} with root privileges.
\texttt{mail} then 
performs port scans of all known hosts 
on the victim's network.

\vspace{-1.5mm}
\subsubsection{Challenges}
\label{sec:motivation:challenges}
APTs stealthily infiltrate their targeted systems
and maintain their presence in victim hosts
for extended periods of time,
exhibiting a unique \emph{low-and-slow} attack pattern.
During the lifecycle of an APT,
it is common for attackers
to leverage various zero-day exploits,
some of which might even be tailored
to the targeted victim systems~\cite{apt1}.
Because of these characteristics,
existing PIDSes are forced to make the following trade-offs:

\noindent\textbf{\textit{Attack Agnosticity}}:
APTs' low-and-slow attack pattern
makes anomaly-based detection difficult,
because attack activity can hide
among a large amount of benign activity
and appear similar to normal behavior
\emph{if execution context is not sufficiently considered}~\cite{unicorn:han:2020}.
For instance,
in our scenario, %
among $32.4$ \emph{million} log entries\remove{captured by DARPA},
we identify approximately only $3,119$ entries %
related to the attack\remove{activity},
which make up merely $0.01\%$ of the entire log.
To circumvent this challenge,
PIDSes
such as Holmes~\cite{holmes:milajerdi:2019}
and RapSheet~\cite{rapsheet:hassan:2020}
use existing threat intelligence knowledge
to manually craft graph-matching rules
that indicate the presence of an APT.
However,
as new exploits continue to surface,
they must constantly update their knowledge base
to include additional rules.
By construction,
they will \emph{always} lag behind
sophisticated adversaries
launching previously unknown attacks.

\noindent\textbf{\textit{\change{Attack Reconstruction}}}:
PIDSes such as Unicorn~\cite{unicorn:han:2020}
\change{and ThreaTrace~\cite{wang2022threatrace}}
take an anomaly-based approach
to detecting system activity 
that deviates significantly from known benign behavior.
While %
they require no \emph{a priori} knowledge 
of APT characteristics
(unlike Holmes), %
their detection provides little %
\change{information
to help sysadmins
understand the attack.}
Consequently,
forensic investigation that follows
typically involves prolonged manual inspection
of large provenance graphs.
\change{For example}, 
Unicorn %
reduces a\remove{large} provenance graph
to a compact feature vector
to %
model\remove{long-term} system behavior,
\remove{without incurring large computational and memory overhead.
While the ability to perform behavioral analysis
over a long timescale
facilitates APT detection,
vectorization loses the details needed for explanation.
For example, 
it is challenging 
to identify the exact processes %
responsible 
for the attack.}%
\change{but an anomalous feature vector
corresponds to~an entire provenance graph.
In our example,
sysadmins must sift through
millions of edges
to identify the attack activity.
ThreaTrace, 
on the other hand,
pinpoints only anomalous \emph{nodes}
(\eg the \texttt{Firefox} and
\texttt{clean} process nodes in~\autoref{fig:motivation:theia-gt})
that might be involved in the attack.
While these nodes 
can be useful starting points,
sysadmins still need
to manually trace through thousands of edges
to understand the complete attack story.
ThreaTrace recognizes this limitation
and acknowledges
the gap 
between anomaly-based detection
and attack construction.
\system fills this gap.}

\noindent\textbf{\textit{Scope}}:
PIDSes
such as Winnower~\cite{winnower:hassan:2018}
\change{construct benign graph templates
to highlight anomalous \emph{subgraphs}
that do not fit into the templates.
While this %
facilitates forensic analysis,
Winnower is unsuitable
for APT detection,
because it cannot scale
to large graphs.
Rather,
Winnower focuses on an application-wide scope
and analyzes much smaller provenance graphs
than the ones 
that can realistically describe whole-system activity 
under APTs.
As such,}
we must run at a minimum multiple instances of Winnower
targeting various applications
(\eg \texttt{Firefox} and \texttt{mail})
to potentially detect the APT in our scenario.
In practice,
a workstation could %
run many dozens of applications,
all of which must be individually monitored 
by Winnower,
since we do not know
a priori what application(s)
would be involved in an APT.
However,
even then, %
it is unclear 
whether Winnower's isolated,
\emph{application-centered} approach would be effective.
This is because
inter-process information flows
are critical to detecting %
APTs~\cite{unicorn:han:2020},
but Winnower is oblivious 
to them.
\change{Like Winnower,
SIGL~\cite{sigl:han:2021} limits its detection 
to anomalies 
during software installations; %
therefore,
it also cannot analyze
a provenance graph of millions of edges
in our scenario.
Moreover,
like ThreaTrace,
SIGL pinpoints only anomalous nodes,
thus incapable of reconstructing attack activity.}

\noindent\textbf{\textit{Timeliness}}:
Timely APT detection and forensic analysis
is important\remove{in practice}
to quickly identify the attack
and take remedial actions.
PIDSes
such as Poirot~\cite{poirot:milajerdi:2019}
match complex graph signatures,
each describing
\remove{in detail}%
the behavior of a specific malware program.
This\remove{largely} expedites threat understanding
\emph{after} a threat is matched.
However,
even if we dismiss %
the issue of attack agnosticity,
Poirot's matching process\remove{itself} is slow
and thus unsuitable for \change{run-time} detection,
for two reasons. %
First,
Poirot takes minutes 
to search for \emph{each} signature
in a provenance graph.
Therefore,
the approach \emph{cannot} scale
as the number of signatures grows.
Second,
matching only succeeds
if a malware program
exhibits its complete behavior
as described in the signature.
As such,
Poirot must repeatedly try 
to match the same graph signatures
as the graph evolves over time,
which exacerbates the scaling issue even further.

\subsubsection{\system' Result}
\system accurately identifies the attack
and reconstructs the APT scenario
at run time
without relying on any a priori attack knowledge,
even though
the malicious activity %
blends in
with the benign activity in the background.
Note that the size of the logs 
capturing the benign activity
\remove{(from many hosts within the network)}%
is several orders of magnitude larger.
\autoref{fig:motivation:theia-gt} shows 
the attack summary graph
\emph{automatically generated} by \system
from the original provenance graph
that describes whole-system host behavior
(of all participating hosts).

The provenance data
in this %
scenario is captured~by THEIA~\cite{fazzini2017tagging},
which performs 
\emph{system-wide} audit
to track fine-grained information flow
between kernel-level entities.
\system analyzes
THEIA's audit data 
to monitor \emph{all} applications 
running on victim hosts (\emph{scope}).
\system' model is trained
only on benign system behavior
that is observed
before the APT campaign
is launched 
(\emph{attack agnosticity}).
As the attack slowly unfolds,
\system gradually constructs
the graph we see in~\autoref{fig:motivation:theia-gt},
as highly-anomalous edges
that are deemed relevant
to the attack 
appear in the provenance graph (\emph{timeliness}).
This compact graph
succinctly describes the attack,
\emph{summarizing} the malicious activity
extracted from the anomalous edges
for clarity.
The original graph contains $32.4$ million edges
and $690K$ nodes;
in contrast,
\system' summary graph
contains only
$29$ edges and $20$ nodes.
More importantly,
\system' output almost perfectly aligns
with the ground truth of our experiment,
which is provided by DARPA
alongside the dataset~\cite{darpa-e3}.
This helps sysadmins
quickly understand the APT attack (\change{\emph{attack reconstruction}}).

%% file: threat.tex
Similar to prior PIDSes~\cite{unicorn:han:2020, provdetector:wang:2020, holmes:milajerdi:2019, poirot:milajerdi:2019},
our work considers
attackers
attempting to\remove{infiltrate}%
\remove{and} take control of a system
\remove{(or a network of systems)}%
and maintain a persistent presence
by \eg exploiting software vulnerabilities
and deploying communication backdoors.
However,
we do not consider hardware-level,
side-channel, or covert-channel attacks,
since their behavior 
is typically not \emph{explicitly} captured
\remove{(or not \emph{explicitly} captured)}%
by kernel-level audit systems.
\system is an \emph{anomaly-based} detection system;
therefore,
we further assume that
host systems are not under the influence
of an attacker
when \system learns 
from provenance graphs of benign system execution
and that \system \emph{thoroughly} observes system activity
during this initial learning period.
\change{If system behavior changes in the future
(or if \system did not fully observe all benign behavior), 
\emph{concept drift} might~occur~\cite{tsymbal2004problem}.
While we exclude concept drift from our threat model,
as is standard in anomaly-based detectors~\cite{sigl:han:2021, wang2022threatrace},
we show empirically 
how \system can mitigate this issue
in~\autoref{sec:evaluation:detection}}.

Our trusted computing base (TCB) includes
the underlying %
OS,
the audit framework,
and \system' analysis code,
which is also standard
among existing PIDSes.
As such, 
we do not consider kernel-level attacks
and assume the use of existing system hardening techniques
to mitigate any potential audit framework compromise~\cite{bates2015trustworthy,camflow:pasquier:2017}.

Finally, 
we assume the integrity of the output data
(\ie provenance graphs)
from the audit framework.
Existing secure provenance systems~\cite{bates2015trustworthy,camflow:pasquier:2017}
and tamper-evident logging techniques~\cite{DBLP:conf/ndss/PaccagnellaDH0F20, DBLP:conf/ccs/PaccagnellaLT020}
can ensure log integrity
and detect any malicious interference with provenance logs.

%% file: framework.tex
\begin{figure*}[!t]
	\centering
	\includegraphics[width=\textwidth]{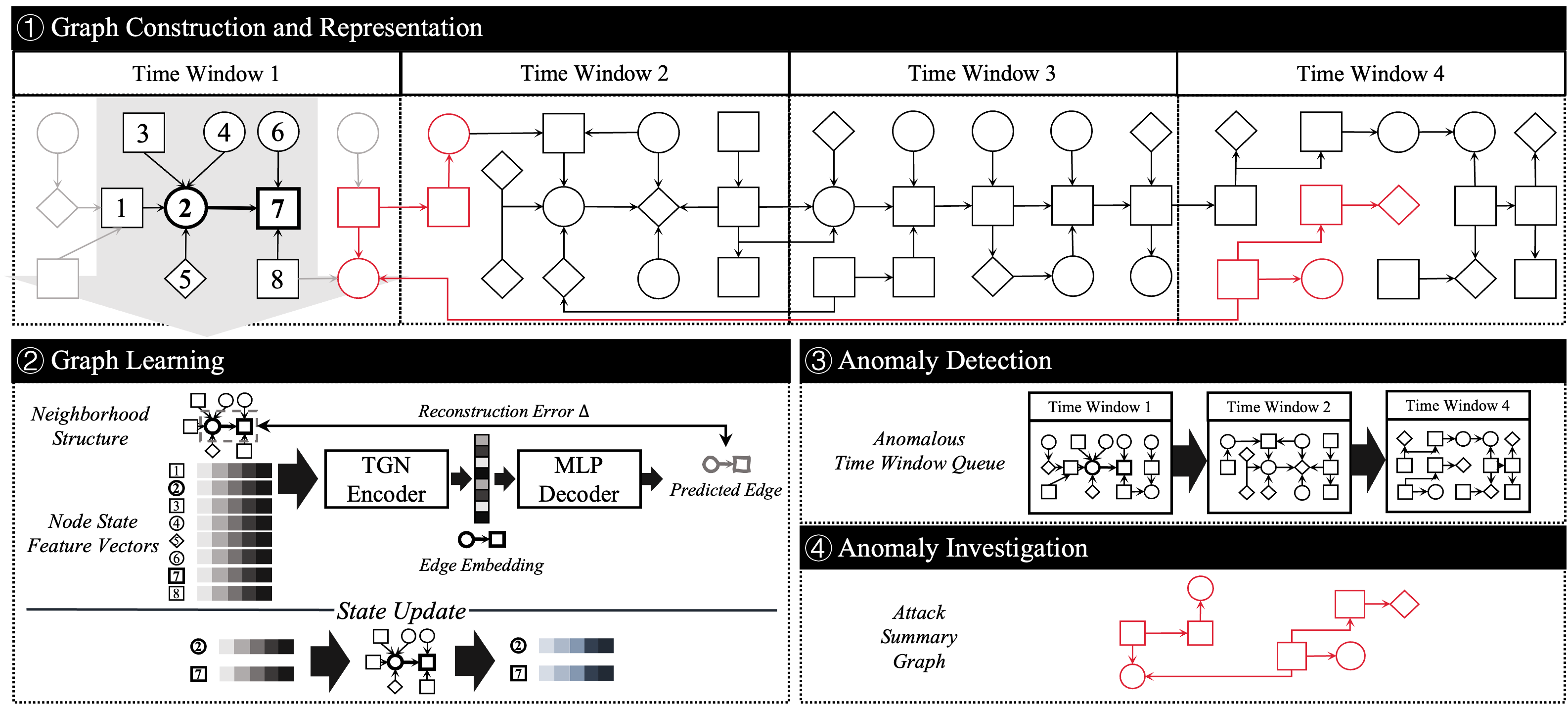}
	\caption{Overview of \system' architecture.}
	\label{fig:arch}
	\vspace{-10pt}
\end{figure*}

\system is an anomaly-based
intrusion detection and attack investigation system.
It leverages state-of-the-art deep graph learning
and community discovery 
through causal dependencies
in a provenance graph
to (1) detect anomalous system behavior
\emph{without} prior knowledge
of any specific attack characteristics,
and (2) correlate detected anomalies 
based on
information flows between kernel objects.
\system provides concise and meaningful \emph{summary graphs}
for labor-saving human-in-the-loop forensic analysis.
\autoref{fig:arch} depicts 
\system' architecture %
consisting of four major components:

\noindgras{\circled{1} Graph Construction and Representation
(\autoref{sec:framework:building}).}
\system analyzes a graph
in a \emph{streaming} fashion,
chronologically ingesting edges
as they appear in the graph.

\noindgras{\circled{2} Graph Learning
(\autoref{sec:framework:learning}).}
When a new edge
(\eg the bold edge \circled{2} $\rightarrow\fbox{7}$
in~\autoref{fig:arch})
appears
in the graph,
\system uses an encoder-decoder architecture
to \emph{reconstruct} the edge.
The encoder takes as input
the neighborhood structure 
around the edge
and
the \emph{states}
of the nodes in the neighborhood.
A node's state is
a feature vector
associated with each node
that describes
the \emph{history} of
the changes in
the node's neighborhood.
The decoder then
reconstructs the edge
from the edge embedding
output by the encoder.
The difference between
the original edge
and the reconstructed edge
is called the \emph{reconstruction error}.
In the training phase,
\system simultaneously
trains the encoder and the decoder
to minimize reconstruction errors
of benign edges.
During deployment,
reconstruction errors
of individual edges
are used as the basis
for anomaly detection and investigation.
Additionally,
\system updates the states
of the source and destination nodes
(node \circled{2} and $\fbox{7}$)
of the new edge.

\noindgras{\circled{3} Anomaly Detection
(\autoref{sec:framework:detection}).}
\system constructs
\emph{time window queues}
to detect anomalies
during deployment.
To do so,
\system identifies
a set of \emph{suspicious nodes}
in each time window
based on the edges' reconstruction errors.
Two time windows
with overlapping suspicious nodes
are enqueued together.
When a new time window
is added to a queue,
\system updates
the \emph{anomaly score}
of the queue,
based also on
reconstruction errors.
If the \remove{anomaly}score
exceeds a threshold,
\remove{(as determined during \emph{training}),}
\system considers
the \remove{time window }queue
to be anomalous
and triggers an alert.
Thus,
\system performs anomaly detection
periodically
at intervals of a time window.
In~\autoref{fig:arch},
\system detects
an anomalous queue
consisting of
time windows 1, 2, and 4.

\noindgras{\circled{4} Anomaly Investigation
(\autoref{sec:framework:investigation}).}
To help sysadmins 
reason about the alarm,
\system automatically generates 
compact \emph{attack summary graphs}
from anomalous time window queues.
This involves identifying
communities of edges
with high reconstruction errors
\remove{and further reducing}%
\remove{the size of the communities}%
to improve legibility.
Graph reduction is necessary,
because unlike images and texts, %
graphs %
are hard
to visualize and interpret
even by human experts~\cite{DBLP:journals/corr/abs-2203-09258}.
In~\autoref{fig:arch},
sysadmins need only
to understand a \remove{single, }small,~\emph{summarized} graph
from \system,
instead of tracing through 
a much larger one\remove{provenance graph}
in the anomalous time window queue
that triggered the alarm.

\vspace{-2mm}
\subsection{Graph Construction and Representation}
\label{sec:framework:building}
\vspace{-2mm}

\begin{table}[t]
 \caption{System entities, their attributes, and dependency relationships.}
 \label{tab:framework:entities}
 \begin{tabularx}{\columnwidth}{|c|c|l|l|}
 \hhline{|-|-|-|-|}
 \textbf{Subject} & \textbf{Object} & \textbf{Relationships} & \textbf{Entity Attributes} \\ \hhline{|-|-|-|-|}
& Process & Start, Close, Clone & Image pathname \\  \hhline{|~|-|-|-|} 
& File & Read, Write, Open, Exec & File pathname \\ \hhline{|~|-|-|-|} 
\multirow{-3}{*}{Process} & Socket & Send, Receive & Src/Dst IP/port \\ \hhline{|-|-|-|-|}
\end{tabularx}
\vspace{-4mm}
\end{table}

\system constructs
a whole-system provenance graph
from audit data
collected by logging infrastructures,
such as Windows ETW, Linux Audit, and CamFlow
(\autoref{sec:motivation:provenance}).
\system considers three types of kernel objects
and nine types of interactions (\ie system events).
\system transforms each event
into a directed, time-stamped edge,
in which
the source node represents
the \emph{subject} of the event
and the destination node
the \emph{object} being acted upon.
\autoref{tab:framework:entities} shows
the types of relationships (\ie interactions)
between kernel subjects and objects
and the node attributes
we consider. %

\system
encodes a node's feature %
using a hierarchical feature hashing technique~\cite{Zhang2020DynamicMA}
based on node attributes.
Hierarchical feature hashing %
\remove{is a \emph{hashing trick}~\cite{weinberger2009feature},}%
\remove{which is a variation of the well-known}%
\remove{\emph{kernel trick}~\cite{theodoridis2006pattern}.}%
\remove{This hashing trick }projects
high-dimensional input vectors
into a lower-dimensional feature space
while preserving the hierarchical similarity
between the original input.
As a result,
two files located in the same parent directory
(\eg \texttt{/var/log/wdev} and \texttt{/var/log/xdev}),
for example,
are mapped closer in the feature space
than a file in a different directory
(\eg \texttt{/home/admin/profile}).

To perform hierarchical feature hashing,
\system encodes a node's attribute
multiple times,
each at a different level of hierarchy.
\remove{to capture the hierarchical information}%
\remove{within the attribute.}%
For example,
for a file node with a pathname \texttt{/home/admin/clean},
\system creates three \emph{substrings}
of the pathname attribute:
\texttt{/home}, 
\texttt{/home/admin},
and \texttt{/home/admin/clean};
\change{for a socket node
with an IP address \texttt{161.116.88.72},
it creates four substrings:
\texttt{161}, \texttt{161.116},
\texttt{161.116.88}, and \texttt{161.116.88.72}.}
\system then projects each substring $s$
into the feature space.
The $i$-th dimension
of $s$' feature vector
is computed by
\(
\phi_{i}(s)=\sum_{j: h(s_{j}) = i} \mathcal{H}(s_j)
\)
where $s_j$ is a character in the substring,
$h$ is a hash function
that maps each character
to one of the dimensions in the feature space,
and $\mathcal{H}$ is another hash function
that hashes a character 
to $\{\pm 1\}$.
Therefore,
we add $\mathcal{H}(s_j)$
to dimension $i$
in $s$' feature vector
if $h(s_{j})$ is $i$.
$\phi(s)$ is the feature vector
of the substring $s$.
The attribute $a$'s feature vector
is the sum of the feature vectors of all its substrings,
\(  
  \Phi(a)=\sum_{j} \phi (s_j)
\)
where $\phi(s_j)$ represents 
each substring's encoded vector,
and $\Phi(a)$ represents 
the final encoding of a node's attribute.

\change{Hierarchical feature hashing
assumes that two kernel entities
of similar semantics
have similar hierarchical features.
While this is often the case,
it is possible that an adversary
attempts to manipulate an entity's attribute
to~evade detection.
However,
\system' graph learning (\autoref{sec:framework:learning})
will %
update these initial feature vectors 
based on %
temporal \emph{and} structural equivalence,
which is hard to manipulate,
to overcome such issues.
\system can also leverage %
other node embedding techniques~\cite{sigl:han:2021},
but %
\emph{all} approaches \reviewchange{make some} assumptions
about the similarity between two system entities.}

\vspace{-2mm}
\subsection{Graph Learning}
\label{sec:framework:learning}
\vspace{-2mm}

Node featurization in~\autoref{sec:framework:building}
captures only attributes
of system entities,
without considering any
\emph{structural}
(\ie \emph{interactions} between an entity
and other entities)
or 
\emph{temporal}
(\ie \emph{sequences} of events involving an entity)
relationships
between individual entities
and the rest of the provenance graph.
This is unfortunate,
because
the \emph{evolving} provenance graph itself,
which describes the \emph{dynamic behavior} of a system,
clearly manifests such relationships.
More importantly,
these relationships
provide rich contextual information
that enables us to model
distinctive baseline (benign) system behavior
and distinguish anomalies from the baseline.

For example,
process injection leads to arbitrary code execution
in the address space 
of a legitimate process.
While malicious execution
is masked under the legitimate process
(\ie the attribute of the process 
remains the same),
under the influence of the adversary,
the compromised process
would exhibit interactions
that deviate from its normal activity
(\eg accessing privileged system resources
that the process typically does not require).
These interactions
are reflected 
as anomalous \emph{structural relationships}
in a provenance graph.

Temporal information
can further reveal subtle behavioral differences;
these differences are hard,
if not impossible,
to identify
if one looks at
only \emph{static snapshots} of 
a dynamic provenance graph.
For example,
a DDoS attack
that quickly overwhelms a victim system
with a large number of network connections
may result in the same graph structure
as an unattacked system
handling the same number of connections
over a reasonable period of time.
Without taking %
\emph{temporal relationships} into account,
it is difficult to detect the attack
by comparing \emph{only} graph structures.

\system learns \emph{both}
temporal and structural relationships
in a provenance graph.
\remove{At a high level,}%
\system' graph learning follows
an encoder-decoder architecture~\cite{kazemi2020representation}.
When a new edge $e_{t}$ appears 
in the streaming \remove{provenance }graph $G_{t}$
at time $t$,
the encoder
embeds \remove{the edge}%
\change{$e_{t}$}
into a latent representation
based on \emph{the state 
of its %
neighborhood
immediately before $t$}
(which we denote as $t^{-}$).
That is,
the edge embedding
summarizes graph features
in $G_{t^{-}} = G_{t} - e_{t}$.
The decoder then
takes as input the edge embedding from the encoder
and predicts the type of the edge
as a probability distribution,
\ie the probability of the edge $e_{t}$
belonging to each of the nine possible types
(\autoref{sec:framework:building}).

\system simultaneously trains
the encoder and the decoder
using \emph{only benign provenance graphs}.
The goal of training is to
minimize the difference
between the actual edge type
(when a new edge appears in the graph)
and the type
predicted by the decoder 
from its embedding.
We call this difference
the \emph{reconstruction error}.
At test time,
the decoder assigns
a small reconstruction error
to an edge
if its embedding
encodes graph structures
that \emph{resemble the structural context
observed from benign system activity
in a similar temporal context}.
Otherwise, 
a large reconstruction error
is assigned,
the magnitude of which
depends on
the extent of the deviation
in \emph{both} contexts.

\noindgras{Encoder.}
\system' encoder uses
a \emph{temporal graph network} (TGN)~\cite{rossi2020temporal}
architecture
to encode provenance graph features
into edge embeddings.
At time $t$,
\system generates
an edge embedding
$\mathbf{z}$ for
the new edge $e_{t}$
using a graph neural network (GNN) based model
called UniMP~\cite{Shi2021MaskedLP}:
\remove{which takes the following input:}
\begingroup
\setlength\abovedisplayskip{1.5pt}
\setlength\belowdisplayskip{1.5pt}
\begin{equation*}
  \label{eq:gnn}
  \mathbf{z} =\mathrm{GNN}(\mathbf{s}_{t^-}, \mathbf{e}, \mathbf{t})
\end{equation*}
\endgroup
$\mathbf{s}_{t^{-}}$ denotes the \emph{state}
of the graph structure
surrounding $e_{t}$ 
at $t^{-}$.
In \system,
a graph structure's state
is represented by
the states of all the nodes
in the structure.
Each node state
is a feature vector
that describes the history
of graph changes involving the node.
When a new node appears
in the graph,
its state is initialized to
a feature vector with all zeros,
because there is no historical information
on the node.
As new edges 
change the node's
\emph{neighborhood}
$\mathcal{N}$,
\system \remove{simultaneously }%
updates the node state
(which we \remove{will }discuss later). 
$\mathbf{s}_{t^{-}}$
thus describes
the states of $e_{t}$'s
source and destination nodes
$v_{src}$ and 
$v_{dst}$, 
as well as the states of
sampled nodes in $\mathcal{N}_{v_{src}}$ 
and $\mathcal{N}_{v_{dst}}$.
$\mathbf{e}$ represents
the edges in $\mathcal{N}_{v_{src}}$ 
and $\mathcal{N}_{v_{dst}}$
from sampled nodes.
Each edge is encoded
as a concatenation of
the source and destination node's
feature embedding (\autoref{sec:framework:building})
and the one-hot encoding
of the edge type.
$\mathbf{t}$ is a vector
of timestamps
corresponding to the edges 
in $\mathbf{e}$.
$(\mathbf{e}, \mathbf{t})$
provides the structural context
of $e_{t}$,
\remove{which is }similar to the \emph{only} information
that prior \remove{existing }PIDSes~\cite{unicorn:han:2020, DBLP:conf/ccs/PasquierHMBHEBS18} use
to learn provenance graphs.

\noindgras{State Update.}
\remove{As mentioned above,}%
\system needs to update
the states of \remove{both }$v_{src}$ and $v_{dst}$,
since their neighborhoods have changed.~To do so,
\system trains 
a gated recurrent unit (GRU) model~\cite{Cho2014LearningPR}:
\begingroup
\setlength\abovedisplayskip{1.5pt}
\setlength\belowdisplayskip{1.5pt}
\begin{align}
    \label{eq:update}
    \mathbf{s}_{t}(v_{src})&=\mathrm{GRU}(\mathbf{s}_{t-}(v_{src}),e_t)\nonumber\\
    \mathbf{s}_{t}(v_{dst})&=\mathrm{GRU}(\mathbf{s}_{t-}(v_{dst}),e_t)\nonumber
\end{align}
\endgroup
Note that
the new edge $e_t$'s information
is propagated to $v_{src}$'s and $v_{dst}$'s
updated states
$\mathbf{s}_{t}(v_{src})$
and $\mathbf{s}_{t}(v_{dst})$,
so that \emph{future} edge embeddings
(of new edges appearing after $t$)
can incorporate $e_t$
if $e_t$ is in their %
neighborhoods.
However,
$e_t$'s information is \emph{not} propagated to
the \emph{current} edge embedding
$\mathbf{z}$
at $t$,
because $e_t$ %
should not be leaked to the decoder
from $\mathbf{z}$
when the decoder is used
to predict $e_t$.
We refer interested readers
to Rossi \etal~\cite{rossi2020temporal}
for technical details
on TGNs.

\noindgras{Decoder.}
\system' decoder uses a
multilayer perceptron (MLP)
to predict the type of the edge
connecting $v_{src}$ and $v_{dst}$.
In other words,
the decoder
learns to \emph{reconstruct}
$e_t$ from the encoder's output
$\mathbf{z}$,
\remove{Therefore,}%
\remove{the encoder}\change{which} provides
both structural and temporal contexts
for the decoder
\remove{in edge reconstruction.}%
\change{to reconstruct edges.}
The dimension of the MLP's last (\ie output) layer is nine,
which is the number of all possible edge types
(\autoref{sec:framework:building}).
The decoder outputs
a vector $\mathbf{P}(e_t)$
of the probabilities of $e_t$
being each of the nine types:
\begingroup
\setlength\abovedisplayskip{1.5pt}
\setlength\belowdisplayskip{1.5pt}
\begin{equation*}
  \label{eq:gnn}
  \mathbf{P}(e_t) =\mathrm{MLP}(\mathbf{z})
\end{equation*}
\endgroup
During training,
\system minimizes
the \emph{reconstruction error} (RE)
between $\mathbf{P}(e_t)$
and the observed edge type $\mathbf{L}(e_t)$
from benign provenance graphs:
\begingroup
\setlength\abovedisplayskip{1.5pt}
\setlength\belowdisplayskip{1.5pt}
\begin{equation*}
  \label{eq:loss_function}
  \mathrm{RE}=\mathrm{CrossEntropy}(\mathbf{P}(e_t),\mathbf{L}(e_t))
\end{equation*}
\endgroup
$\mathbf{L}(e_t)$ is an one-hot vector
where the probability
of $e_t$'s actual edge type is 1
and the rest 0.
At test time,
\system assigns low REs
to edges 
whose structural and temporal contexts
are similar to those
learned from benign graphs
but high REs
if they deviate significantly from
known normal system behavior.
As we see next,
\system uses
these suspicious edges
to detect (\autoref{sec:framework:detection})
and investigate (\autoref{sec:framework:investigation})
anomalies.

\vspace{-2mm}
\subsection{Anomaly Detection}
\label{sec:framework:detection}
\vspace{-2mm}

At a high level,
\system performs anomaly detection 
at the level of \emph{time windows}.
A time window $\mathcal{T}$ contains all system events
(\ie provenance edges)
whose timestamps fall
within a specific period of wall-clock time.
Within a time window,
\system identifies
a set of \emph{suspicious nodes} $\mathcal{S}$
based on
the reconstruction errors
of graph edges (\autoref{sec:framework:learning})
and the \emph{rareness} of the nodes.
\system then incrementally builds
\emph{queues of time windows}
based on each time window's $\mathcal{S}$.
More specifically,
time windows within a queue $q$
are correlated
through their suspicious nodes;
therefore,
a queue captures the activity 
of suspicious nodes
over time and between each other.
\system can construct
many time window queues,
and each time window can
belong to multiple queues
as long as correlation exists
between time windows
in those queues.
\system assigns an \emph{anomaly score}
to each queue
and flags a queue as anomalous
if its anomaly score
is above an anomaly threshold.
Consequently, 
all time windows 
in an anomalous queue
are deemed abnormal.

\system analyzes time window queues,
instead of only individual edges
or individual time windows,
because \system is cognizant of
the distinctive characteristics of modern attacks.
Advanced adversaries today frequently
leverage the ``low-and-slow'' attack pattern
(\autoref{sec:motivation:example}),
so that
it is difficult
to distinguish malicious events
from benign 
but unusual
system activities
in the background.
However,
while these unusual %
activities
are typically discrete,
events belonging to an attack
are %
connected
via information flow
over a long time period~\cite{holmes:milajerdi:2019}.
Time window queues
provide \system
with additional context
necessary to distinguish attack behavior
from unusual but benign activity
and further,
to reconstruct series of events 
constituting the attack (\autoref{sec:framework:investigation}).

In the remainder of this section,
we describe in detail
the process of
identifying suspicious nodes
within a time window 
(\autoref{sec:framework:detection:nodes}),
constructing queues of time windows
(\autoref{sec:framework:detection:windows}),
and identifying abnormal queues
(\autoref{sec:framework:detection:queues}).

\subsubsection{Identifying Suspicious Nodes}
\label{sec:framework:detection:nodes}

\system considers a node 
in a time window $\mathcal{T}$ to be suspicious
if the node satisfies the following two properties:

\noindgras{Anomalousness}: 
A node is anomalous
if it is a source or a destination node
of an edge
that has a reconstruction error
(\autoref{sec:framework:learning})
greater than a \emph{reconstruction threshold}.
\system computes a reconstruction threshold
$\sigma_{\mathcal{T}}$
for each time window $\mathcal{T}$
based on the reconstruction errors
of all the edges in $\mathcal{T}$.
In our experiments,
$\sigma_{\mathcal{T}}$ is $1.5$ standard deviations (SDs)
above the mean
of all reconstruction errors in a time window.

\noindgras{Rareness}:
A node is rare
if its corresponding system entity
does not appear frequently in a benign execution.
We use the \emph{inverse document frequency} (IDF)~\cite{church1999inverse}
to compute a node's rareness.
That is, 
for a given node $v$,
we calculate:
\begingroup
\setlength\abovedisplayskip{1.5pt}
\setlength\belowdisplayskip{1.5pt}
\begin{equation*}
\label{eq:idf}
  \mathrm {IDF}(v)=\mathrm{ln}(\frac{N}{N_{v}+1} )
\end{equation*}
\endgroup
where $N$ is the total number of time windows
and $N_{v}$ the number of time windows
that contain the node $v$.
\remove{A node's IDF is calculated
solely from the \emph{training data}.}%
A node $v$ receives the maximum IDF
if it does not exist in the \remove{training data}past,
\ie $N_{v} = 0$.
Note that
the higher a node's IDF,
the \emph{rarer} it is.
\system considers a node to be rare
if its IDF value is above 
the \emph{rareness threshold} $\alpha$.
\remove{which is set to be
$1.0$ SD
above the mean of all IDFs in training.}%
\system identifies
a set of suspicious nodes $\mathcal{S}_{\mathcal{T}}$
satisfying both \emph{anomalousness} and \emph{rareness} %
for each time window $\mathcal{T}$. %

\change{Prior work, such as
NoDoze~\cite{nodoze:hassan:2019} and PrioTracker~\cite{priotracker:liu:2018},
also explored frequency-based approaches
to measuring rareness.
However, \system's approach is novel, because
(1) \system combines frequency with graph learning to identify suspicious nodes,
while prior work either considers only frequency or 
with node fan-outs, which discounts important structural and temporal anomalies;
and
(2) \system leverages frequency in the context of anomaly detection
while both systems focus only on forensic analysis.
\remove{(\ie backtracking from a known point of intrusion).}}

\vspace{-2.5mm}
\subsubsection{Constructing Queues of Time Windows}
\label{sec:framework:detection:windows}

\system constructs time window queues iteratively,
as new time windows appear
in a streaming provenance graph.
For a new time window $\mathcal{T}_{new}$,
\system either assigns $\mathcal{T}_{new}$
to one or more existing queues
or creates a new queue 
containing only $\mathcal{T}_{new}$.
That is,
$\mathcal{T}_{new}$ is appended
to an existing queue $q$ if:
\begingroup
\setlength\abovedisplayskip{1.5pt}
\setlength\belowdisplayskip{1.5pt}
\begin{equation*}
\exists \mathcal{T} \in q: \mathcal{S}_{\mathcal{T}_{new}} \cap \mathcal{S}_{\mathcal{T}} \neq \emptyset
\end{equation*}
\endgroup
where $\mathcal{T}$ represents any time window in $q$.
If $\mathcal{T}_{new}$ is not correlated
to any existing queues
through suspicious nodes,
$\mathcal{T}_{new}$ itself becomes the start
of a new queue. %

\vspace{-2.5mm}
\subsubsection{Detecting Anomalous Queues}
\label{sec:framework:detection:queues}

The anomaly score of a queue $q$
is the \emph{product}
over the anomaly scores
of all the time windows $\mathcal{T}_{i}$ in the queue:
\begingroup
\setlength\abovedisplayskip{1.5pt}
\setlength\belowdisplayskip{1.5pt}
\begin{equation*}
  \mathrm{AnomalyScore}(q) = \prod_{i=1}^{n} \mathrm{AnomalyScore}(\mathcal{T}_{i})
\end{equation*}
\endgroup
The anomaly score
of a time window $\mathcal{T}$
is the mean
of the reconstruction errors
of the edges in $\mathcal{T}$
whose reconstruction errors
are above the reconstruction threshold $\sigma_{\mathcal{T}}$
(\autoref{sec:framework:detection:nodes}).

\system incrementally updates
the anomaly score
of a queue
at run time
when a new time window is appended 
to the queue.
Each time a queue's anomaly score
is updated,
\system compares the updated anomaly score
with the anomaly threshold $\beta$
to determine
whether the queue 
is abnormal.
\system uses benign \emph{validation} data
to set $\beta$ after model training.
\remove{$\beta$ is the \emph{largest}
anomaly score among
all the queues in the validation graphs.}%
If a queue's anomaly score
at run time (during test)
exceeds $\beta$,
the queue
(and thus all time windows in the queue)
are considered to be anomalous.

\vspace{-2.2mm}
\subsection{Anomaly Investigation}
\label{sec:framework:investigation}
\vspace{-2.3mm}
  
While anomalous time window queues
(\autoref{sec:framework:detection})
significantly reduce 
the size of the \remove{provenance }graph
that sysadmins must inspect
\remove{when \system raises }\change{in case of} an intrusion alarm,
\remove{these queues}\change{they} can still contain
thousands of nodes and edges.
\remove{depending on system workloads.}%
\remove{As a result,
manual post-detection forensic analysis
would remain labor-intensive.}%
To further lessen the burden
on the analyst,
\system automates
the attack investigation process
by constructing \emph{candidate summary graphs}
from anomalous time window queues.
\system does so
without relying on any prior attack knowledge;
therefore, 
its ability to reconstruct 
more precise attack footprints
is \emph{not} limited
to 
previously known attacks.

Given an anomalous queue $q$,
\system first applies standard 
provenance graph reduction techniques~\cite{xu2016high}
to reduce the size of the graph in $q$
without changing its semantics.
For example,
\system merges edges
from the same source and destination nodes
if \remove{the edges}\change{they} are of the same type.
\change{\system' graph reduction
does \emph{not} affect anomaly detection,
because \system performs reduction
only \remove{to simplify the reconstructed attack graph}%
\emph{after} it detects 
an anomalous %
queue.}

Similar to prior work~\cite{DBLP:conf/acsac/PeiGSM0ZSZX16},
we observe that
attack activities typically form
a dense community of nodes
that are connected via
edges of \emph{high} reconstruction errors,
separating them from
other,
non-attack-related nodes.
Thus,
after graph reduction,
\system leverages the \emph{community discovery algorithm Louvain}~\cite{Blondel2008FastUO} 
to identify 
those communities.

To do so,
\system first constructs
a weighted graph $\mathcal{G}_{q}$
from all the anomalous queues
by adding to $\mathcal{G}_{q}$ 
any edge $e$
that has a $\mathrm{RE}_{e}$
greater than the reconstruction threshold
$\sigma_{\mathcal{T}}$
(\autoref{sec:framework:detection:nodes})
of its corresponding time window $\mathcal{T}$:
\begingroup
\setlength\abovedisplayskip{1.5pt}
\setlength\belowdisplayskip{1.5pt}
\begin{equation*}
\mathcal{G}_{q} = \{e: \mathrm{RE}_{e} > \sigma_{\mathcal{T}}, \forall e \in \mathcal{T},  \forall \mathcal{T} \in q\}
\end{equation*}
\endgroup
$\mathcal{G}_{q}$ is
the input graph to Louvain,
and the reconstruction error $\mathrm{RE}_{e}$ of each edge
in $\mathcal{G}_{q}$
is the weight of the edge.

Then, \remove{the }Louvain\remove{ algorithm} starts with
each node in $\mathcal{G}_{q}$
representing a separate community.
For every node $v \in \mathcal{G}_{q}$,
\remove{the algorithm}\change{it} moves $v$
from its current community
to one of its neighboring communities
that leads to the largest improvement
(if any)
of \emph{modularity},
which measures the degree of
connection density within communities
compared to that between communities:
\begingroup
\setlength\abovedisplayskip{1.5pt}
\setlength\belowdisplayskip{1.5pt}
\begin{equation*}
  \mathrm{Modularity} =\sum_{c}\left[\frac{\sum_{in}}{2m}-\left(\frac{\sum_{tot}}{2m}\right)^{2}\right]
\end{equation*}
\endgroup
where $\sum_{in}$ is 
the sum of the $\mathrm{RE}$s
of the edges in the community $c$,
and $\sum_{tot}$ is the sum 
of the $\mathrm{RE}$s
of $c$'s~neighboring edges
\change{(\ie edges with
one of their incident nodes
inside $c$ 
and the other one outside)}.
\remove{\footnote{An edge is a community's}%
\remove{	neighboring edge if one of its nodes}%
\remove{	(source or destination)}%
\remove{	is inside the community }%
\remove{	and the other one is not.}}%
$m$ is the sum of the $\mathrm{RE}$s
of all edges in $\mathcal{G}_{q}$.
Note that $v$ remains 
in its community
if relocating it achieves no modularity gain.
\remove{The }Louvain \remove{algorithm}%
runs this process repeatedly
until modularity no longer increases.

\change{To the best of our knowledge,
\system is the first~to
\emph{bridge graph learning and community detection 
in the context of anomaly detection}.
For example,
HERCULE~\cite{DBLP:conf/acsac/PeiGSM0ZSZX16}
also leverages community detection
but pre-defines
29 edge features %
to cluster edges.
In contrast,
\system avoids
manual feature engineering
by using \emph{learned}
anomaly scores.}

The resulting communities 
are then simplified
to create
\emph{candidate summary graphs}.
These \remove{summary }graphs
\remove{can }concisely describe
malicious behavior
that spans long time periods
and involves multiple stages
of an attack kill-chain~\cite{holmes:milajerdi:2019},
although
sometimes,
they might also represent
abnormal but otherwise benign system activity
(due to the nature 
of \emph{anomaly-based} detection~\cite{DBLP:journals/datamine/AkogluTK15}).
However,
as we see in~\autoref{sec:evaluation:explainability}, 
sysadmins can easily
dismiss benign candidate graphs
(\ie false alarms)
and quickly identify
the attacker's footprints
from small 
but attack-revealing graphs
without the need to backtrack~\cite{backtracking:king:2003}
or forward-track~\cite{DBLP:conf/ndss/KingMLC05}
the entire provenance graph.
In fact,
analyzing those small graphs is the only time 
\system requires expert knowledge
in its \emph{entire} intrusion detection pipeline.
\remove{Compared to}\change{Unlike} prior systems~\cite{holmes:milajerdi:2019, poirot:milajerdi:2019, rapsheet:hassan:2020}
that require expert-crafted attack signatures
and manual exploration of the whole provenance graph,
\system greatly reduces human~involvement.
\remove{Note that }%
Sysadmins still have the option
to inspect the entire \remove{provenance }graph
for further, in-depth analysis,
\emph{but only if they choose to do~so}.

In~\autoref{sec:evaluation:detection}, 
we show how sysadmins
can update \system' model
based on \emph{benign} candidate graphs
to continuously improve
the quality of
detection and investigation.

%% file: evaluation.tex
We implemented a \system prototype
in Python.
We use
scikit-learn~\cite{sklearn}
to implement
hierarchical feature hashing (\autoref{sec:framework:building})
and PyG~\cite{Fey/Lenssen/2019}
\remove{a PyTorch-based
graph neural networks library,}%
to implement \system'
graph learning framework (\autoref{sec:framework:learning}).
\remove{The core anomaly investigation algorithm,}%
Louvain
is implemented
using NetworkX~\cite{hagberg2008exploring}.
\remove{a Python graph analysis library.}
Finally,
we use GraphViz~\cite{ellson2001graphviz}
to visualize summary graphs
for manual inspection.

We evaluate \system
on \change{eight} publicly available datasets,
analyzing %
kernel-level provenance data
that captures
whole-system behavior of various platforms
(namely Linux, FreeBSD, and Android)
with and without attacks.
\autoref{sec:evaluation:datasets}
describes the experimental datasets
in detail.
All experiments are performed on a server
running CentOS 7.9
with 2.20GHz 20-core Intel Xeon Silver 4210 CPU
and 64 GB of memory.
Unless otherwise stated,
we set the following hyperparameters
in all the experiments
except those in~\autoref{sec:evaluation:hyperparameters},
where we examine
the effect of hyperparameters
on \system' performance:
node feature embedding dimension $|\Phi|$ = $16$,
node state dimension $|\mathbf{s}(v)| = 100$,
neighborhood size $|\mathcal{N}| = 20$,
edge embedding dimension $|\mathbf{z}| = 200$,
and time window length
\change{$|\mathbf{tw}| = 15$ minutes}.
Our evaluation focuses on
answering the following research questions:
\\
\textbf{Q1.} Can \system accurately detect anomalies
in a running system under attack,
especially when \change{they} \remove{the anomalies}%
are low-and-slow
\remove{(such as the case in}%
\change{\eg}
the APT \remove{example}in~\autoref{sec:motivation:example})
and thus difficult to detect? (\autoref{sec:evaluation:detection})\\
\textbf{Q2.} How does \system compare
to state-of-the-art? %
(\autoref{sec:evaluation:compare})\\
\textbf{Q3.} How do hyperparameters
affect \system' detection
and run-time performance? (\autoref{sec:evaluation:hyperparameters})\\
\textbf{Q4.} Can \system accurately
reconstruct attack behavior
from the original provenance graph?
\remove{thus providing an explanation for its detection?}%
(\autoref{sec:evaluation:explainability})\\
\textbf{Q5.} What is \system' end-to-end performance?
(\autoref{sec:evaluation:performance})

\subsection{Datasets}
\label{sec:evaluation:datasets}
\vspace{-1mm}

We obtain our experimental datasets
from two sources,
Manzoor~\etal~\cite{streamspot-data}
and DARPA\change{~\cite{darpa-e3, darpa-e5, darpa-optc}}.
They are 
the few \emph{open-source} datasets
widely used in evaluating provenance-based \remove{security}%
systems~\cite{unicorn:han:2020, holmes:milajerdi:2019, Manzoor2016FastMA, poirot:milajerdi:2019,Yu2019NeedleIA,Berrada2019AggregatingUP,Berrada2020ABF,Xie2021PGaussianPG,Hossain2020CombatingDE}.
\autoref{tab:evaluation:datasets}
summarizes the statistics
of the %
graphs
in those datasets.

\begin{table}[t]
\caption{Summary of the experimental datasets. }
\label{tab:evaluation:datasets}
\resizebox{\columnwidth}{!}{%
  \begin{tabular}{|l|l|l|l|l|}
    \hline
    \textbf{Dataset} &
    \textbf{\# of Nodes} &
    \textbf{\begin{tabular}[c]{@{}c@{}}\# of Edges \\ (in millions)\end{tabular}} &
    \textbf{\begin{tabular}[c]{@{}c@{}}\# of Attack \\ Edges\end{tabular}} &
   \textbf{\begin{tabular}[c]{@{}c@{}}\% of Attack \\ Edges\end{tabular}} \\ %
    \hline
    Manzoor~\etal & 999,999 & 89.8 & 2,842,345 & 3.165\% \\ %
    DARPA-E3-THEIA & 690,105 & 32.4 & 3,119 & 0.010\% \\ %
    DARPA-E3-CADETS & 178,965 & 10.1 & 1,248 & 0.012\% \\ %
    DARPA-E3-ClearScope  & 68,549 & 9.7 & 647 & 0.006\% \\ %
    DARPA-E5-THEIA & 739,329 & 55.4 & 86,111 & 0.156\% \\ %
    DARPA-E5-CADETS  & 90,397  & 26.5  & 793 & 0.003\% \\ %
    DARPA-E5-ClearScope  & 91,475  & 40.0 & 4,044 & 0.010\% \\ %
    \change{DARPA-OpTC} & \change{9,485,265} & \change{75.0} & \change{33,504} & \change{0.045}\% \\ %
    \hline
  \end{tabular}
}
\vspace{-3mm}
\end{table}

\subsubsection{Manzoor~\etal Dataset}
\label{sec:evaluation:datasets:streamspot}

\begin{table}[t]
	\caption{Characteristics of the Manzoor~\etal dataset.}
	\label{tab:evaluation:dataset:streamspot}
	\footnotesize
		\centering
		\begin{tabularx}{\columnwidth}{|X|X|X|X|}
			\hline
			\textbf{Scenarios} &
			\textbf{\# of Graphs} &
			\textbf{\begin{tabular}[c]{@{}l@{}}Average \# \\ of Nodes\end{tabular}} &
			\textbf{\begin{tabular}[c]{@{}l@{}}Average \# \\ of Edges\end{tabular}}
			\\
			\hline
			YouTube & 100 & 8,292 & 113,229 \\
			Gmail & 100 & 6,827 & 37,382 \\
			Video Game & 100 & 8,831 & 310,814 \\
			Attack & 100 & 8,891 & 28,423 \\
			Download & 100 & 8,637 & 112,958 \\
			CNN & 100 & 8,990 & 294,903 \\ \hline
		\end{tabularx}
\vspace{-4mm}
\end{table}

This dataset contains provenance graphs
captured by SystemTap~\cite{jacob2008systemtap}
from six activity scenarios
in a controlled lab environment.
Five of \remove{the scenarios}\change{them}~(\ie watching YouTube\remove{ videos},
checking Gmail,
playing a video game,
downloading files,
and browsing cnn.com)
contain only benign \remove{system }activity.
The attack scenario
involves a drive-by download
from a malicious URL
that exploits a Flash vulnerability,
which allows the attacker
to gain root access.
\remove{to the host.}%
Manzoor~\etal repeatedly ran each scenario
to generate~$100$ graphs per scenario.
\autoref{tab:evaluation:dataset:streamspot}
details the graph statistics.

\remove{We use this dataset}%
\change{This dataset allows us}
to demonstrate \system' high efficacy
on traditional,
``smash-and-grab'' attacks,
where the attacker
quickly subverts a system.
\remove{This dataset}\change{It} also allows us
to fairly compare \system
with Unicorn~\cite{unicorn:han:2020}
\change{and ThreaTrace~\cite{wang2022threatrace},
two state-of-the-art PIDSes}
that perform \remove{real-time,}%
\remove{anomaly-based intrusion detection}%
\change{anomaly detection}
on whole-system provenance graphs.
\change{Both systems used
this dataset for their own evaluations.}
However,
it is difficult to demonstrate
\system' ability to reconstruct attack activity
with this dataset,
as the fine-grained attack ground truth
(\ie the exact attack procedure)
is not public and thus unknown to us.
We use DARPA datasets
(\autoref{sec:evaluation:datasets:darpa})
for such evaluation.

\noindgras{Data Labeling.}
\remove{As shown in~\autoref{tab:evaluation:dataset:streamspot},}
We label the \texttt{Attack} scenario graphs~as attack
and the remaining graphs as benign.
Due to the lack of \remove{precise }attack knowledge,
we use a single time window (\autoref{sec:framework:detection})
for each graph
(\ie the time window queue length $|q|$ is \remove{always }1).
Thus,
an attack graph corresponds to
a single \emph{attack time window}.
From each benign scenario,
we use only one graph
to train \system
and 24 graphs
as \emph{validation} data
to configure \remove{anomaly }detection thresholds (\autoref{sec:framework:detection})
\remove{so as not }to \change{not} introduce bias 
in their selection~\cite{arp2022and}.
We use the remaining benign graphs
($75$ for each scenario)
and all $100$ attack graphs
as test data. \remove{to compare with Unicorn}%
\remove{\change{and ThreaTrace}.}%

\subsubsection{DARPA Datasets}
\label{sec:evaluation:datasets:darpa}
\change{We use datasets from DARPA's
Transparent Computing (TC)
and Operationally Transparent Cyber (OpTC) programs.
TC} \remove{program }organized
\remove{a number of}\change{several} adversarial engagements
that simulated real-world APTs \remove{attacks}%
on enterprise networks.
During the engagements,
\remove{DARPA's}\change{a} red team launched~a series of \remove{APT }attacks
towards an enterprise's security-critical~services
(\eg web, email, and SSH servers)
while engaging in benign activities
such as browsing websites,
checking emails,
and SSH log-ins.
A separate team deployed
various provenance capture systems
(\eg CADETS, ClearScope, and THEIA)
on different platforms
to record whole-system \change{host} activity.
\remove{of the entire network.}%
The provenance data
from the third (E3) and the fifth (E5) engagement
is publicly available~\cite{darpa-e3,darpa-e5}.

\change{The OpTC dataset contains 
benign activities of $500$ Windows hosts
over %
seven days
and additional three days 
of a mixture of benign and APT activities.
The red team simulated a three-day long
APT attack
using a number of known CVEs
on a small subset
of hosts. %
The large scale of this dataset
(with its total size in the order of
a few dozen TBs)
enables us to evaluate \system 
under a more ``real'' setting, 
where the amount of test data is much larger
than that of training data.
Specifically,
we randomly select six hosts
and use only one day of the benign data
from them
for training,
one additional benign day %
for validation, %
but all three attack days
\emph{from all hosts}
for testing.}
\autoref{tab:evaluation:datasets} details
the \remove{statistics of the graphs}\change{graph statistics}
from different provenance systems.
\autoref{tab:evaluation:tc:scenarios}
in~\refappendix{sec:appendix:darpa}
summarizes all DARPA attacks. %
\remove{in these datasets.}%

We use DARPA datasets
to show that \system can
(1) accurately detect anomalies
even though \remove{these anomalies}\change{they}
are hidden
among a large amount of benign activity
across a long time span,
and (2) precisely distill
the original provenance graph
(that describes both benign and attack activity)
into a compact attack summary graph
\remove{without any \emph{a priori} knowledge of the attack,}%
\change{without prior attack knowledge,}
even though attack activity
is several orders of magnitude rarer
(see~\autoref{tab:evaluation:datasets}).
Moreover,
\change{we use (3) %
the TC dataset to compare \system with Unicorn and ThreaTrace,
and (4) the OpTC dataset to
demonstrate that \system can be realistically 
deployed in a large-scale network of systems.}
Our motivating example (\autoref{sec:motivation:example})
uses %
the E3 dataset.

\noindgras{Data Labeling.}
Unlike Manzoor~\etal,
DARPA provides attack ground truth,
which enables us to %
label \emph{individual} \remove{provenance }nodes and edges
related to the attack.
\remove{in each dataset.}%
\remove{This in turn}%
\remove{allows us to }\change{Thus, we can} manually compare
\system' reconstructed attack graph
with the ground-truth \remove{attack }graph.
It is worth noting that
the ground truth %
is used only by us
to verify \system' efficacy;
\system does \emph{not} leverage any attack knowledge
in its own analysis.

\remove{The \change{TC} engagements lasted for days,
but the attacker was not active every day.
We call the days with no APT activity \emph{benign days}
and the ones with APT activity \emph{attack days}.}%
\change{%
In both TC and OpTC,
attack activity occurred
only in a subset of time windows
within an attack day.}
For instance,
in our motivating example (\autoref{sec:motivation:example}),
the ground truth shows
some attack activity
on April $10^{th}$, 2018
at $13$:$41$
when the attacker attempted
to manipulate Firefox.
The next attack activity
occurred almost an hour later.
As such,
we mark the time window
that includes the Firefox event
as an \emph{attack time window}.
Since each time window is $15$-minute long
in our experiments,
the next several time windows
are therefore \emph{benign time windows},
until the attack activity resumes.

\autoref{tab:evaluation:tc:datasplit}
in~\refappendix{sec:appendix:darpa}
summarizes specific
benign and attack days
we use for training, validation, and detection.

\vspace{-2.2mm}
\subsection{Detection Performance}
\label{sec:evaluation:detection}
\vspace{-2.2mm}

\begin{table}[t]
\caption{\system' experimental results.}
\label{tab:evl:exp}
\resizebox{\columnwidth}{!}{%
  \centering
  \begin{tabular}{|l|c|c|c|c|c|c|c|c|}
    \hline
    \textbf{Datasets} & \textbf{TP} & \textbf{TN} & \textbf{FP} & \textbf{FN} & \textbf{Precision} & \textbf{Recall} & \textbf{Accuracy} & \textbf{AUC} \\ \hline
    Manzoor \etal & 100 & 375 & 0 & 0 & 1.000  & 1.000 & 1.000 & 1.000 \\ \hline
    E3-THEIA & 9 & 216 & 2 & 0 & 0.818 & 1.000 & 0.991 & 0.995  \\ \hline
    E3-CADETS & 4 & 174 & 1 & 0 & 0.800 & 1.000 & 0.994 & 0.997 \\ \hline
    E3-ClearScope & 5 & 112 & 2 & 0 & 0.714  & 1.000 & 0.983 & 0.991 \\ \hline
    E5-THEIA & 2 & 173 & 1 & 0 & 0.667 & 1.000 & 0.994 & 0.997 \\ \hline
    E5-CADETS & 7 & 238 & 9 & 0 & 0.438 & 1.000 & 0.965 & 0.982 \\ \hline
    E5-ClearScope & 10 & 217 & 5 & 0 & 0.667  & 1.000  & 0.978  & 0.989   \\ \hline
    \change{OpTC} & \change{22} & \change{1210} & \change{16} & \change{0} & \change{0.579} & \change{1.000} & \change{0.987} & \change{0.993} \\ \hline
    \end{tabular}
}
\vspace{-3mm}
\end{table}

\begin{table}[t]
\caption{\system' adjusted experimental results.}
\label{tab:evl:exp_adjusted}
  \resizebox{\columnwidth}{!}{%
    \centering
    \begin{tabular}{|l|c|c|c|c|c|c|c|c|}
      \hline
      \textbf{Datasets} & \textbf{TP} & \textbf{TN} & \textbf{FP} & \textbf{FN} & \textbf{Precision} & \textbf{Recall} & \textbf{Accuracy} & \textbf{AUC} \\ \hline
      Manzoor \etal & 100 & 375 & 0 & 0 & 1.000  & 1.000 & 1.000 & 1.000 \\ \hline
      E3-THEIA  & \textbf{10} & \textbf{216} & \textbf{1} & \textbf{0} & \textbf{0.909}  & \textbf{1.000}  & \textbf{0.996} & \textbf{0.998} \\ \hline
      E3-CADETS & 4 & 174 & 1 & 0 & 0.800 & 1.000 & 0.994 & 0.997 \\ \hline
      E3-ClearScope & 5 & 112 & 2 & 0 & 0.714  & 1.000 & 0.983 & 0.991 \\ \hline
      E5-THEIA & 2 & 173 & 1 & 0 & 0.667 & 1.000 & 0.994 & 0.997 \\ \hline
      E5-CADETS & \textbf{16} & \textbf{238} & \textbf{0} & \textbf{0} & \textbf{1.000} & \textbf{1.000} & \textbf{1.000} & \textbf{1.000} \\ \hline
      E5-ClearScope & 10 & 217 & 5 & 0 & 0.667  & 1.000  & 0.978  & 0.989 \\ \hline
      \change{OpTC} & \change{\textbf{32}} & \change{\textbf{1210}} & \change{\textbf{6}} & \change{\textbf{0}} & \change{\textbf{0.842}} & \change{\textbf{1.000}} & \change{\textbf{0.995}} & \change{\textbf{0.998}} \\
      \hline
      \end{tabular}
  }
  \vspace{-3mm}
  \end{table}

To evaluate \system'
detection performance,
we replay test data
in each dataset
as if \system was monitoring
the behavior of the host system
\change{as it runs.}
Model training is performed offline
using only benign data.
Note that
this experimental setup
automatically ensures
two desiderata of PIDSes
introduced in~\autoref{sec:introduction}:
\emph{scope}
and \emph{attack agnosticity}.

\autoref{tab:evl:exp} shows
the precision, recall, accuracy,
and area under ROC curve (AUC)
results for all datasets.
We compute these metrics
based on \emph{time windows}.
As mentioned in~\autoref{sec:evaluation:datasets},
we manually label
each time window in a provenance graph
as either \emph{benign} or \emph{attack}
according to the ground truth.
If \system marks a \emph{benign} time window
as anomalous
(\ie if \system mistakenly includes a benign time window
in an anomalous queue),
we consider the time window to be a \emph{false positive} (FP).
On the other hand,
if \system correctly marks an \emph{attack} time window
as anomalous,
it is counted as a \emph{true positive} (TP).
\emph{False negatives} (FN) and \emph{true negatives} (TN)
are calculated in a similar fashion.
\autoref{tab:evl:exp} also shows the number of
TP, TN, FP, and FN time windows.

We see in \autoref{tab:evl:exp}
that \system can accurately detect all attacks\remove{in our experiments},
achieving $100\%$ recall.
\remove{However,}%
\system reports \remove{false positives}\change{FPs}
(which lead to lower precision)
in a subset of experiments
for several reasons.
First,
\system continues
to assign high reconstruction errors
to edges
whose nodes \remove{(\ie entities)}%
were under the \change{attacker's} influence
\remove{of the attacker,}%
even \emph{after} the attacker
stops \emph{actively} manipulating
\remove{these entities}\change{them}.
\system still considers
these entities
to be compromised,
because \system
remembers the \emph{history}
of their states (\autoref{sec:framework:learning}),
part of which indeed
involves the attacker.
However,
in the ground truth,
entities that remain active
after the attack
are often dismissed,
since they are no longer
part of the attack.
For example,
in E5-CADETS,
the attacker exploited
a vulnerable \texttt{Nginx} process
to download and execute a malicious payload.
\remove{on the victim machine.}%
Once the payload was executed,
subsequent attack activity
no longer involved \texttt{Nginx},
but \remove{the }\texttt{Nginx} \remove{server }continued to serve benign requests.
Any entity,
once compromised by an attacker,
should be considered~problematic\remove{(\ie its behavior cannot be trusted)}.
We manually
identify these ``fake''
\remove{false positives}\change{FPs}
(\ie processes that potentially remain 
under an attacker's control 
but whose subsequent behavior
is not part of the ground truth),
and we show the adjusted results
in~\autoref{tab:evl:exp_adjusted}.
Notice \remove{that}%
\remove{the results}%
\remove{improve significantly for}%
\change{the significant improvement for}
\change{E3-THEIA, E5-CADETS, and OpTC}.

Second,
\system assigns high reconstruction errors
to \emph{novel} activities
of \emph{new} applications
that were introduced
\emph{only} in the test data.
Since their behavior
is completely unknown,
it is \remove{indeed}abnormal to \system,
albeit non-malicious.
\change{This is an example of
\emph{concept drift}~\cite{tsymbal2004problem},
where new benign behavior does not 
fit into the underlying statistical properties
learned by the model.}
For example,
in E5-ClearScope,
we test \system
\remove{on the time windows}%
on May $15^{th}$
and $17^{th}$, $2019$
when the attack took place.
\system reports FPs on both days.
Upon \remove{further }inspection
of the candidate summary graphs
from \system (\autoref{fig:evaluation:benign_community}),
we easily conclude
that all the FPs are caused by
the behavior of \texttt{screencap},
which does not appear
in the training data.
We next show
how sysadmins
can effectively
mitigate FPs by
incrementally retraining
\system' model.

\begin{table}[t]
	\caption{E5-ClearScope's May ${17}^{th}$ detection performance
	with and without retraining based on May ${15}^{th}$ FPs.}
	\label{tab:evl:retrain}
	\resizebox{\columnwidth}{!}{%
		\small
		\centering
		\begin{tabular}{|l|c|c|c|c|c|c|c|c|}
			\hline
			\textbf{E5-ClearScope} & \textbf{TP} & \textbf{TN} & \textbf{FP} & \textbf{FN} & \textbf{Precision} & \textbf{Recall} & \textbf{Accuracy} & \textbf{AUC} \\ \hline
			Without retraining & 6 & 87 & 2 & 0 & 0.750 & 1.000 & 0.979 & 0.989 \\ \hline
			With retraining & 6 & 89 & 0 & 0 & 1.000 & 1.000 & 1.000 & 1.000  \\ \hline
		\end{tabular}
	}
	\vspace{-4mm}
\end{table}

\noindgras{Model Retrain.}
\system makes model retraining
more practical,
because it enables
sysadmins
to quickly identify
false alarms
by providing them with
compact candidate summary graphs
to inspect (\autoref{sec:evaluation:explainability}).
To \emph{update} the model,
we repeat the same training process
(\autoref{sec:framework:learning})
on the existing model
using only
the provenance data from
the FP time windows.
\autoref{tab:evl:retrain} shows
the experimental results
on E5-ClearScope
before and after
we identify FP time windows
and update the model.
More specifically,
in addition to
the original training data,
we further train the model
on the FP data
on May $15^{th}$,
which was previously used
as part of the test data.
We then evaluate
the updated model
on the May $17^{th}$'s data,
which is the remaining test data.
For fair comparison,
\autoref{tab:evl:retrain}
reports the results before model update
only on May $17^{th}$
(while \autoref{tab:evl:exp_adjusted}
reports the results on both days).
\remove{We see that }\system can continuously learn
from FPs %
to \change{address concept drift and}
\remove{reduce the possibility of}avoid making similar mistakes
in the future.

In practice,
\system' model
should be regularly updated
as new benign behavior emerges.
Note that retraining potentially
breaks the assumptions made in our threat model (\autoref{sec:threat}),
since benign training data \remove{is no longer}\change{may not be}
captured in a controlled environment
where the absence of an attacker is guaranteed.
\remove{This has security implications,}%
\remove{as attackers now have opportunities}%
\change{As such, attackers may}
\remove{to }exploit retraining
to poison the \remove{ML }model~\cite{Zgner2018AdversarialAO,Zugner2019AdversarialAO}. %
Detecting and preventing~model poisoning~\cite{Zhang2020GNNGuardDG}
is \change{further discussed in~\autoref{sec:discussion},
along with other possible evasion strategies,
but it is}
beyond the scope of this work.
In summary, 
while \system supports %
retraining,
we leave its thorough exploration and evaluation %
to future work.

\vspace{-2mm}
\subsection{Comparison Study}
\label{sec:evaluation:compare}
\vspace{-2mm}

Fairly comparing \system
with state-of-the-art PIDSes 
is hard
for several reasons.
First,
the majority of \remove{existing }PIDSes are signature-based~\cite{alsaheel2021atlas,holmes:milajerdi:2019,poirot:milajerdi:2019,rapsheet:hassan:2020,sleuth:hossain:2017},
while \system detects \remove{system }\emph{anomalies}.
The performance of signature-based PIDSes
depends \remove{largely }on the quality of the signatures,
which are often proprietary knowledge
unavailable to the public.
Comparison between signature-
and anomaly-based PIDSes
can easily be biased
by manipulating \remove{the number of }signatures
that can be matched to \change{the} attack. \remove{ activity}%
\remove{in test data.}%
Therefore,
we exclude signature-based PIDSes
for comparison.
Second,
most anomaly-based PIDSes~\cite{sigl:han:2021,Xiong2022ConanAP} %
are closed-source
and evaluated using private datasets.
\remove{While w}\change{W}e attempt to re-implement some PIDSes \remove{ourselves}%
based on published descriptions,
\remove{to the best of our ability,}%
\change{but} it is challenging to verify correctness
with no access to datasets
\remove{for us }to reproduce
the original \remove{experimental }results.
For example,
we re-implemented ProvDetector~\cite{provdetector:wang:2020}
but \remove{we }are unable to
compare it against \system
due to unreasonably long run time
on even the smallest DARPA dataset
(we attribute this outcome to our lack of skills, 
not to the original authors).
Similarly,
ShadeWatcher~\cite{zengy2022shadewatcher}
is not \emph{fully} open-source.
Specifically,
we confirm with the authors
that a major component of ShadeWatcher
\remove{which extracts high-level behavioral interactions}%
\remove{from provenance graphs,}%
is proprietary.
Unfortunately,
we are unable to replicate the algorithm
from the description
in its publication alone.
Last but not least,
PIDSes might use
different metrics to report
their detection performance,
further complicating comparison
and %
giving a misleading impression
of %
performance.
\change{We further discuss the issues of evaluating
PIDSes in general in~\autoref{sec:discussion}}
and leave benchmarking PIDSes to future work.

Due to these difficulties,
we choose Unicorn~\cite{unicorn:han:2020} 
\change{and ThreaTrace~\cite{wang2022threatrace}
as the primary PIDSes
for comparison,
because they are anomaly-based,
open-source,
and evaluated by the authors
using both Manzoor~\etal and 
a subset of DARPA datasets.}
Similarly, %
StreamSpot~\cite{Manzoor2016FastMA}
and Frappuccino~\cite{frappuccino:han:2017}
are also
open-source anomaly detection systems.
\remove{that analyze streaming provenance graphs.}%
However,
\change{Unicorn has been shown to 
outperform these systems~\cite{unicorn:han:2020}.
Our own evaluation of StreamSpot on DARPA's TC datasets
confirmed prior performance analyses by others:
\emph{StreamSpot cannot detect any anomalies
in all TC datasets.}
Therefore,
due to space constraints,
we will not further discuss StreamSpot or Frappuccino.}

\subsubsection{Unicorn}
\label{sec:evaluation:compare:unicorn}
Unicorn builds a behavioral model
of a system
by featurizing an evolving provenance graph
into a series of fixed-size, incrementally-updatable graph sketches.
Each sketch represents a \emph{snapshot}
describing the \emph{entirety}
of the graph from the very beginning of system execution
till the point where the snapshot is taken.
The frequency of generating a new sketch
is a hyperparameter,
determined by the number of new edges
streamed to the graph.
At test time,
Unicorn can quickly generate
and update graph sketches of the system being monitored
and compare them with
known benign sketches in the model
to perform \change{run-time} detection.

We use the same evaluation protocol 
as \remove{in the Unicorn's paper}\change{in Unicorn}~\cite{unicorn:han:2020} 
to ensure fairness.
Specifically,
Unicorn computes evaluation metrics
at the \emph{graph} level,
instead of the finer-grained time-window level
(as in \system).
That is, 
\remove{for a given provenance graph,}%
Unicorn classifies~the \emph{entire} graph
as benign or containing an attack,
and uses it as a single data point
to calculate detection performance.
To adopt Unicorn's way
of computing metrics,
we consider an entire graph
to be malicious
\remove{as long as}\change{if}
\system marks \emph{at least one} time window
\remove{in the graph}%
as \remove{an }attack.\remove{time window.}
We do not need to modify the experimental results
for the Manzoor~\etal dataset,
because each graph is already 
a single time window.
Since Unicorn was not originally evaluated
on the DARPA E5 datasets,
we will not compare Unicorn
on these datasets for fairness
(because extensive hyperparameter tuning
might be needed for Unicorn to produce
the best results).

\begin{table}[t]
\caption{Comparison study between Unicorn and \system.}
\label{tab:evl:cmp_u_and_k}
\resizebox{\columnwidth}{!}{%
  \scriptsize
  \begin{tabular}{|l|l|c|c|c|}
  \hline
  \textbf{Datasets}              & \textbf{System} & \textbf{Precision} & \textbf{Recall} & \textbf{Accuracy} \\ \hline %
  \multirow{2}{*}{Manzoor \etal}  & Unicorn	  & 0.98 			& 0.93			& 0.96	\\ \cline{2-5} %
  							 & \system          & \textbf{1.00}               & \textbf{1.00}            & \textbf{1.00}       \\ \hline %
  \multirow{2}{*}{E3-CADETS}      & Unicorn         & 0.98               & 1.00            & 0.99            \\ \cline{2-5} %
                                 & \system          & \textbf{1.00}               & 1.00            & \textbf{1.00}      \\ \hline %
  \multirow{2}{*}{E3-THEIA}      & Unicorn         & 1.00               & 1.00            & 1.00             \\ \cline{2-5} %
                                 & \system          & 1.00               & 1.00            & 1.00        \\ \hline %
  \multirow{2}{*}{E3-ClearScope} & Unicorn         & 0.98               & 1.00            & 0.98            \\ \cline{2-5} %
                                 & \system          & \textbf{1.00}               & 1.00            & \textbf{1.00}          \\ \hline %
  \end{tabular}
  }
  \vspace{-4mm}
  \end{table}

\noindgras{Experimental Results.}
\autoref{tab:evl:cmp_u_and_k} shows that
\system either outperforms Unicorn
or achieves equally high performance
in all datasets.
By comparing~\autoref{tab:evl:cmp_u_and_k}
with~\autoref{tab:evl:exp_adjusted},
we also see that
a coarse-grained, graph-level evaluation
can be misleading,
since the detection system
might not accurately or completely
identify the entirety of attack activity.

Unlike \system,
Unicorn does not support
fine-grained detection
or automated post-detection investigation.
We notice
a time lag between the first occurrence of a malicious event
and Unicorn's detection of system anomalies,
\remove{This lag }\change{which} results in additional graph sketches.
Since a graph sketch
is a vectorized graph representation
that describes an entire evolving graph,
these additional sketches
could represent tens of
thousands of \emph{more} graph elements \remove{\footnote{Unicorn by default sets}%
\remove{the frequency of new sketch generation to}%
\remove{$3,000$ edges.}%
\remove{Therefore, for example,}%
\remove{additional two sketches means that}%
\remove{sysadmins must inspect \emph{additional} $6,000$ edges.}%
\remove{The amount of time lag varies in different experiments.}}%
that sysadmins must inspect
\emph{on top of} the sketch
that actually contains attack activity.
Consequently,
when Unicorn raises an alarm,
attack activity can be hidden
anywhere within the graph,
requiring sysadmins
to blindly backtrack the graph
to reason about the alarm.
In contrast,
\system not only
produces fewer false alarms,
but also
creates compact \emph{summary graphs}
that highlight possible attack footprints,
all without any human intervention
(\autoref{sec:evaluation:explainability}).

\begin{table}[t]
\caption{Comparison study between ThreaTrace and \system.}
\label{tab:evl:cmp_t_and_k}
\resizebox{\columnwidth}{!}{%
  \scriptsize
  \begin{tabular}{|l|l|c|c|c|}
  \hline
  \textbf{Datasets}              & \textbf{System} & \textbf{Precision} & \textbf{Recall} & \textbf{Accuracy} \\ \hline %
  \multirow{2}{*}{Manzoor \etal}  & ThreaTrace	  & 0.98 			& 0.99			& 0.99	\\ \cline{2-5} %
  							 & \system          & \textbf{1.00}               & \textbf{1.00}            & \textbf{1.00}       \\ \hline %
  \multirow{2}{*}{E3-CADETS}      & ThreaTrace         & 0.90               & 0.99            & \change{0.99}            \\ \cline{2-5} %
                                 & \system          & \change{\textbf{1.00}}               & \change{0.95}            & \change{0.99}      \\ \hline %
  \multirow{2}{*}{E3-THEIA}      & ThreaTrace         & 0.87               & 0.99            & \change{0.99}             \\ \cline{2-5} %
                                 & \system          & \change{\textbf{1.00}}               & \change{0.95}            & \change{0.99}        \\ \hline %
  \multirow{2}{*}{E5-CADETS} & ThreaTrace         & 0.63               & 0.86            & \change{0.97}            \\ \cline{2-5} %
                                 & \system          & \change{\textbf{1.00}}               & \change{0.85}            & \change{\textbf{0.98}}          \\ \hline %
  \multirow{2}{*}{E5-THEIA} & ThreaTrace         & 0.70               & 0.92            & \change{0.99}            \\ \cline{2-5} %
                                 & \system          & \change{\textbf{1.00}}               & \change{0.92}            & \change{0.99}          \\ \hline %
  \end{tabular}
  }
  \vspace{-4mm}
  \end{table}

\vspace{-2mm}
\change{\subsubsection{ThreaTrace}
\label{sec:evaluation:compare:threatrace}
ThreaTrace builds a
model for each type of nodes
in a provenance graph to detect
\emph{anomalous nodes}.
We use both the Manzoor~\etal dataset
and a subset of DARPA datasets
used by ThreaTrace
for fair comparison.
ThreaTrace %
converts 
its node-level detection
to graph-level
for the Manzoor~\etal dataset,
since node-level ground truth is unavailable.
It considers a graph
to be anomalous if the number of
anomalous nodes %
exceeds a predefined threshold.
\system and ThreaTrace are thus
directly comparable
on this dataset.
For DARPA datasets,
we adopt ThreaTrace's
way of computing metrics
and use anomalous nodes
in suspicious time windows
to %
compute precision, recall,
and accuracy.

\noindgras{Experimental Results.}
\autoref{tab:evl:cmp_t_and_k} shows that
\system achieves comparable
performance to ThreaTrace
in all datasets.
We note that ThreaTrace authors
manually label as anomalous
both the nodes in the ground truth
\emph{and their $2$-hop
ancestor and descendant nodes},
even though the neighboring nodes
were \emph{not} involved in an attack.
More concerningly,
benign nodes mistakenly detected
by ThreaTrace as anomalous
are \emph{not}
considered to be FPs
\emph{as long as any of their
$2$-hop neighbors 
are labeled
as anomalous}.
Thus,
a benign node
as far as \emph{$4$ hops away
from a true anomalous node}
in the ground truth
can be misclassified by ThreaTrace
but not reported as a FP.
This %
labeling approach
likely leads to favorable
precision and recall,
but even then,
\system outperforms
ThreaTrace in most cases.
We further discuss the issues of
benchmarking PIDSes in~\autoref{sec:discussion}.

Unlike \system, 
ThreaTrace cannot reconstruct a complete attack story.
While ThreaTrace's node-level detection
can facilitate attack comprehension
to some extent,
this approach is impractical 
when the graph is large and the number of FP nodes is high. 
For example,
ThreaTrace identified
over $63K$ FP nodes
in the E5-THEIA experiment
(after using the aforementioned
labeling strategy),
which would undoubtedly overwhelm 
human analysts.
This limitation
is explicitly recognized
by the ThreaTrace authors.}

\input{hyper-param-figures}

\vspace{-2mm}
\subsection{Hyperparameter Impact on Performance}
\label{sec:evaluation:hyperparameters}
\vspace{-2mm}

In previous sections,
we evaluate \system with
a set of fixed hyperparameters.
\remove{($|\Phi|$ = $16$,
$|\mathbf{s}(v)| = 100$,
$|\mathcal{N}| = 20$,
$|\mathbf{z}| = 200$,
\change{and
$|\mathbf{tw}| = 15$ minutes},
see~\autoref{sec:evaluation}).}%
\remove{In this section,}%
Here,
we vary each \remove{hyperparameter }independently 
and report its impact on 
\emph{detection} and \emph{run-time} performance.
We show detailed results for \remove{the }E3-THEIA \remove{dataset }here
and include results for all \remove{DARPA }TC datasets
in \refappendix{sec:appendix:hyper}
due to space constraints.

\noindgras{Node Embedding Dimension ($|\Phi|$).}
Node embedding encodes
initial node features.
We see in~\autoref{fig:evaluation:node_emb_dim_detection-theia3}
that a relatively small dimension
is sufficient to encode
these features.
A large dimension leads to 
\emph{sparse} features,
which could severely affect 
detection performance
and incur large memory overhead
(\autoref{fig:evaluation:memuse_node_feature_theia3}).
On the other hand,
if $|\Phi|$ is too small,
we instead increase the probability
of hash collision
in hierarchical feature hashing (\autoref{sec:framework:building}).
We find $|\Phi| = 16$
to be the ideal dimension 
across all datasets (\autoref{fig:evaluation:node_feature_detection-auc-all}).

\noindgras{Node State Dimension ($|\mathbf{s}(v)|$).}
A node's state captures
the temporal evolution
of a node's neighborhood
over time.
\remove{As we see i}In~\autoref{fig:evaluation:node_state_detection-theia3},
when $|\mathbf{s}(v)|$ is too small, 
\system has difficulties in
retaining information about past events.
On the other hand,
if $|\mathbf{s}(v)|$ grows too large,
detection performance degrades,
because states might contain
outdated history 
irrelevant to current events~\cite{unicorn:han:2020}.
$|\mathbf{s}(v)|$ also
influences memory overhead (\autoref{fig:evaluation:memuse_node_state_theia3}),
because a state vector is associated
with \emph{each} node.
When $|\mathbf{s}(v)| = 100$,
across \emph{all} TC datasets
(\autoref{fig:evaluation:node_state_detection-auc-all}),
\system can
fully contextualize 
a new event
using a node's past interactions
with other entities,
while incurring small
run-time overhead.

\noindgras{Neighborhood Sampling Size ($|\mathcal{N}|$).}
A node's neighborhood captures 
the structural role of a node, 
so that two nodes 
with a similar neighborhood
likely have the same structural role~\cite{grover2016node2vec}.
Too small of a neighborhood sampling size
makes it difficult for \system
to understand a node's structural role.
However,
as we continue to increase $|\mathcal{N}|$,
detection performance no longer improves (\autoref{fig:evaluation:neighbor_size_detection-theia3}).
This is because the majority of the nodes
in a dataset
have fewer neighboring nodes
than $|\mathcal{N}|$.
For example,
in E3-THEIA,
about $97\%$ of the nodes have a neighborhood size of
$20$ or less.
As such,
increasing $|\mathcal{N}|$ above $20$
has little to no effect.
This also explains
why the additional memory overhead
we incur is not proportional
to the increase in $|\mathcal{N}|$
(\autoref{fig:evaluation:memuse_neighbor_size_theia3}).
We find $|\mathcal{N}| = 20$
to be ideal among all datasets 
(\autoref{fig:evaluation:neighbor_size_detection-auc-all}).

\noindgras{Edge Embedding Dimension ($|\mathbf{z}|$).}
The edge embedding $\mathbf{z}$ encodes
both the state and the structural information 
of the graph
surrounding an edge.
With increasing $|\mathbf{z}|$,
the edge embedding can better retain
temporal and structural information
for the %
decoder
to reconstruct an edge.
However,
an overly large $|\mathbf{z}|$
complicates the model
and affects
\system' generalization capability. %
\autoref{fig:evaluation:emb_dim_detection-theia3} confirms
our hypothesis:
Within a certain range,
increasing $|\mathbf{z}|$
improves \system' detection performance,
until we reach a point
where the performance starts to degrade.
Memory overhead
(\autoref{fig:evaluation:memuse_edgeemb_theia3})
also increases
as $|\mathbf{z}|$ grows,
as expected.
Across all datasets (\autoref{fig:evaluation:emb_dim_detection-auc-all}),
$|\mathbf{z}| = 200$ gives
the best detection performance.

\change{\noindgras{Time Window Length ($|\mathbf{tw}|$).}
The length of a time window determines
the frequency of \system performing
its anomaly detection algorithm (\autoref{sec:framework:detection}).
Generally,
a longer time window
accumulates a larger number of system events.
Since the amount of benign activity overwhelmingly dominates
that of attack activity (\autoref{sec:motivation:challenges}),
a large time window
can make anomaly detection difficult.
As we see in~\autoref{fig:evaluation:time_window_detection-theia3},
while a time window length between 5-30 minutes
has little influence on detection performance,
when $|\mathbf{tw}|$ is too large (60 minutes),
\system generates more \emph{false negatives},
which leads to low recall (and high precision)
and overall low accuracy.
When $|\mathbf{tw}|$ is small (5 minutes),
we see a slight decline in performance,
because a short time window can limit
\system' ability to accurately contextualize
an event.
However,
it is unnecessary to use small time windows
just to improve detection \emph{timeliness},
because APT actors only \emph{slowly}
infiltrate their target systems (\autoref{sec:motivation:challenges}).
For example,
in E3-THEIA,
the attacker performed
two adjacent attack activities
in a kill-chain
almost one hour apart (\autoref{sec:evaluation:datasets:darpa}).
We find $|\mathbf{tw}| = 15$ minutes
to be ideal among all datasets (\autoref{fig:evaluation:time_window_detection-auc-all}).
\autoref{fig:evaluation:memuse_time_window_theia3}
shows that increasing the time window length
only slightly increases
memory overhead,
even when the length is large.
This is because \system
processes a provenance graph
in a \emph{streaming} fashion
and does not keep the entire graph 
in memory.}

Note that
CPU utilization is consistently
less than $1\%$
in all E3-THEIA experiments.
Varying hyperparameter values
only slightly impacts CPU utilization.
\autoref{fig:evapre-90th-cpu-all}
shows the $90^{th}$ percentile CPU utilization
for all TC experiments.

\begin{table}
\caption{Statistics of attack summary graphs.}
\label{tab:evaluation:suspicious_subgraphs}
	\resizebox{\columnwidth}{!}{%
		\begin{tabular}{|l|c|c|c|c|}
			\hline
			\textbf{Dataset} &
			\textbf{\# of Nodes} &
			\textbf{\# of Edges} &
			\textbf{\begin{tabular}[c]{@{}c@{}}\# of Edges in \\ Time Windows\end{tabular}} &
			\textbf{Reduction}  \\
			\hline
			E3-THEIA &
			20 & 31 &
			3,393,536 & 109,469X \\
			\hline
			E3-CADETS &
			18 & 26 &
			115,712 & 4,450X \\
			\hline
			E3-ClearScope &
			10 & 16 &
			210,944 & 13,184X \\
			\hline
			E5-THEIA &
			11 & 17 &
			826,368 & 48,610X \\
			\hline
			E5-CADETS &
			11 & 17 &
			351,232 & 20,661X \\
			\hline
			E5-ClearScope &
			10 & 10 &
			344,064 & 34,406X  \\
			\hline
			\change{OpTC} &
			\change{77} & \change{101} &
			\change{1,065,984} & \change{10,554X} \\
			\hline
		\end{tabular}
	}
\vspace{-4mm}
\end{table}

\vspace{-3mm}
\subsection{\change{Attack Reconstruction}}
\label{sec:evaluation:explainability}
\vspace{-3mm}

\change{The ability
to reconstruct complete but 
concise attack stories} 
is a \emph{first-order} design metric
in \system.
It is particularly important
for anomaly detection systems,
especially the ones (like \system)
that leverage deep learning.
This is because %
\change{attack reconstruction}
(1) establishes trust on the decisions,
(2) facilitates the necessary human-in-the-loop component
in \remove{reasoning about}\change{understanding} system anomalies,
and~(3) expedites the process
of identifying and reducing FPs
(\autoref{sec:evaluation:detection})~\cite{han2021deepaid}.

In the DARPA datasets,
\system is able to reconstruct
the true attack activity
describing the APT,
while reporting
only a couple of benign candidate graphs
(\autoref{sec:framework:investigation}).
\autoref{tab:evaluation:suspicious_subgraphs}
shows the size
of the attack summary graph
that \system generates
from an anomalous time window queue
in each DARPA dataset
\change{(the OpTC dataset contains
three APT scenarios,
while each experiment in TC
contains only one)}.
We see that
candidate summary graphs
are small
(due to graph reduction,
see~\autoref{sec:framework:investigation}).
In fact,
compared to
the size of
the anomalous time window queues
from which
they are generated,
the size of
attack summary graphs
is up to \emph{five orders of magnitude} smaller.
For example,
in E3-THEIA,
\system achieves
$109,469$X edge reduction,
narrowing down
the total number of edges
that require manual inspection
from $3.4$ million
in anomalous time window queues
to only $31$.
This means that
sysadmins can
quickly and easily
reason about
candidate summary graphs,
eliminate the benign ones,
and identify true attack activity.
In the remainder
of this section,
we use an attack
and a benign
summary graph
to illustrate how
\system' ability
to \remove{provide as explanation}\change{construct}
concise graphs
enables effective and efficient attack investigation.
\remove{\refappendix{sec:appendix:usecase} and \refappendix{sec:appendix:candidates} have more candidate graph examples 
from each DARPA dataset
in our experiment.}
Due to space constraints,
we provide full graph results
in~\refappendix{sec:appendix:attack_examples},
\refappendix{sec:appendix:benign_examples},
and a separate document~\cite{kairos-supp}.

\noindgras{The Attack Summary Graph.}
\autoref{fig:motivation:theia-gt}
shows a candidate summary graph
from \system
that describes
APT activity
in \remove{the }E3-THEIA\remove{ experiment}.
This graph
and DARPA's ground truth
match almost perfectly,
even though
a small number of
perhaps extraneous graph elements
\remove{which are }not mentioned
in the ground truth
(colored blue\remove{ in~\autoref{fig:motivation:theia-gt}})
are \remove{also }included
in the graph.
However,
notice that these graph elements
are closely connected
to system entities that are indeed
under the influence of the attacker.

\system also misses
several entities
(colored in pink and dashed)
explicitly mentioned in the ground truth.
For example,
the socket nodes and the edges
describing the communications
between the compromised \texttt{Firefox}
process and
two malicious IP addresses
are not included.
This is because
in general,
it is common for
a \texttt{Firefox} process
to read from and write to
an external IP.
As such,
it is difficult to classify
those behaviors
without providing
\eg a complete allowlist/blocklist.
However,
\system accurately identifies
\texttt{Firefox}'s \emph{anomalous behavior}
(colored red\remove{in~\autoref{fig:motivation:theia-gt}})
as a result of
these communications
with the malicious IPs.
Therefore,
sysadmins familiar with the system environment
can easily verify
the presence and the progression
of an attack,
even without the missing components.
\change{Note that %
graph reduction (\autoref{sec:framework:investigation})
does not lead to missing entities;
instead, these are the result of low $\mathrm{RE}$s
during anomaly~detection.}

\begin{figure}
	\centering
	\includegraphics[width=\linewidth]{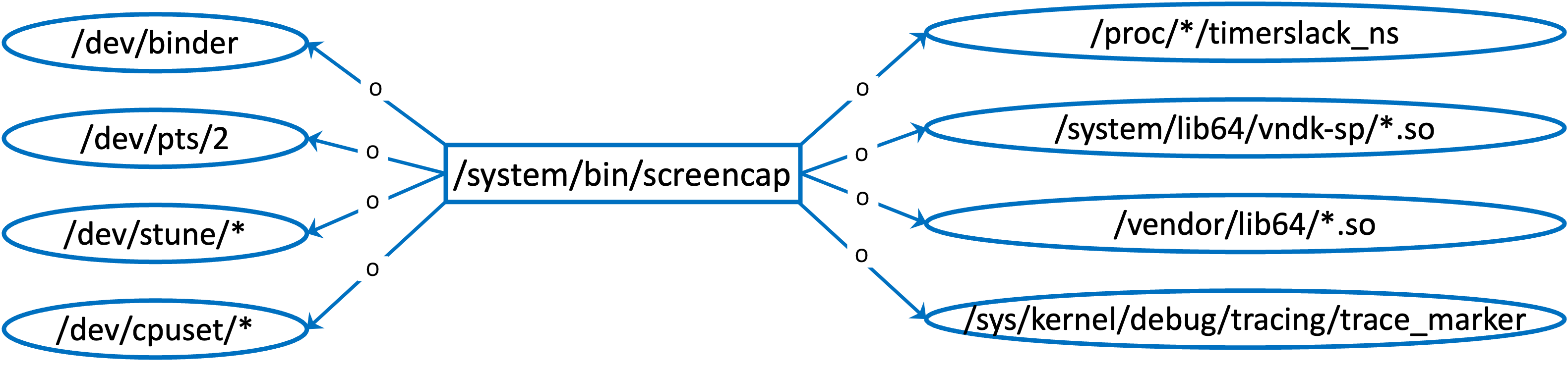}
	\caption{A benign candidate summary graph.}
	\label{fig:evaluation:benign_community}
\vspace{-4mm}
\end{figure}

\noindgras{The Benign Summary Graph.}
\autoref{fig:evaluation:benign_community} shows
a benign candidate summary graph.
Unlike attack graphs,
benign graphs from \system
typically have one or two process nodes
that are \emph{hubs},
forming small ``communities''
with well-defined behavioral boundaries
from other processes.
These graphs
are isolated by \system,
often because they represent
benign but uncommon activity.
\remove{on the system.}%
For example,
as discussed in~\autoref{sec:evaluation:detection},
\texttt{screencap}
in \autoref{fig:evaluation:benign_community}
never appeared in training,
thus resulting in
relatively high reconstruction errors
among its edges.
However,
it is easy for sysadmins
to quickly disregard
this benign candidate 
summary graph,
because
it is small and well-structured.

\input{runtime-figures}

\vspace{-2mm}
\subsection{End-to-end Performance}
\label{sec:evaluation:performance}
\vspace{-2mm}

\system processes a streaming provenance graph
at regular intervals of a time window
and raises an intrusion alert 
when an anomalous time window queue 
is detected (\autoref{sec:framework}).
We show in \autoref{fig:evaluation:timewindow_cost_time} 
the time it takes for \system 
to process $15$-minute time windows 
on \remove{the }E3-THEIA. \remove{dataset.}%
Throughout this experiment,
\system takes \emph{at most} $228.8$ seconds
(or $25.4\%$ of 15 minutes)
to process a single time window 
(which contains about %
$2.5M$ edges), %
well below the duration of a time window.
The \emph{median} size
of time windows in our dataset
has $57K$ edges, %
which takes only $11.6$ seconds
to compute.
We cannot clearly
show the execution time of
many time windows
in~\autoref{fig:evaluation:timewindow_cost_time},
because it takes
only about \emph{one} second (or less)
to compute each,
as they contain fewer than $10K$ edges.
\system' computational cost 
is proportional 
to the number of graph elements in a time window.
\autoref{tab:evaluation:edge_process_cost_time} summarizes
time window execution times
for all DARPA datasets.
\change{Compared to StreamSpot~\cite{Manzoor2016FastMA},
which processes around $14K$ edges per second,
\system incurs slightly higher latency,
processing about $11K$ edges per second.
However,
\system significantly outperforms StreamSpot 
in detection accuracy (\autoref{sec:evaluation:compare}).
Moreover, 
prior work~\cite{unicorn:han:2020} has shown that
a provenance capture system typically generates
fewer than $10K$ edges per second,
even when the host system is busy.
As such,
\system can just as easily 
process a streaming provenance graph
without ``falling behind''.
Furthermore,
as we discuss
in~\autoref{sec:evaluation:hyperparameters},
\system' time window approach,
similar to batch processing 
implemented in Unicorn~\cite{unicorn:han:2020}
and ThreaTrace~\cite{wang2022threatrace},
\remove{and in fact StreamSpot~\cite{Manzoor2016FastMA},}%
does not affect its detection timeliness.}
\remove{We note that \change{OpTC and }THEIA graphs }%
\remove{are significantly larger }%
\remove{than the graphs }%
\remove{generated by the other systems.}%
\remove{For example,}%
\remove{E5-THEIA captures }%
\remove{approximately $7.6$ times more nodes}%
\remove{than E5-CADETS}%
\remove{and $9.1$ times more than E5-ClearScope}%
\remove{(\autoref{tab:evaluation:datasets}).}%
\remove{In sum,}%
\remove{we demonstrate that}%
\change{Therefore,}
\system can effectively monitor 
a host system
at run time.

%% file: hyper-param-figures.tex
\begin{figure*}[htbp]
  \centering
  \subfigure[%
  $|\Phi|$]{
    \begin{minipage}[t]{0.163\paperwidth}
    \centering
      \label{fig:evaluation:node_emb_dim_detection-theia3}
      \includegraphics[width=0.163\paperwidth]{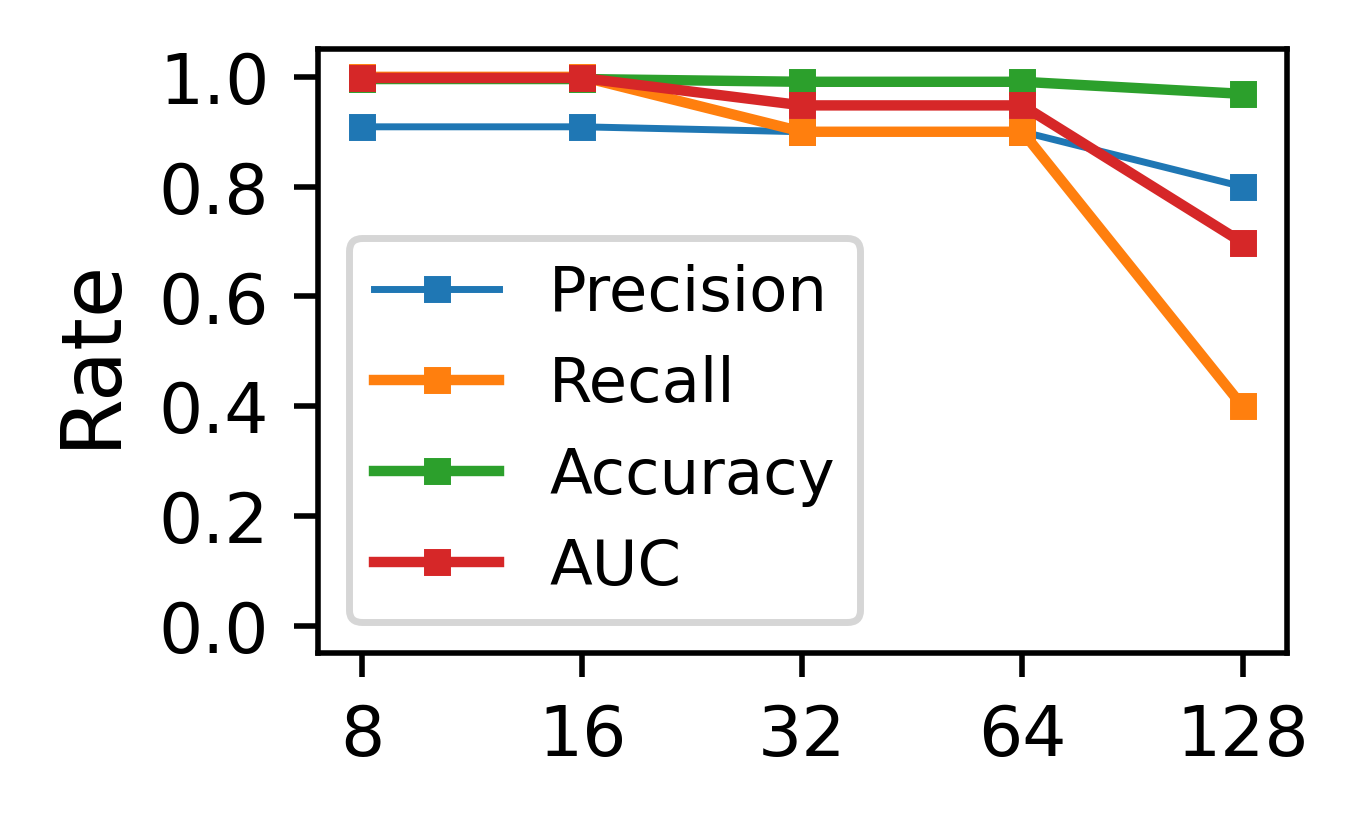}
    \end{minipage}
    }%
  \subfigure[%
  $|\mathbf{s}(v)|$]{
  \begin{minipage}[t]{0.15\paperwidth}
  \centering
		\label{fig:evaluation:node_state_detection-theia3}
		\includegraphics[width=0.15\paperwidth]{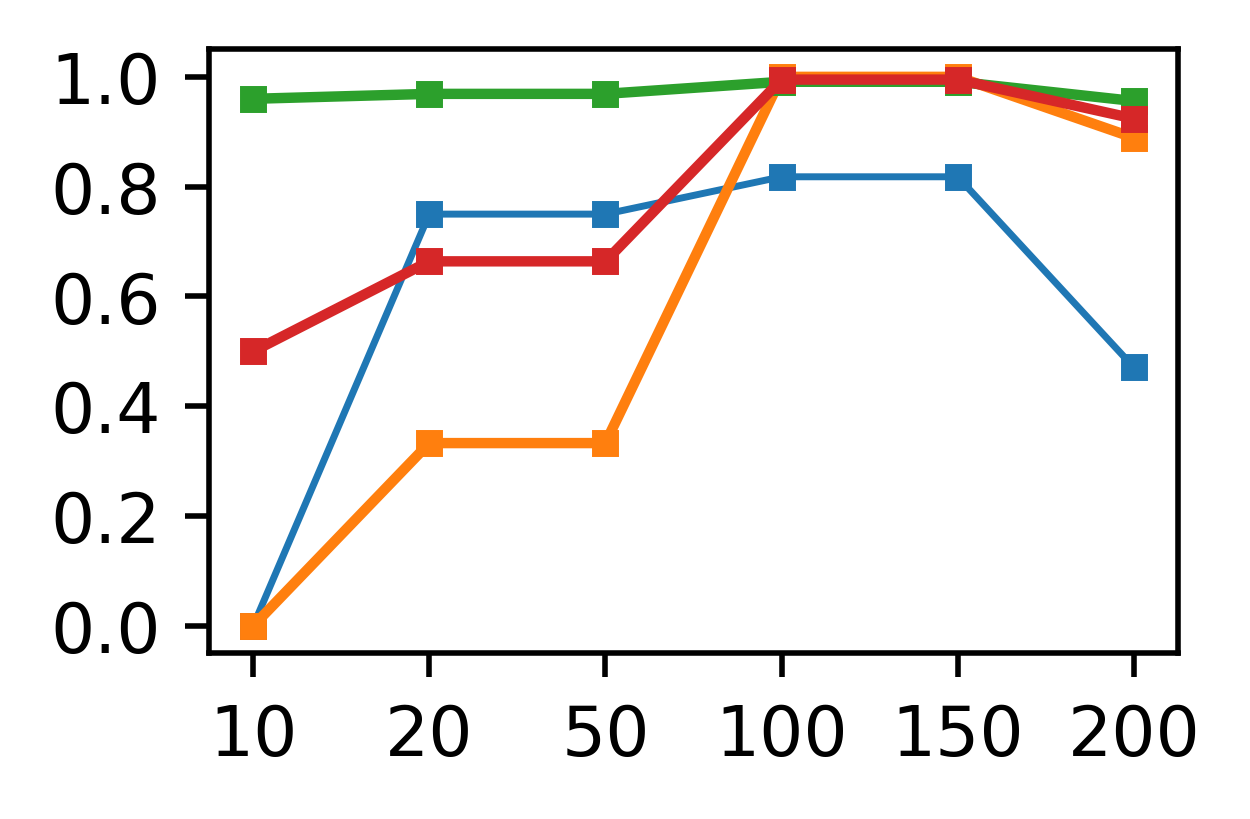}
  \end{minipage}%
  }%
  \subfigure[%
  $|\mathcal{N}|$]{
  \begin{minipage}[t]{0.15\paperwidth}
  \centering
		\label{fig:evaluation:neighbor_size_detection-theia3}
		\includegraphics[width=0.15\paperwidth]{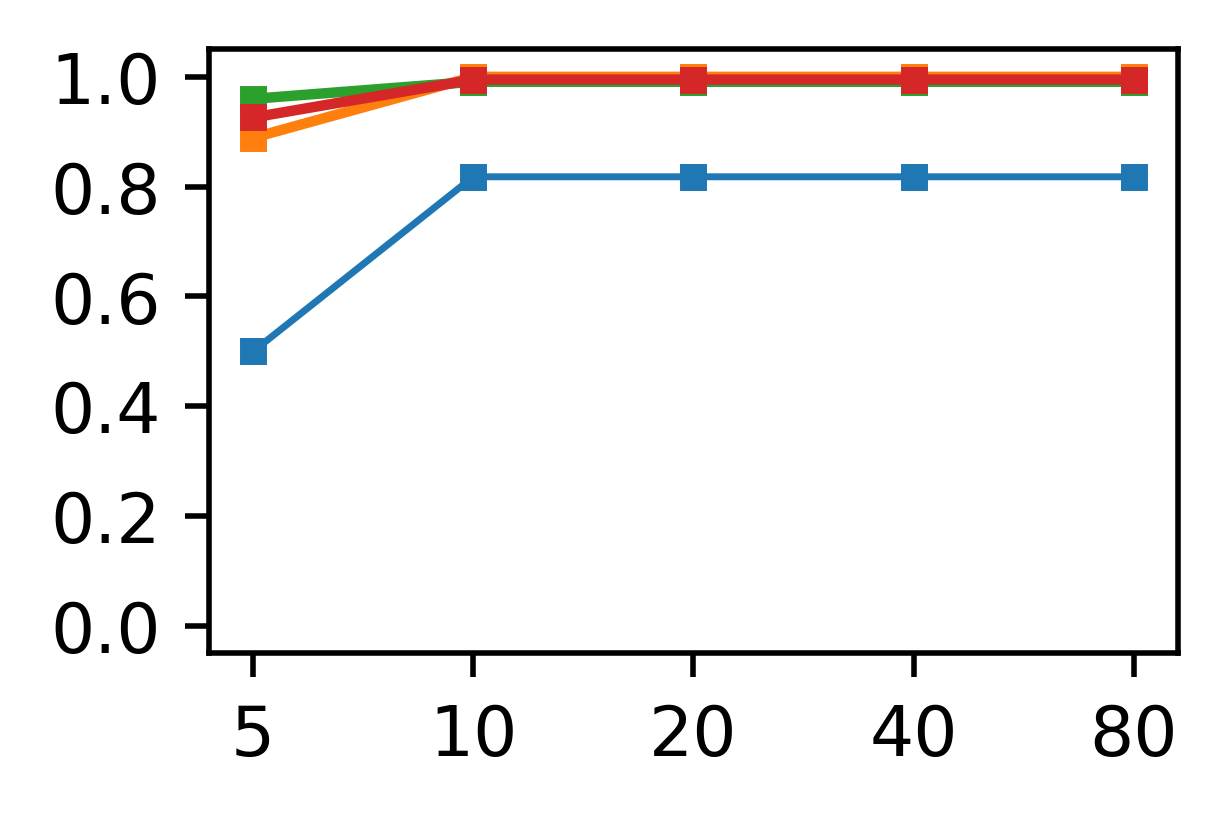}
  \end{minipage}%
  }%
  \subfigure[%
  $|\mathbf{z}|$]{
  \begin{minipage}[t]{0.15\paperwidth}
  \centering
		\label{fig:evaluation:emb_dim_detection-theia3}
		\includegraphics[width=0.15\paperwidth]{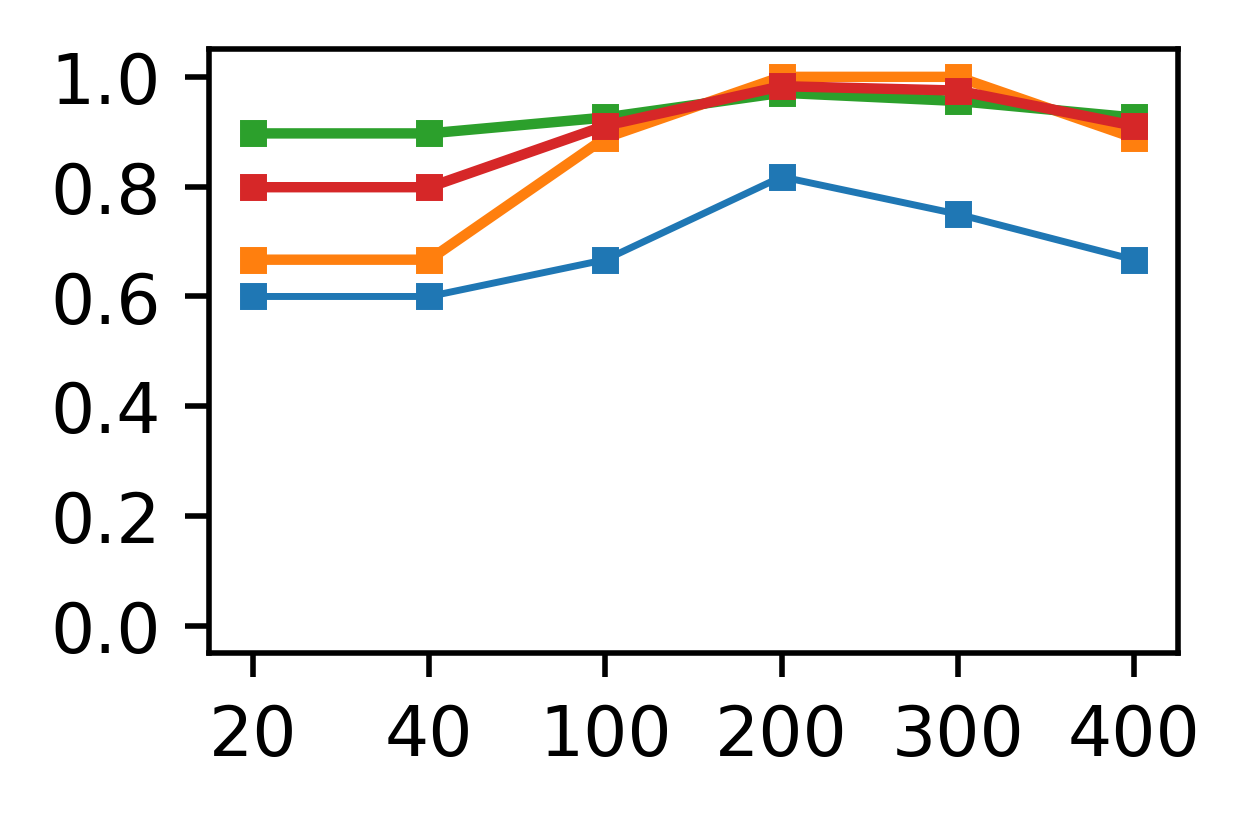}
  \end{minipage}
  }%
  \subfigure[%
  $|\mathbf{tw}|$]{
  \begin{minipage}[t]{0.15\paperwidth}
  \centering
		\label{fig:evaluation:time_window_detection-theia3}
		\includegraphics[width=0.15\paperwidth]{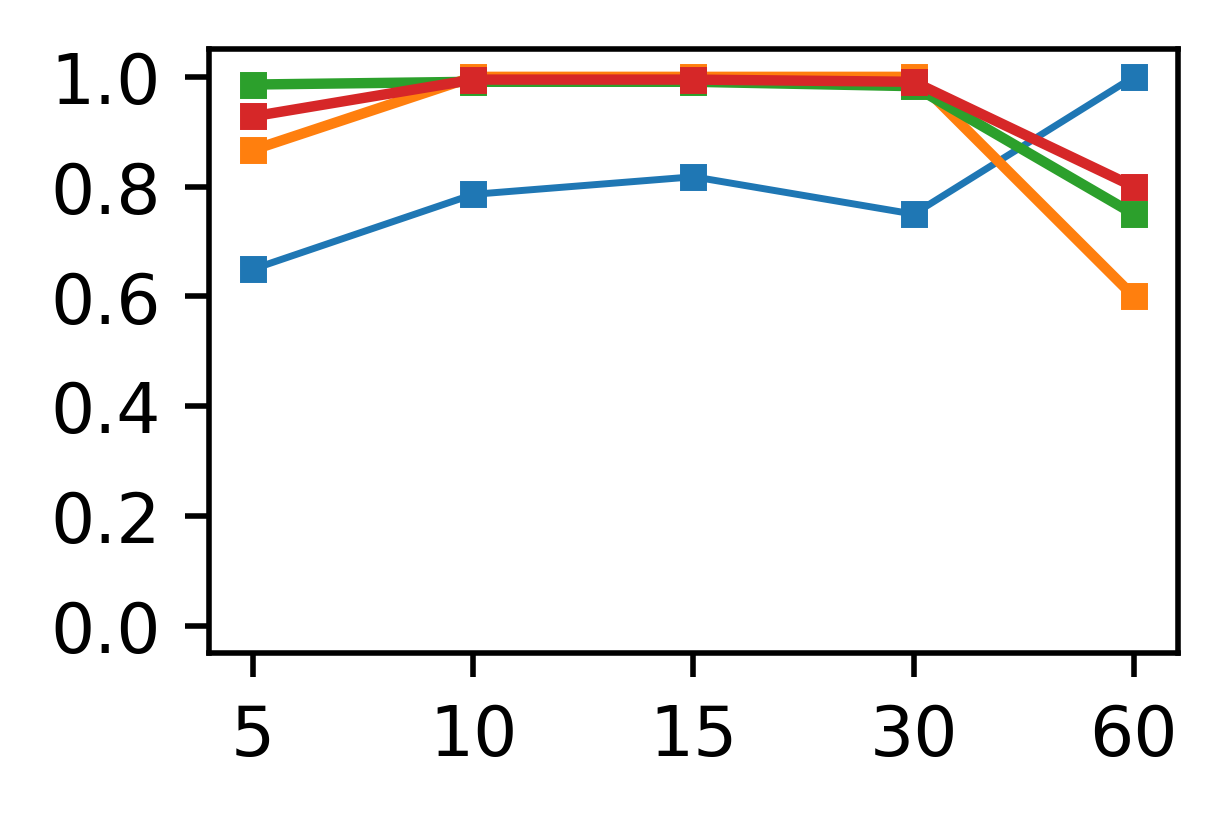}
  \end{minipage}
  }%
  \centering
	\caption{Detection performance (precision, recall, accuracy, and AUC) 
	on E3-THEIA.
	We vary one hyperparameter and fix the others.}
	\label{fig:evapre-theia3}
  \vspace{-4mm}
  \end{figure*}

\begin{figure*}[t]
  \centering
  \subfigure[%
  $|\Phi|$]{
    \begin{minipage}[t]{0.156\paperwidth}
    \centering
      \label{fig:evaluation:memuse_node_feature_theia3}
      \includegraphics[width=0.156\paperwidth]{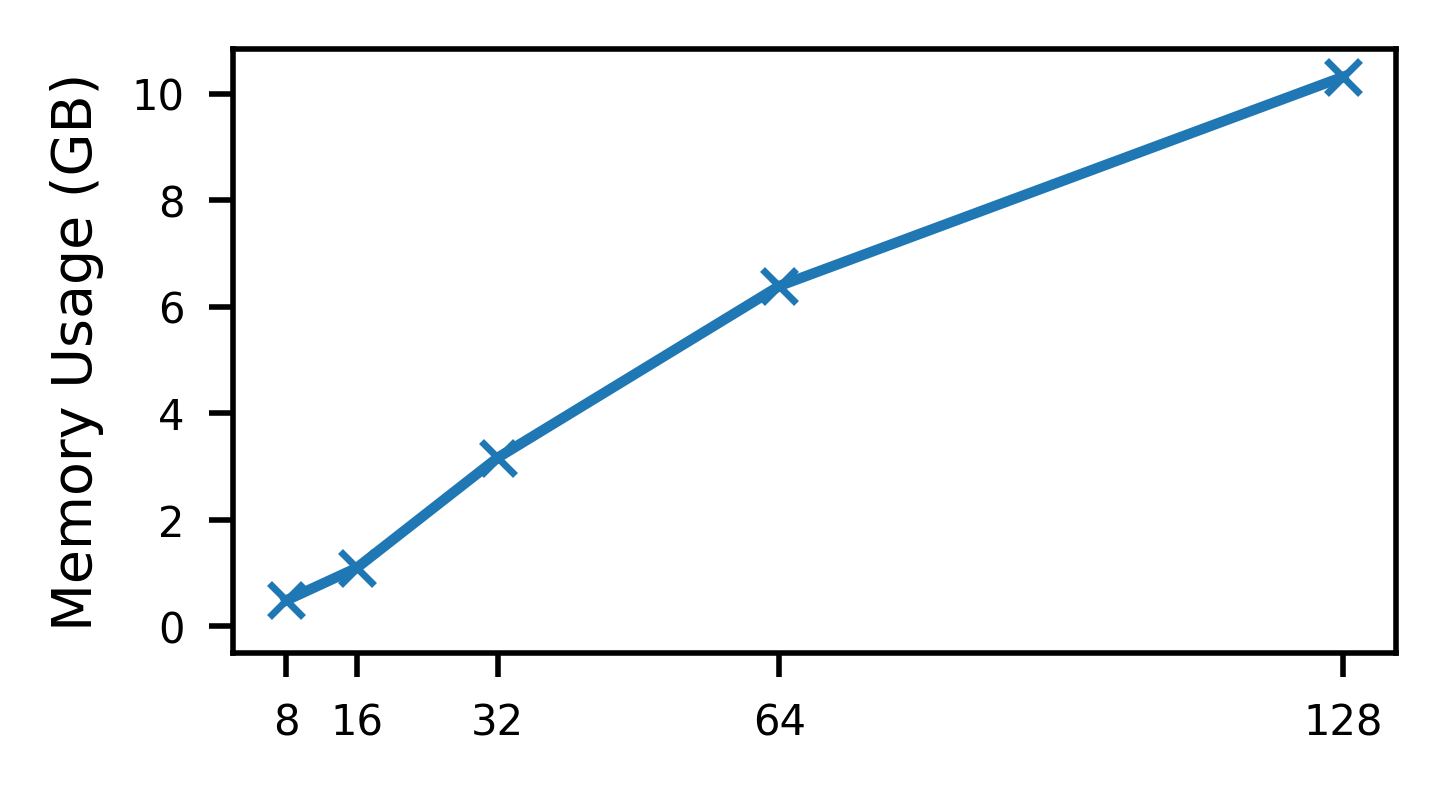}
    \end{minipage}
    }%
  \subfigure[%
  $|\mathbf{s}(v)|$]{
  \begin{minipage}[t]{0.15\paperwidth}
  \centering
    \label{fig:evaluation:memuse_node_state_theia3}
    \includegraphics[width=0.15\paperwidth]{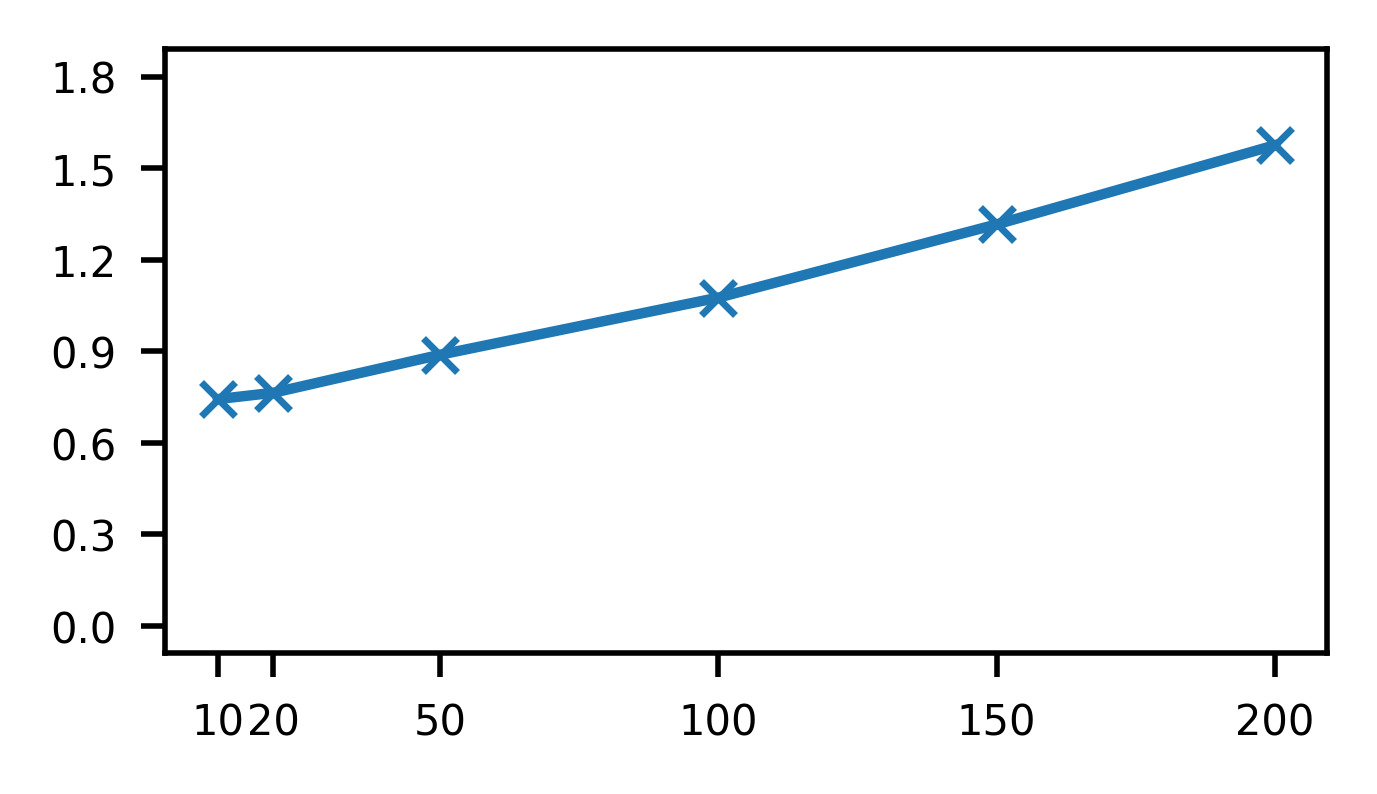}
  \end{minipage}%
  }%
  \subfigure[%
  $|\mathcal{N}|$]{
  \begin{minipage}[t]{0.15\paperwidth}
  \centering
    \label{fig:evaluation:memuse_neighbor_size_theia3}
    \includegraphics[width=0.15\paperwidth]{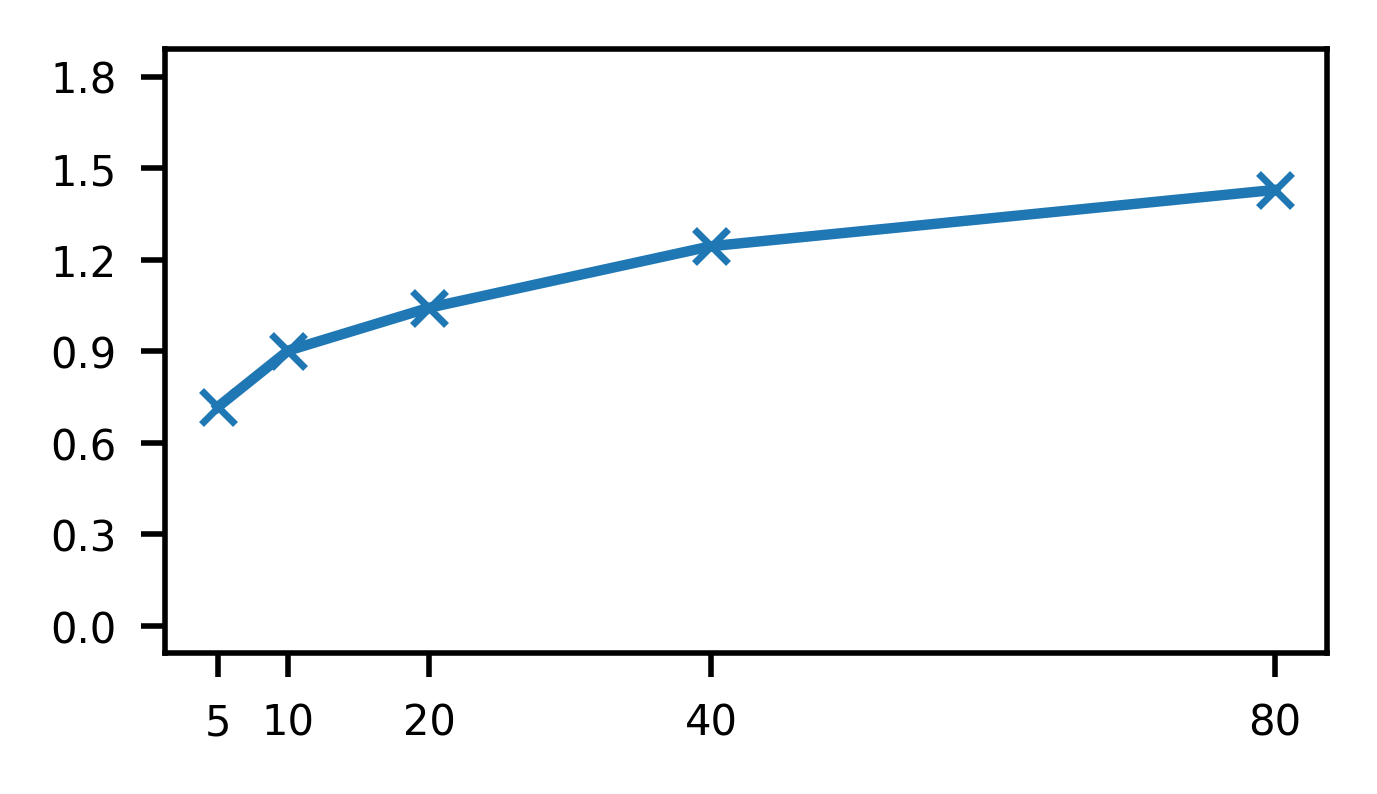}
  \end{minipage}%
  }%
  \subfigure[%
  $|\mathbf{z}|$]{
  \begin{minipage}[t]{0.15\paperwidth}
  \centering
    \label{fig:evaluation:memuse_edgeemb_theia3}
    \includegraphics[width=0.15\paperwidth]{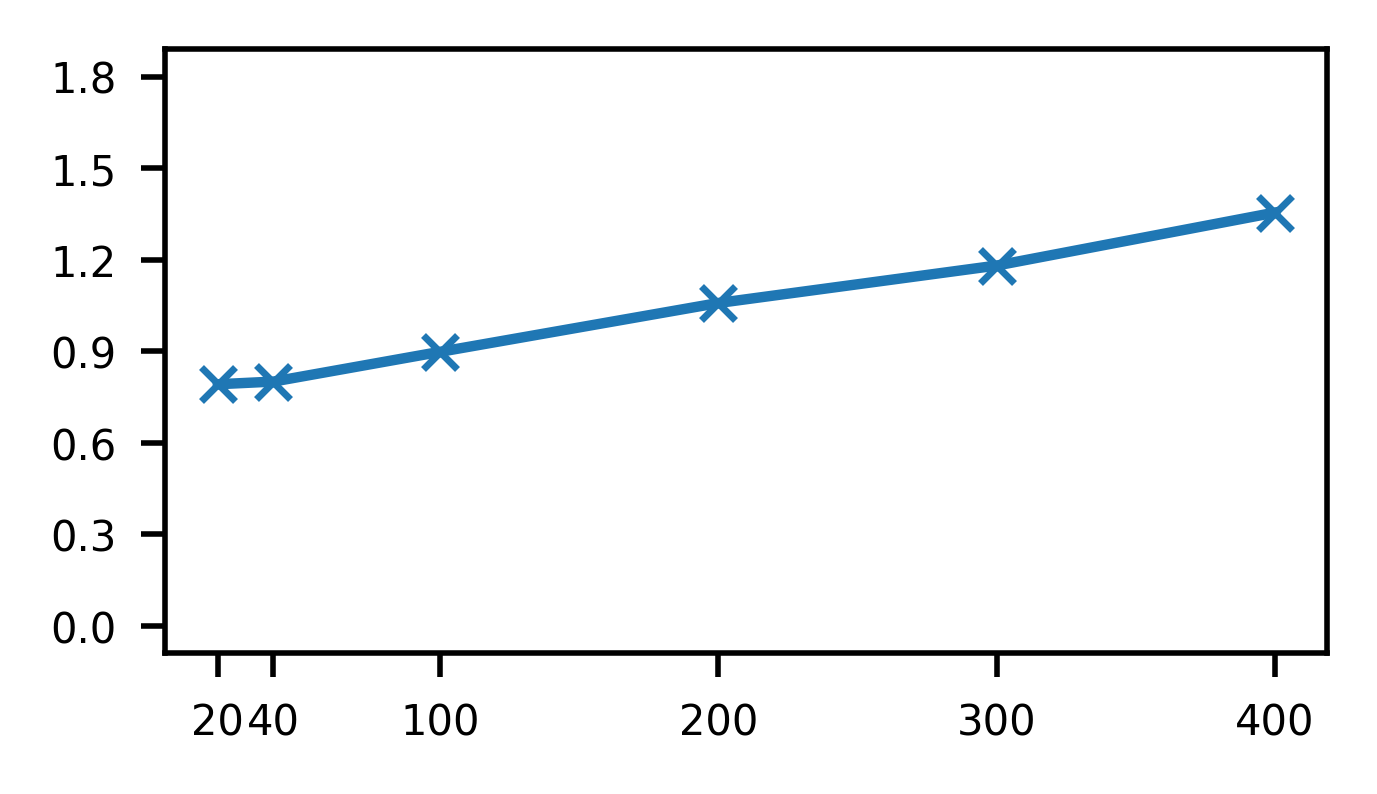}
  \end{minipage}
  }%
  \subfigure[%
  $|\mathbf{tw}|$]{
  \begin{minipage}[t]{0.15\paperwidth}
  \centering
    \label{fig:evaluation:memuse_time_window_theia3}
    \includegraphics[width=0.15\paperwidth]{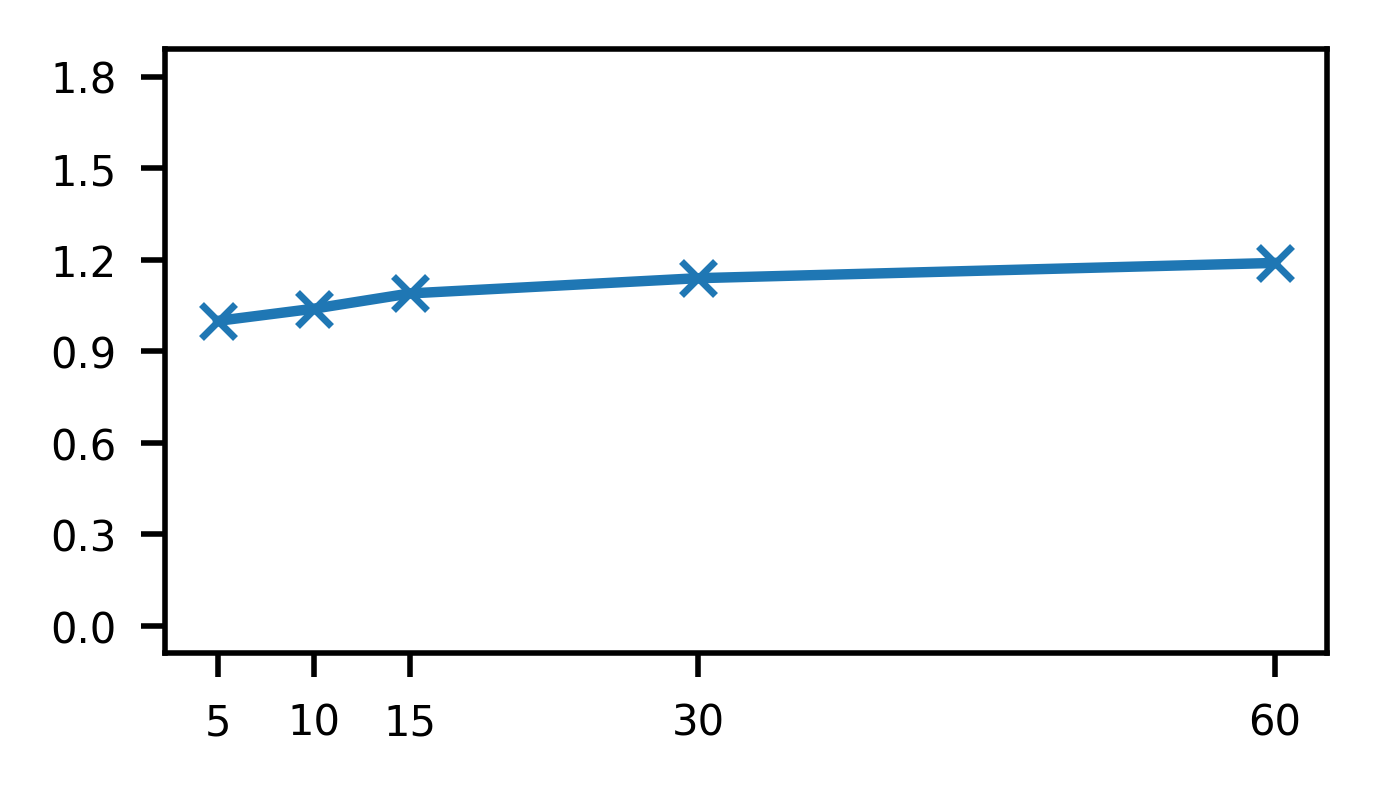}
  \end{minipage}
  }%
  \centering
	\caption{Average memory usage
		on E3-THEIA with varying hyperparameter values.
		We vary one hyperparameter and fix the others.}
	\label{fig:memuse-theia3}
\vspace{-4mm}
\end{figure*}

%% file: runtime-figures.tex
\begin{figure}[t]
	\centering
	\includegraphics[width=\linewidth]{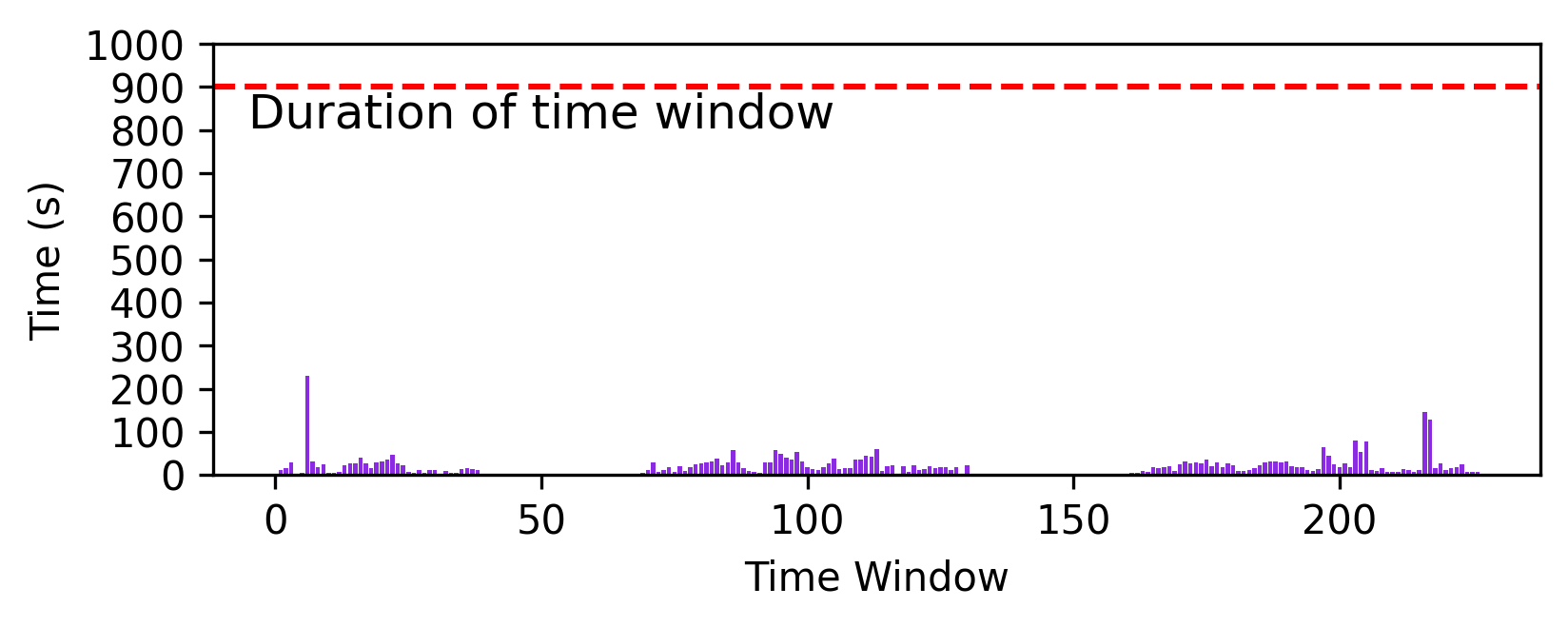}
	\caption{End-to-end time window performance in E3-THEIA.
	Each bar represents the time it takes
	to process the graph in a single time window.}
	\label{fig:evaluation:timewindow_cost_time}
\vspace{-3.5mm}
\end{figure}

\begin{table}[t]
\caption{Summary of execution time.}
\label{tab:evaluation:edge_process_cost_time}
\resizebox{\columnwidth}{!}{%
  \begin{tabular}{|l|l|l|l|l|}
    \hline
    \textbf{Dataset} & \textbf{Min (s)} & \textbf{Median (s)} & \textbf{$90^{th}$ Percentile (s)} & \textbf{Max (s)} \\
    \hline
    DARPA-E3-THEIA & 0.9  & 12.1 & 35.2 & 228.8 \\
    DARPA-E3-CADETS & 1.3 & 2.5 & 4.9 & 19.7 \\
    DARPA-E3-ClearScope  & 0.1 & 3.8 & 4.9 & 19.7 \\
    DARPA-E5-THEIA & 3.6 & 38.3 & 124.7 & 376.2 \\
    DARPA-E5-CADETS & 0.5 & 7.2 & 12.1 & 16.3 \\
    DARPA-E5-ClearScope & 0.1 & 8.2 & 40.1 & 68.8 \\ 
    \change{DARPA-OpTC} & \change{3.7} & \change{19.8} & \change{35.7} & \change{111.7} \\
    \hline
  \end{tabular}
}
\vspace{-5mm}
\end{table}

%% file: discussion.tex
\vspace{-2mm}
\noindgras{Data Poisoning.}
If attackers can poison training data
to include malicious activity,
which is then learned by a machine learning model,
future attacks will remain undetected.
Data poisoning poses a serious threat
to \emph{all} anomaly-based intrusion detection systems.
To the best of our knowledge,
SIGL~\cite{sigl:han:2021} is the only PIDS 
that has evaluated its robustness 
against data poisoning in depth,
but its detection algorithm
works at a much smaller scale.
Others,
unfortunately,
have much more limited evaluation,
if at all.
For example, 
ShadeWatcher~\cite{zengy2022shadewatcher},
which has a system-wide scope equivalent to that of \system,
attempts to evaluate data poisoning using the DARPA datasets.
The authors use one day of the attack data
during training 
and show that ShadeWatcher can detect an attack 
on the second day.
We perform similar evaluation 
and obtain equally good results
showing that detection performance
is barely affected.
However, 
such evaluation is misleading,
because attack activities performed on one day 
will be different on another day.
\remove{A real attacker would use a different}%
\remove{strategy}%
\remove{to evade detection.}%
To properly evaluate robustness, 
we need carefully-crafted, open-source datasets.
Creating such datasets is beyond the scope of this paper.

\noindgras{Evasion.}
An adversarial attacker
with some knowledge of an intrusion detection system
can introduce noise 
\change{or mimic benign system behavior}
during an attack
to mislead the detection system.
While evasion attacks~\cite{Zgner2018AdversarialAO,Zugner2019AdversarialAO,wang2019attacking},
\change{such as mimicry attacks~\cite{DBLP:conf/ccs/WagnerS02},}
are a threat
to all PIDSes,
\change{evading deep graph learning based
systems like \system
is nontrivial.
\system differentiates 
between benign and malicious activity
based on both structural and temporal interactions
between system entities.
Therefore,
to mimic benign behavior,
the attacker must carefully orchestrate
attack activity such that
a malicious process interacts
with a similar set of system objects
in a similar sequential order
while ensuring that 
the actual attack logic remains unchanged.
This requires the attacker to have 
a great knowledge of 
the target system's
benign behavior
and likely the inner workings 
of the trained model.
Even then,
prior work~\cite{sigl:han:2021} has shown
that existing adversarial attacks on graphs
cannot evade PIDSes,
because provenance graphs
have more structural and temporal constraints
than other types of graphs
(\eg social \remove{and citation }networks).
We further evaluated \system
using an evasion dataset based on
DARPA's E3-THEIA 
published by a recent robustness study~\cite{goyalsometimes}.
\system detected the~camouflaged attack,
but in the summary graph (\autoref{sec:framework:investigation}),
it reported only a small subset 
of the attack activity
described in DARPA's ground truth.
While we at first suspected that
the evasion approach proposed %
in the study was to some extent effective,
upon further inspection,
we discovered that
the published dataset
contains only the attack behavior
identified by \system,
rather than the full attack traces included
in the original DARPA dataset.
\remove{Unfortunately,
we were unable to
generate the entire evasion dataset ourselves
based solely on the description in the paper.}%
This discovery
highlights the importance 
of a meaningful intrusion report;
a tool like \system would have helped 
the authors remedy this issue. %

\noindgras{Limitations of Evaluation.}
We identify three major issues 
in PIDS evaluation \remove{that plague
the current provenance-based intrusion detection research}%
in general.
First, 
there lacks open-source implementation
of published PIDSes for comparison.
Second,
only limited publicly-accessible
datasets exist,
and almost all of them are poorly documented.
It is thus difficult to identify any bias
in datasets that might produce misleading results.
Third,
no single performance metric exists
to ensure meaningful comparison.
PIDSes differ in their detection granularity;
\remove{(\eg node-, edge-, path-, and graph-level detection).}%
\remove{As we see }in~\autoref{sec:evaluation:compare},
ad-hoc conversion for the sake of comparison
\remove{or customized calculation of common metrics}%
\remove{can }inadvertently introduces biases. %
These issues weaken 
the conclusion of our own evaluation 
(and potentially that of others) and significantly hinder
independent reproduction of results.
We encourage the community 
to advocate public releases of software artifacts 
and datasets.
}

%% file: rw.tex
\vspace{-2mm}
Historically,
PIDSes have had to
make trade-offs along four dimensions:
scope, attack agnosticity, timeliness, and %
\change{attack reconstruction}
(\autoref{sec:introduction}).
\system is the first
to reconcile these dimensions
while providing
comparable, if not superior,
detection performance.
It is also the first
to efficiently integrate
the \emph{reduction},
\emph{detection},
and \emph{investigation} layers
of the system auditing stack~\cite{inamsok}
with minimal %
overhead.

\noindgras{Provenance-based Intrusion Detection.}
A number of prior PIDSes
have used signature-based techniques
to match \emph{known} attack behavior
in provenance graphs~\cite{sleuth:hossain:2017,nodoze:hassan:2019,holmes:milajerdi:2019,poirot:milajerdi:2019,rapsheet:hassan:2020}.
However,
these approaches are \emph{not attack agnostic} and
therefore have difficulties in detecting unknown attacks.
Other approaches\remove{~\cite{Manzoor2016FastMA,frappuccino:han:2017,priotracker:liu:2018,winnower:hassan:2018,unicorn:han:2020,provdetector:wang:2020,sigl:han:2021,zengy2022shadewatcher}}
leverage anomaly-based detection techniques,
but they either
(1) fail to scale to the entire system~\cite{winnower:hassan:2018,sigl:han:2021},
(2) %
\change{cannot reconstruct attack stories}~\cite{Manzoor2016FastMA,frappuccino:han:2017,unicorn:han:2020,provdetector:wang:2020,zengy2022shadewatcher},
and/or (3) require offline analysis~\cite{priotracker:liu:2018}.
\system overcomes all these limitations simultaneously,
while achieving similar or better
detection and computational performance.
\remove{It is worth noting that
we cannot provide direct comparison
to many of the prior systems
due to the lack of open-source implementations
and insufficient details to reproduce results.}

\noindgras{Provenance-based Investigation.} %
PIDSes~\cite{holmes:milajerdi:2019, rapsheet:hassan:2020}
have often relied on
known attack signatures
to provide %
\change{attack attribution}
for detection.
Prior anomaly-based PIDSes~\cite{camflow:pasquier:2017,frappuccino:han:2017,Manzoor2016FastMA}
require sysadmins
to manually inspect large anomalous graphs,
thus difficult to use in practice.
Recently,
ShadeWatcher~\cite{zengy2022shadewatcher}
\change{and ThreaTrace~\cite{wang2022threatrace}}
take a step in the right direction,
identifying \emph{individual} anomalies
at the node level.
However,
unlike \system,
\change{they fail to}
reconstruct complete and coherent attack stories
but merely provide
a starting point
for sysadmins
to sift through a large amount of data.
\change{Similarly but perhaps more problematically,
SIGL~\cite{sigl:han:2021} not only 
identifies just anomalous \emph{nodes},
but also has limited \emph{scalability},
which makes it unsuitable to analyze 
provenance graphs of a whole-system scope
to detect advanced attacks.}
Recently, Yang \etal~\cite{yangprographer}
proposed ProGrapher that,
similar to Unicorn,
detects anomalies at the \emph{graph} level. 
To support finer-grained attack investigation,
it ranks graph nodes
based on their degrees of anomalousness,
which is similar to SIGL~\cite{sigl:han:2021}.
Therefore,
post-detection investigation remains labor-intensive.
Note that ProGrapher is closed-source
and reports worse overall detection accuracy than \system.
DepComm~\cite{xu2022depcomm}
partitions a provenance graph
into process-centric communities
based on pre-defined random walk schemes
and extracts for each community
\emph{paths} that describes how
information flows through it.
While paths provide more useful context
for attack investigation than nodes,
DepComm requires point-of-interest events
or attack signatures
from an IDS (\eg Holmes~\cite{holmes:milajerdi:2019})
to reconstruct an attack story.

\noindgras{Provenance Reduction.}
Different techniques~\cite{xu2016high, hossain2018dependence, winnower:hassan:2018}
\remove{(\eg causality preserving reduction (CPR)~\cite{xu2016high},
dependency preserved reduction~\cite{hossain2018dependence},
and deterministic finite automaton induction~\cite{winnower:hassan:2018})}%
have been proposed to reduce the size of provenance graphs.
Reduction is performed
either before intrusion detection
or during attack investigation
to reduce computational and memory overhead~\cite{inamsok}.
For example, 
ShadeWatcher~\cite{zengy2022shadewatcher} performs causality preserving reduction~\cite{xu2016high}
before intrusion detection.
In contrast,
\change{\system leverages reduction techniques
\emph{post-detection}
only to minimize a sysadmin's mental load
but performs detection efficiently at scale
on the \emph{entire} graph.
Thus,
\emph{\system' graph reduction
does not affect detection.}}
\remove{Moreover,}%
\remove{the entire graph is still available }%
\remove{for post-detection, in-depth forensic analysis, }%
\remove{if such a need arises.}%

%% file: conclusion.tex
\vspace{-2mm}
\remove{We present }\system is
the first provenance-based intrusion detection system 
that detects system-wide anomalies
\emph{and} generates succinct attack graphs
to \change{describe} \remove{detected anomalies}them
without \remove{any a priori knowledge 
of attack characteristics.}%
prior attack knowledge.
Our evaluation demonstrates that
\system can effectively monitor
long-running systems at run time,
outperforms the state-of-the-art,
and incurs minimal performance overhead.

%% file: ack.tex
We thank S\&P 2023 and 2024 anonymous reviewers for their insightful
comments.
We acknowledge the support of the Natural Sciences and Engineering Research Council of Canada (NSERC). Nous remercions le Conseil de
recherches en sciences naturelles et en génie du Canada (CRSNG) de son soutien.
This work was partially supported by research funding from the National Research Council Canada (NRC).
This material is based upon work supported 
by the U.S. National Science Foundation under Grant CNS-2245442.
Any opinions, findings, and conclusions or recommendations 
expressed in this material are those of the author(s) 
and do not necessarily reflect the views of the National Science Foundation.

%% file: appendix_darpa_data.tex
\vspace{-10pt}
\autoref{tab:evaluation:tc:scenarios}
summarizes the attack scenarios in the DAPRA datasets.
We describe each attack scenario
in detail
\remove{in~\autoref{sec:appendix:usecase}.}%
in a separate document~\cite{kairos-supp}.
\autoref{tab:evaluation:tc:datasplit}
summarizes the specific \remove{benign and
attack}%
data we use from the datasets
for training, validation, and detection.
Similar to prior work~\cite{zengy2022shadewatcher},
we also perform \emph{noise reduction}
and define an allow-list of trusted data objects
that are removed from the causal analysis.

\begin{table}[!b]
\caption{Overview of APT Scenarios in DARPA Datasets.}
\label{tab:evaluation:tc:scenarios}
\resizebox{\columnwidth}{!}{%
  \begin{tabular}{|l|l|l|l|}
  \hline
  \cellcolor[HTML]{FFFFFF}{\color[HTML]{333333} \textbf{Dataset}} & \textbf{Duration} & \textbf{Platform} & \textbf{Attack Surface} %
  \\ \hline
  E3-THEIA & 02d00h12m & Ubuntu 12.04 x64 & Firefox \\ \hline
  E3-CADETS & 00d00h55m & FreeBSD & Nginx \\ \hline
  E3-ClearScope & 00d01h08m & Android 6.0.1 & Firefox \\ \hline %
  E5-THEIA & 00d00h21m & Ubuntu 12.04 x64  & Firefox \\ \hline
  E5-CADETS & 01d01h14m & FreeBSD 13 & Nginx \\ \hline
  E5-ClearScope & 02d01h02m & Android 8 & Appstarter APK \\ \hline
  \change{OpTC} & \change{02d03h00m} & \change{Windows} & \change{PowerShell} \\ \hline
  \end{tabular}
}
\end{table}

\begin{table}[b]
	\caption{DARPA data used for training, validation, and test. 
		The \textbf{bold} days are attack days in which both benign and attack time windows exist. The remaining days are benign days with only benign time windows. }
	\label{tab:evaluation:tc:datasplit}
\resizebox{\columnwidth}{!}{%
  \begin{tabular}{|l|c|c|c|}
    \hline
    \textbf{Datasets} & 
    \multicolumn{1}{c|}{\textbf{\begin{tabular}[c]{@{}c@{}}Training Data\\      (yyyy-mm-dd)\end{tabular}}} & 
    \multicolumn{1}{c|}{\textbf{\begin{tabular}[c]{@{}c@{}}Validation Data\\      (yyyy-mm-dd)\end{tabular}}} & 
    \multicolumn{1}{c|}{\textbf{\begin{tabular}[c]{@{}c@{}}Test Data\\      (yyyy-mm-dd)\end{tabular}}}                    
     \\ 
     \hline
    E3-THEIA & 
    \begin{tabular}[c]{@{}l@{}}2018-04-03/04/05 %
    \end{tabular} & 
    2018-04-09 & 
    \begin{tabular}[c]{@{}c@{}}\textbf{2018-04-10/12}\\ 2018-04-11%
    \end{tabular} \\ \hline
    E3-CADETS & 
    \begin{tabular}[c]{@{}l@{}}2018-04-02/03/04 %
    \end{tabular} &
     2018-04-05 & 
    \begin{tabular}[c]{@{}l@{}}\textbf{2018-04-06}\\ 2018-04-07 \end{tabular}                                                                                                   \\ \hline
    E3-ClearScope  & \begin{tabular}[c]{@{}l@{}}2018-04-04/05/06 %
    \end{tabular} & 2018-04-07 &
    \begin{tabular}[c]{@{}l@{}}2018-04-10 \\ \textbf{2018-04-11}\end{tabular}                               \\ \hline
    E5-THEIA & \begin{tabular}[c]{@{}l@{}}2019-05-08/09 %
    \end{tabular} & 
    2019-05-11 & 
    \begin{tabular}[c]{@{}l@{}}2019-05-14\\ \textbf{2019-05-15}\end{tabular}                               \\ \hline
    E5-CADETS & \begin{tabular}[c]{@{}l@{}}2019-05-08/09/11 %
    \end{tabular} & 2019-05-12 &
    \begin{tabular}[c]{@{}c@{}}2019-05-15 \\ \textbf{2019-05-16/17} %
    \end{tabular} \\ \hline
    E5-ClearScope & \begin{tabular}[c]{@{}l@{}}2019-05-08/09/11 %
    \end{tabular} & 2019-05-12 &
    \begin{tabular}[c]{@{}c@{}}2019-05-14 \\ \textbf{2019-05-15/17} %
    \end{tabular} \\ \hline
    \change{OpTC} & \change{2019-09-22} & \change{2019-09-23} & \change{\begin{tabular}[c]{@{}l@{}}\textbf{2019-09-23/24/25} %
    \end{tabular}} \\ \hline
    \end{tabular}
}
  \end{table}

%% file: appendix_hyper.tex
\begin{figure*}[htbp]
	\centering
	\subfigure[%
	$|\Phi|$]{
		\begin{minipage}[t]{0.163\paperwidth}
			\centering
			\label{fig:evaluation:node_feature_detection-auc-all}
			\includegraphics[width=0.163\paperwidth]{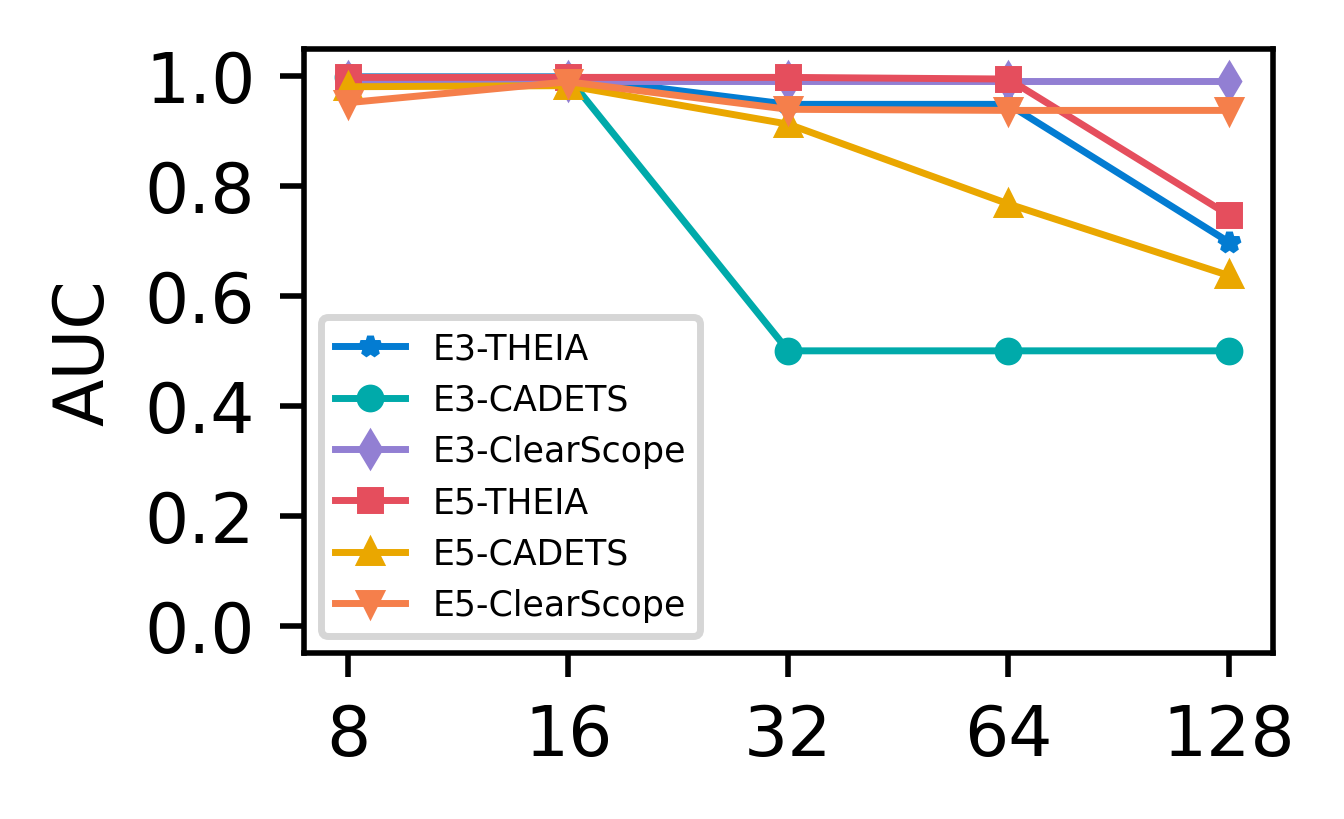}
		\end{minipage}
	}%
	\subfigure[%
	$|\mathbf{s}(v)|$]{
		\begin{minipage}[t]{0.15\paperwidth}
			\centering
			\label{fig:evaluation:node_state_detection-auc-all}
			\includegraphics[width=0.15\paperwidth]{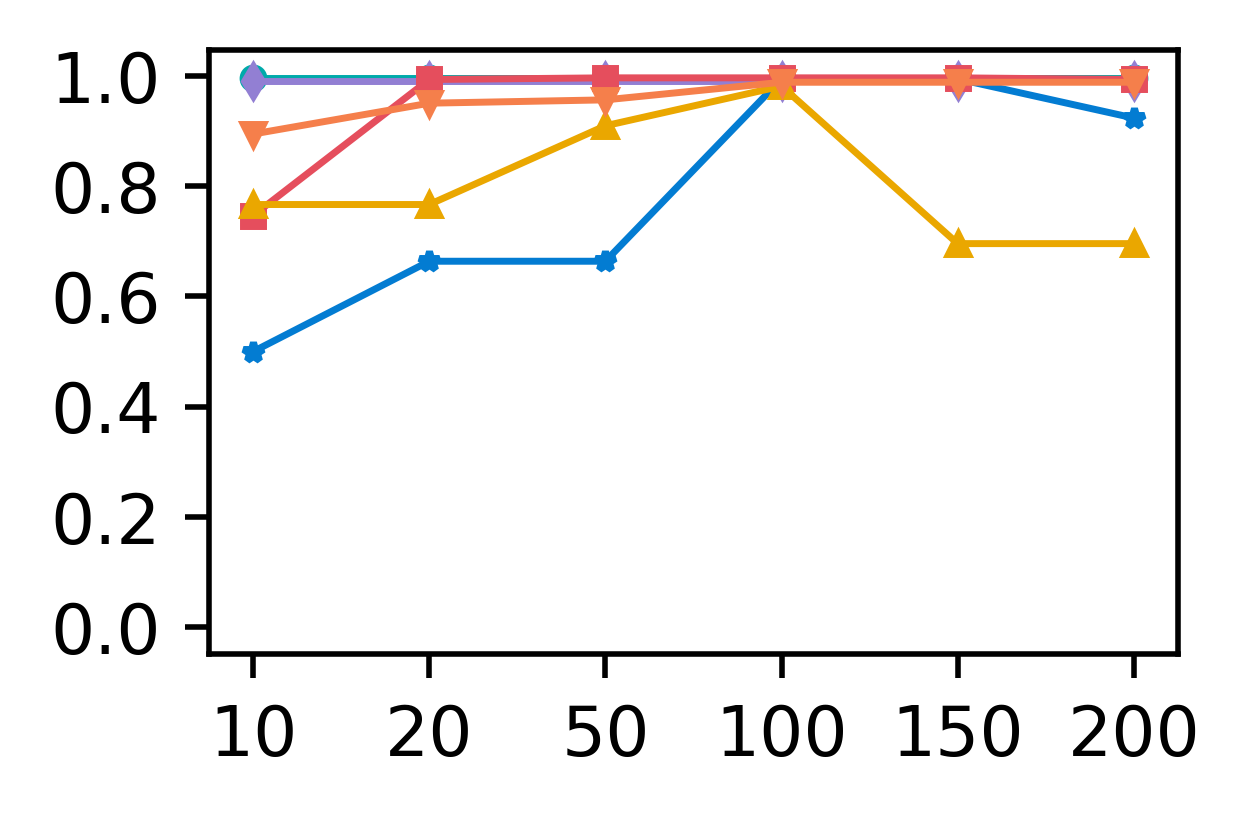}
		\end{minipage}%
	}%
	\subfigure[%
	$|\mathcal{N}|$]{
		\begin{minipage}[t]{0.15\paperwidth}
			\centering
			\label{fig:evaluation:neighbor_size_detection-auc-all}
			\includegraphics[width=0.15\paperwidth]{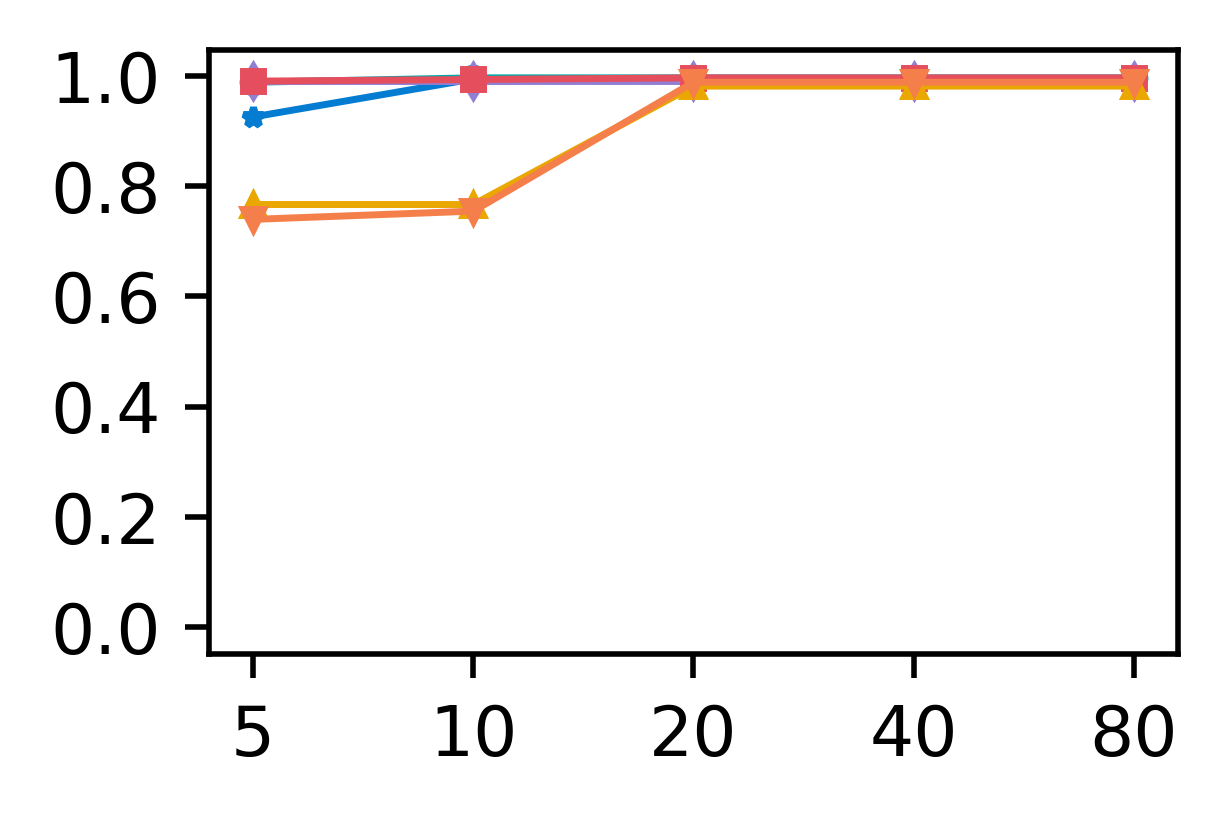}
		\end{minipage}%
	}%
	\subfigure[%
	$|\mathbf{z}|$]{
		\begin{minipage}[t]{0.15\paperwidth}
			\centering
			\label{fig:evaluation:emb_dim_detection-auc-all}
			\includegraphics[width=0.15\paperwidth]{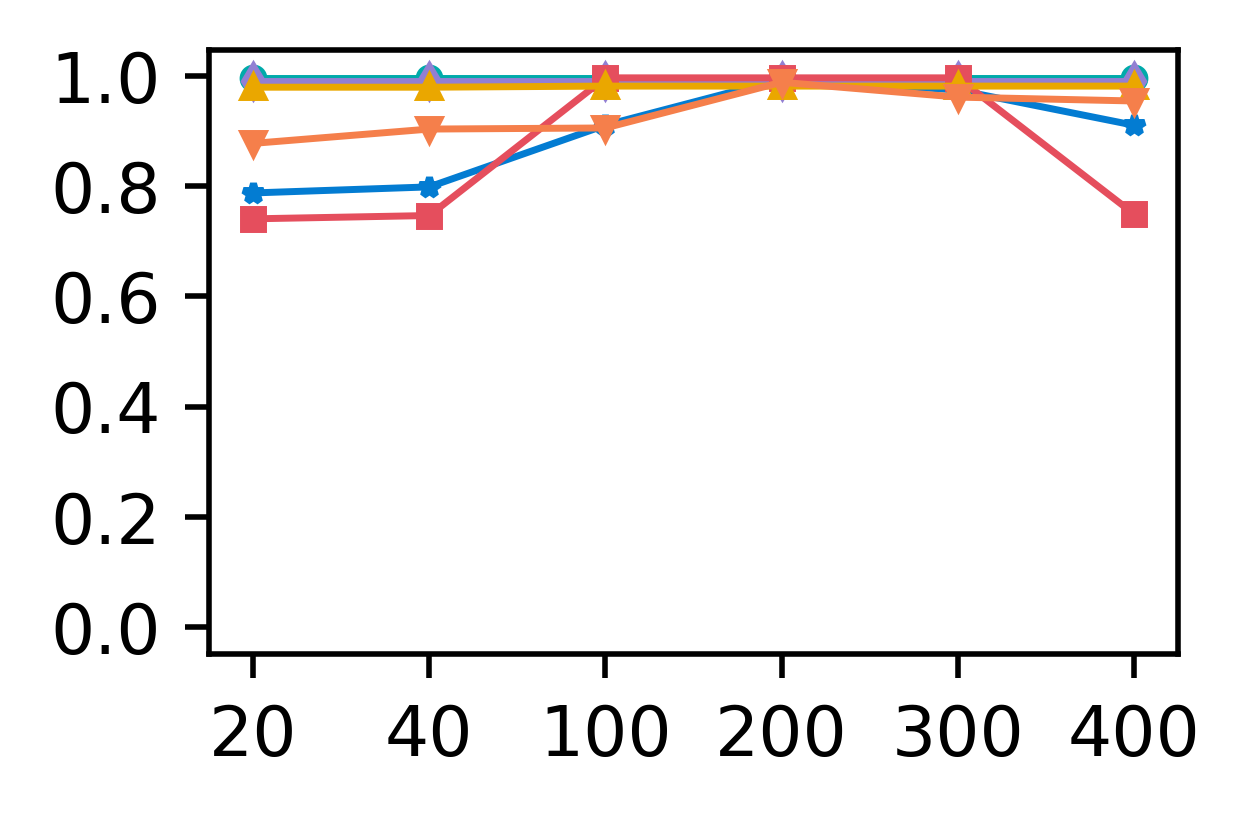}
		\end{minipage}
	}%
	\subfigure[%
	$|\mathbf{tw}|$]{
		\begin{minipage}[t]{0.15\paperwidth}
			\centering
			\label{fig:evaluation:time_window_detection-auc-all}
			\includegraphics[width=0.15\paperwidth]{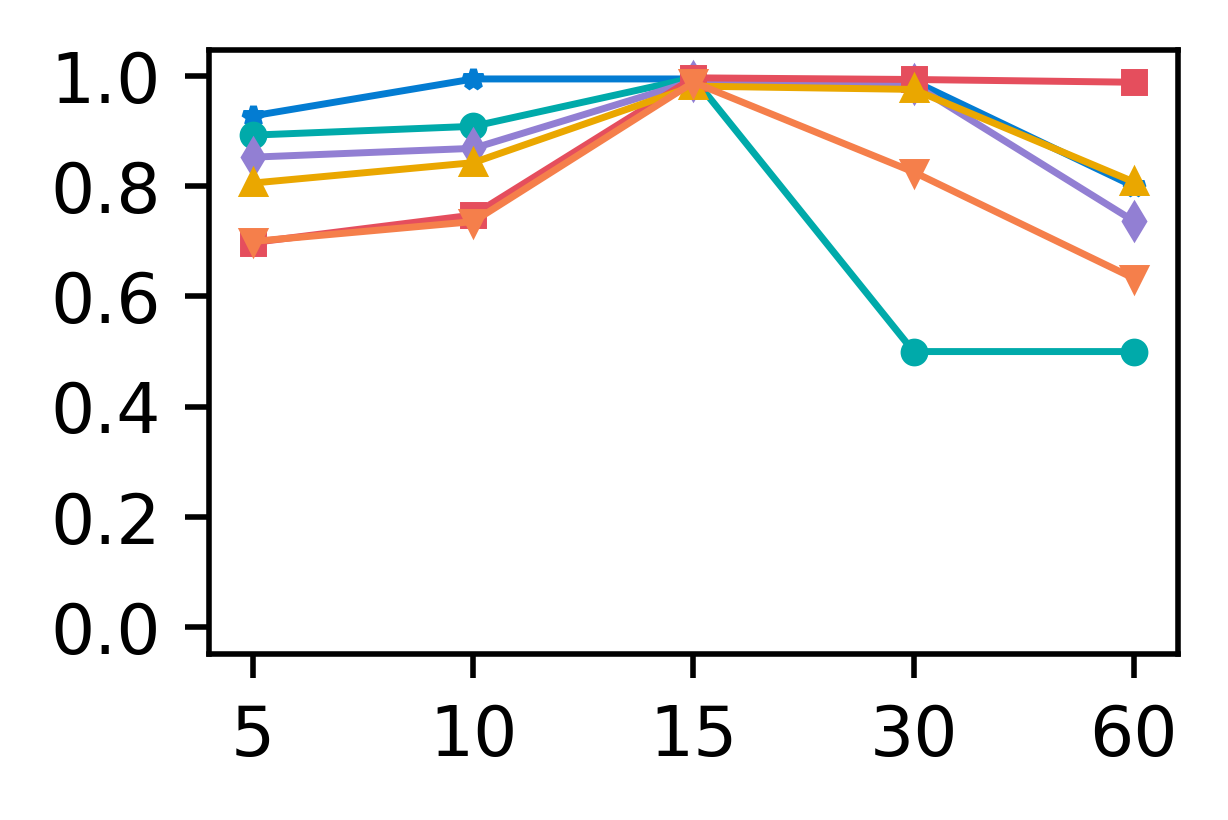}
		\end{minipage}
	}%
	\centering
	\caption{AUC on all DARPA datasets with varying hyperparameter values.}
	\label{fig:evapre-auc-all}
 \vspace{-10pt}
\end{figure*}

\begin{figure*}[t]
	\centering
	\subfigure[%
	$|\Phi|$]{
		\begin{minipage}[t]{0.16\paperwidth}
			\centering
			\label{fig:evaluation:node_feature_memory-usage-all}
			\includegraphics[width=0.16\paperwidth]{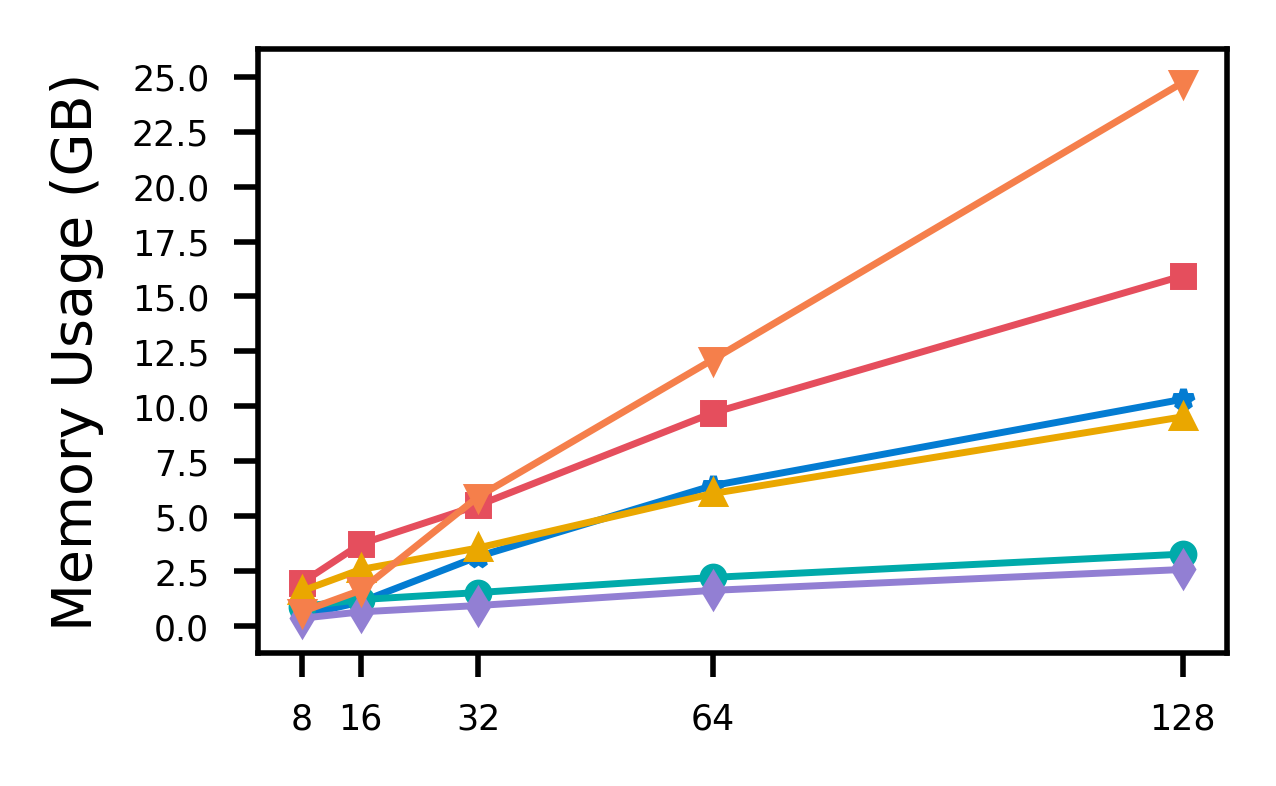}
		\end{minipage}
	}%
	\subfigure[%
	$|\mathbf{s}(v)|$]{
		\begin{minipage}[t]{0.15\paperwidth}
			\centering
			\label{fig:evaluation:node_state_memory-usage-all}
			\includegraphics[width=0.15\paperwidth]{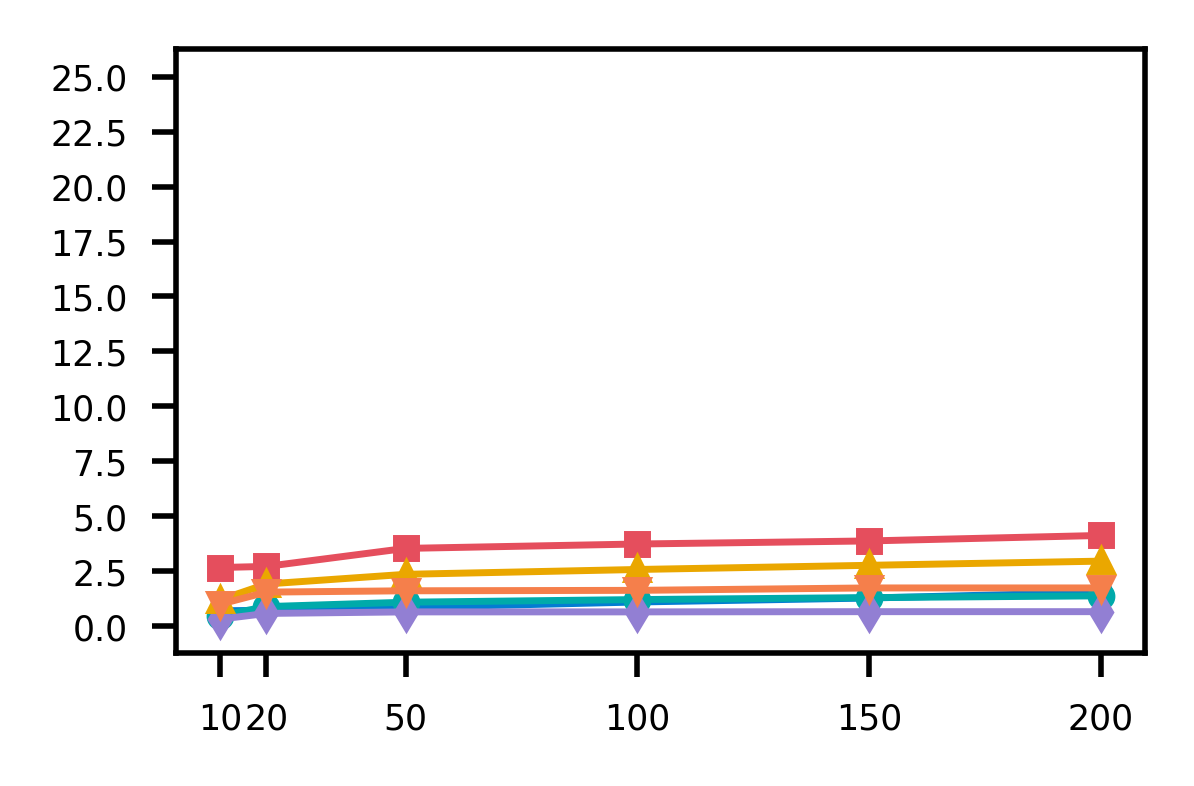}
		\end{minipage}%
	}%
	\subfigure[%
	$|\mathcal{N}|$]{
		\begin{minipage}[t]{0.15\paperwidth}
			\centering
			\label{fig:evaluation:neighbor_size_memory-usage-all}
			\includegraphics[width=0.15\paperwidth]{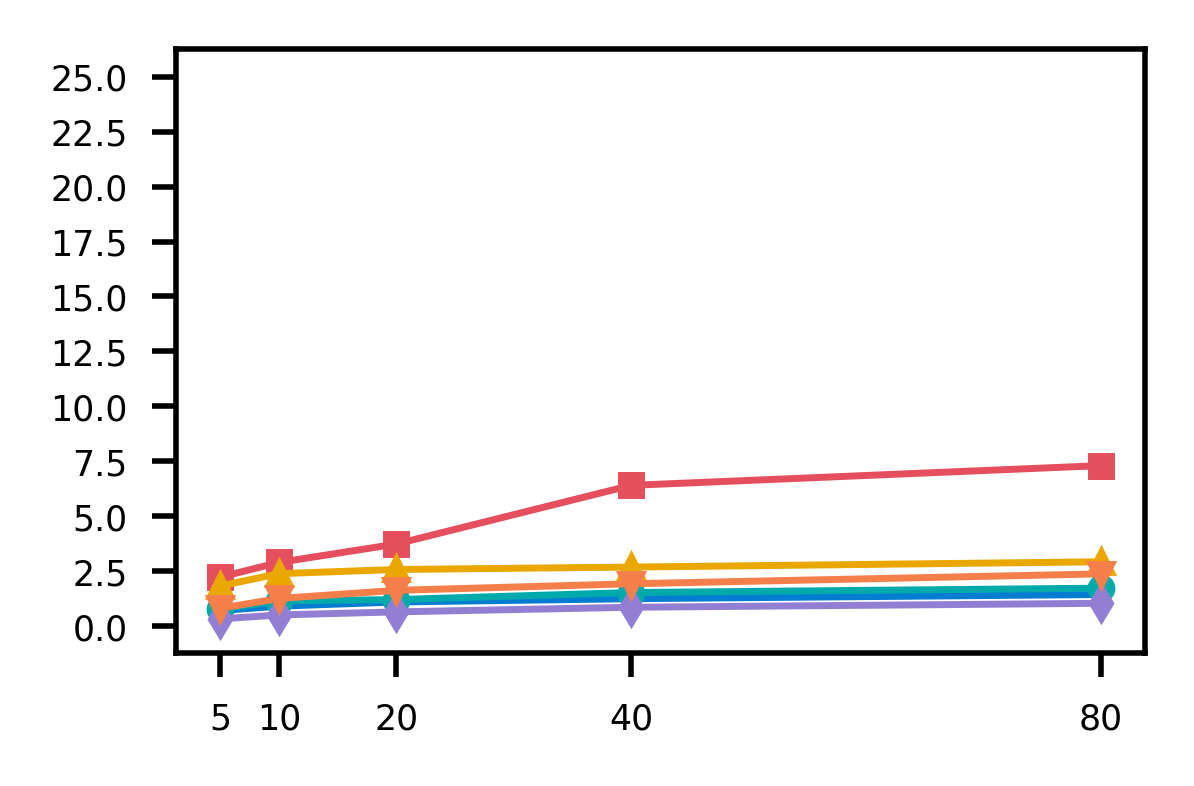}
		\end{minipage}%
	}%
	\subfigure[%
	$|\mathbf{z}|$]{
		\begin{minipage}[t]{0.15\paperwidth}
			\centering
			\label{fig:evaluation:emb_dim_memory-usage-all}
			\includegraphics[width=0.15\paperwidth]{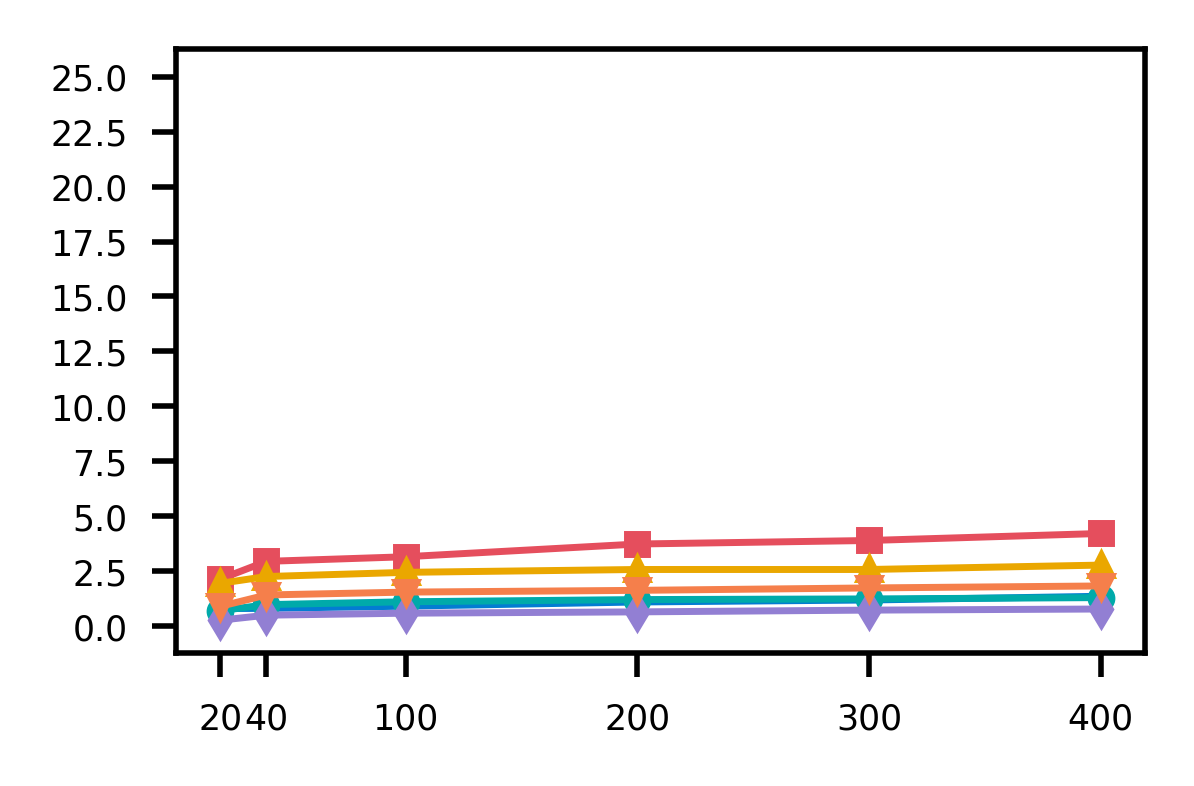}
		\end{minipage}
	}%
	\subfigure[%
	$|\mathbf{tw}|$]{
		\begin{minipage}[t]{0.15\paperwidth}
			\centering
			\label{fig:evaluation:time_window_memory-usage-all}
			\includegraphics[width=0.15\paperwidth]{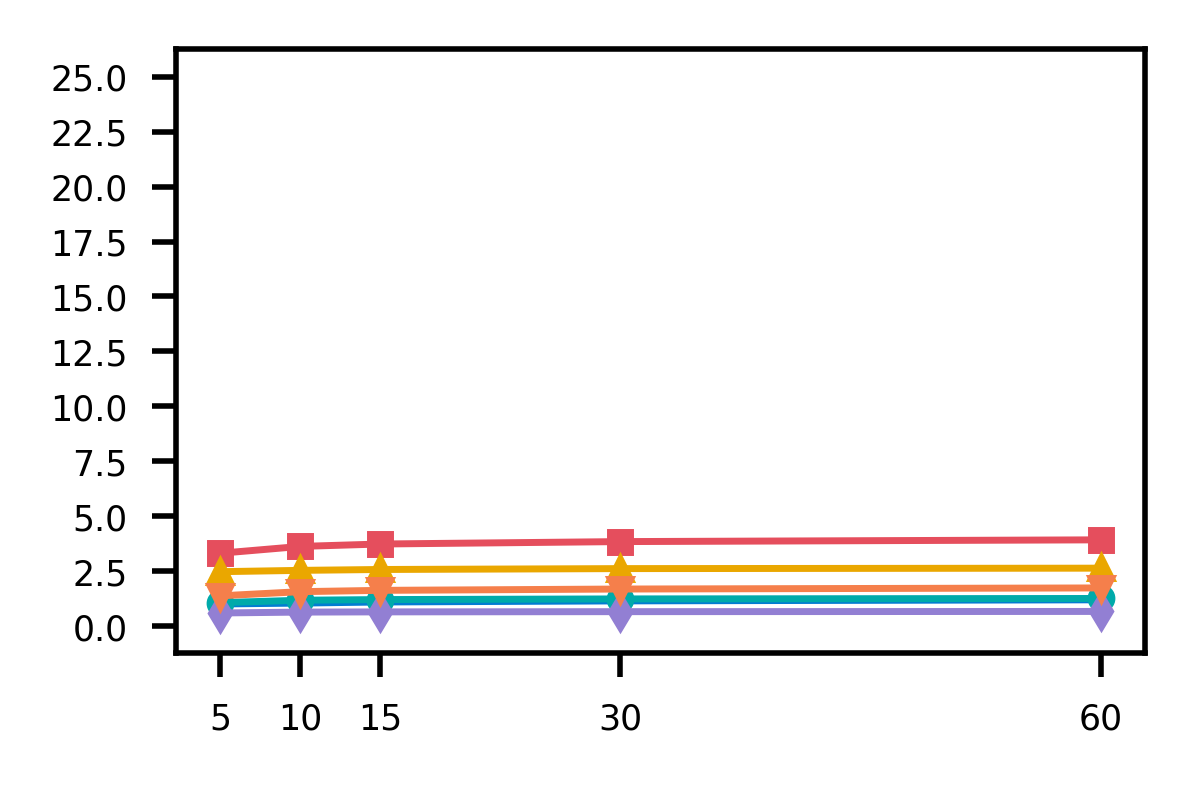}
		\end{minipage}
	}%
	\centering
	\caption{Average memory usage on all DARPA datasets with varying hyperparameter values.}
	\label{fig:evapre-memory-all}
  \vspace{-10pt}
\end{figure*}

\begin{figure*}[t]
	\centering
	\subfigure[%
	$|\Phi|$]{
		\begin{minipage}[t]{0.163\paperwidth}
			\centering
			\label{fig:evaluation:node_feature_cpu-usage-all}
			\includegraphics[width=0.163\paperwidth]{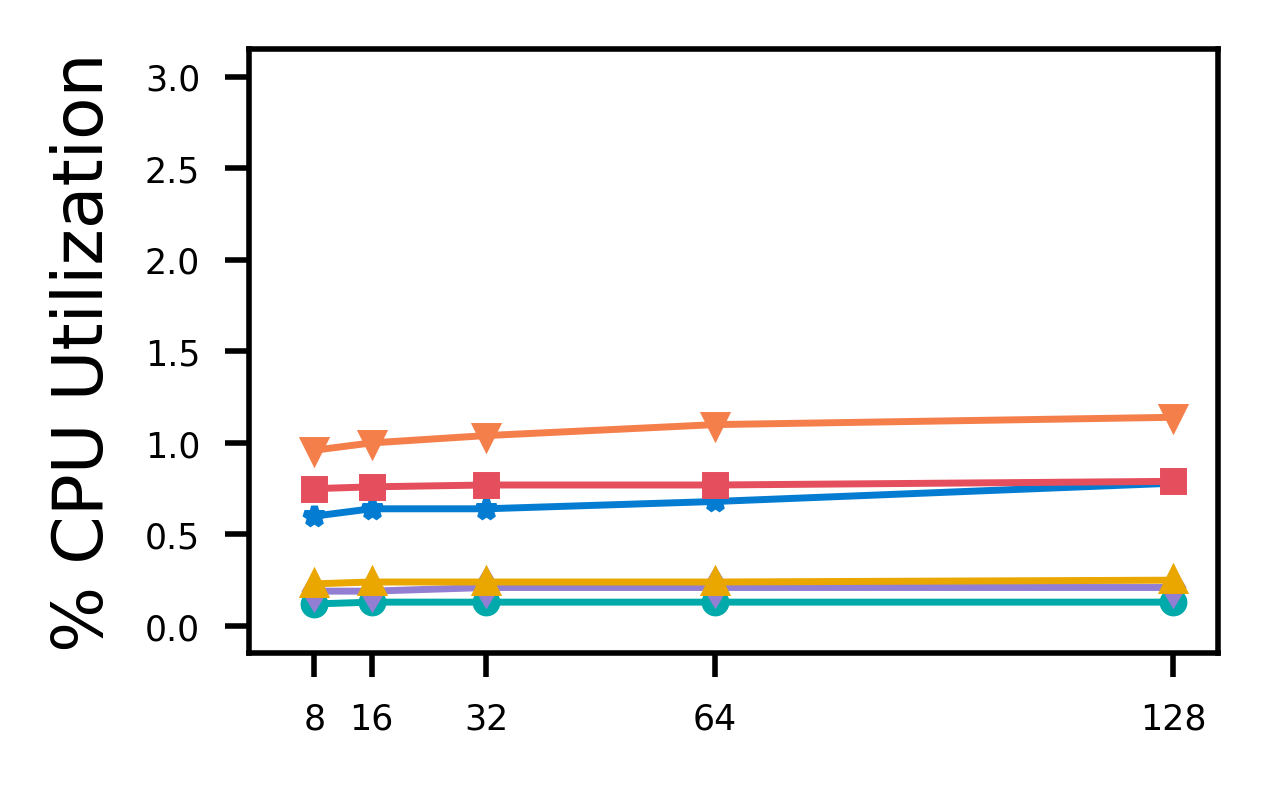}
		\end{minipage}
	}%
	\subfigure[%
	$|\mathbf{s}(v)|$]{
		\begin{minipage}[t]{0.15\paperwidth}
			\centering
			\label{fig:evaluation:node_state_cpu-usage-all}
			\includegraphics[width=0.15\paperwidth]{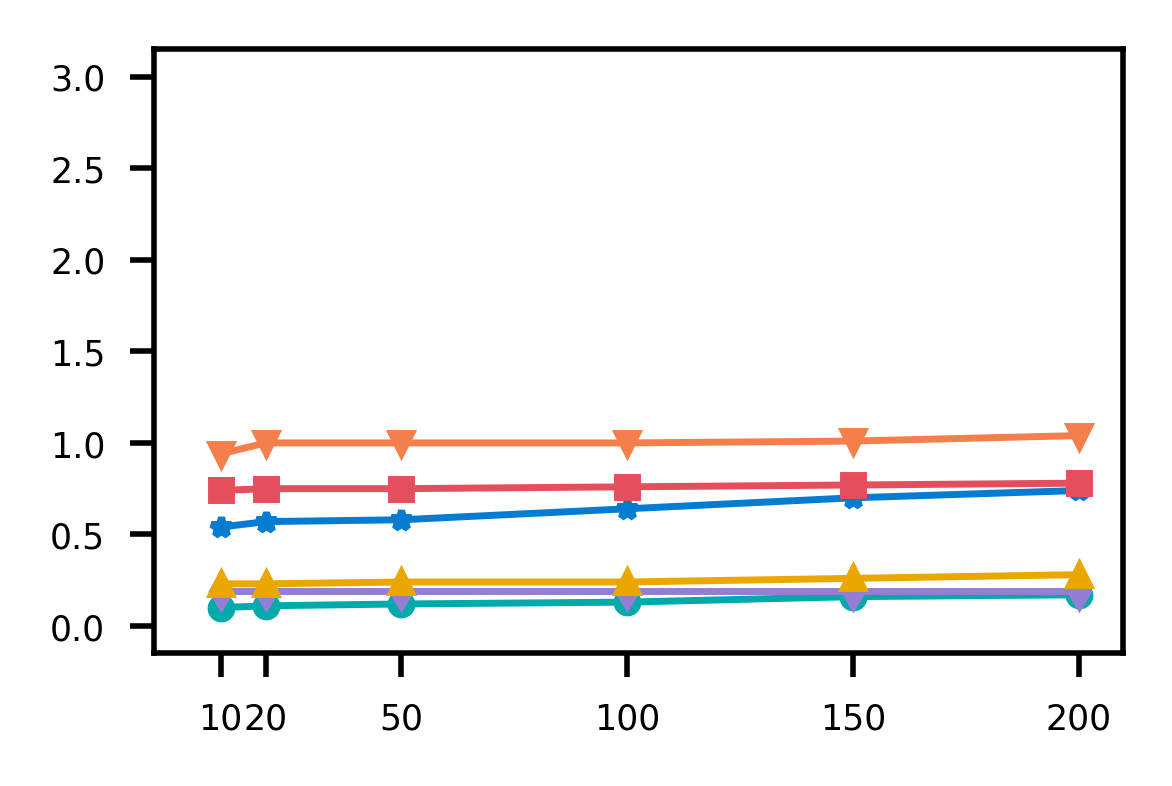}
		\end{minipage}%
	}%
	\subfigure[%
	$|\mathcal{N}|$]{
		\begin{minipage}[t]{0.15\paperwidth}
			\centering
			\label{fig:evaluation:neighbor_size_cpu-usage-all}
			\includegraphics[width=0.15\paperwidth]{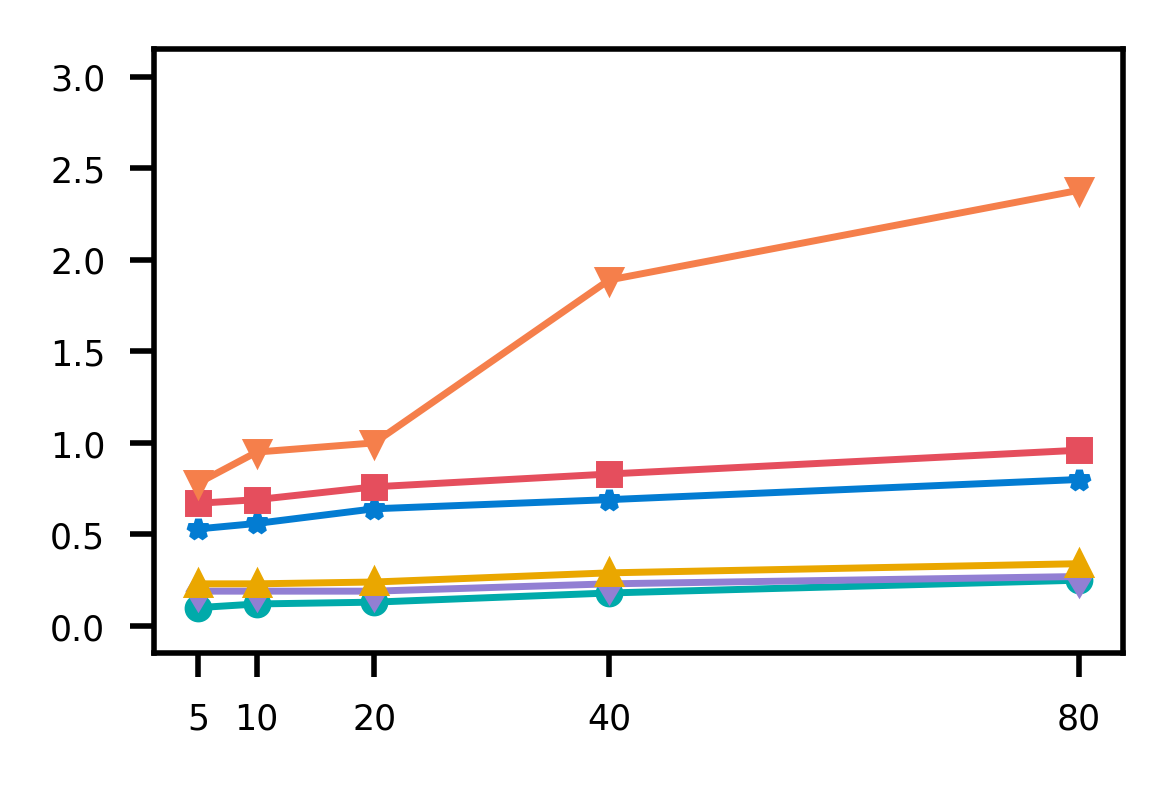}
		\end{minipage}%
	}%
	\subfigure[%
	$|\mathbf{z}|$]{
		\begin{minipage}[t]{0.15\paperwidth}
			\centering
			\label{fig:evaluation:emb_dim_cpu-usage-all}
			\includegraphics[width=0.15\paperwidth]{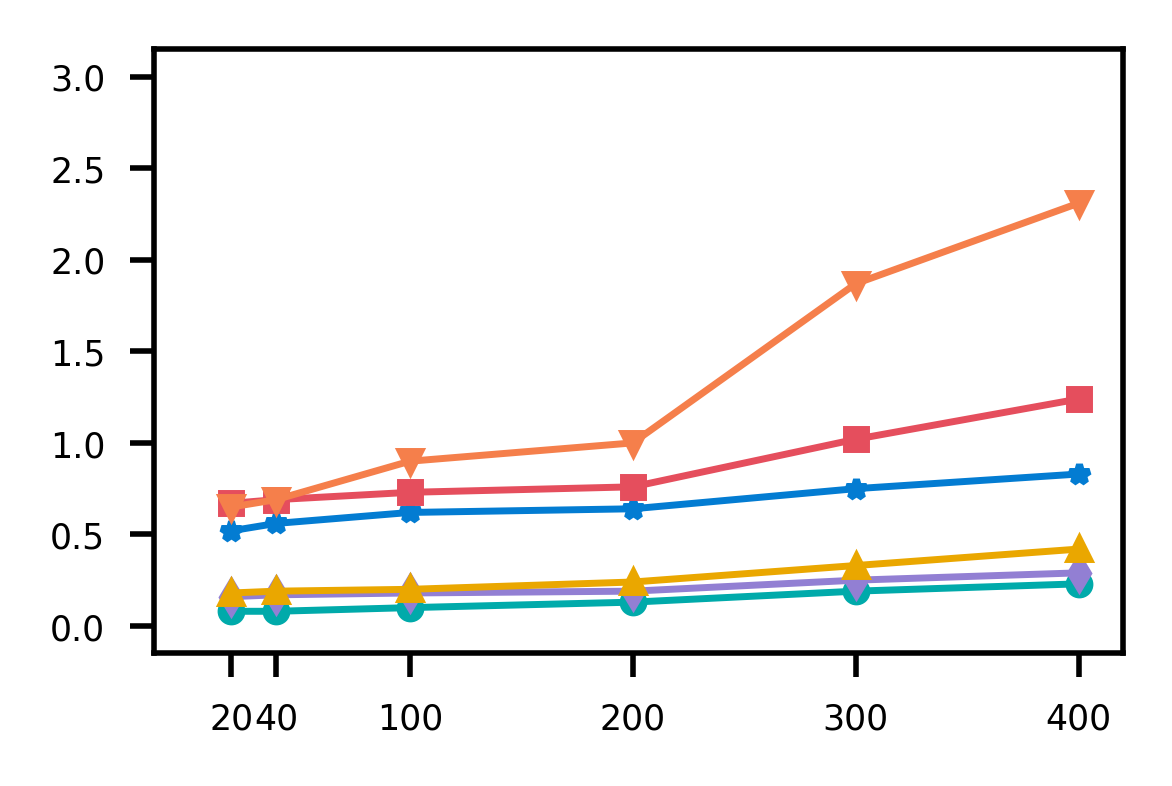}
		\end{minipage}
	}%
	\subfigure[%
	$|\mathbf{tw}|$]{
		\begin{minipage}[t]{0.15\paperwidth}
			\centering
			\label{fig:evaluation:emb_dim_cpu-usage-all}
			\includegraphics[width=0.15\paperwidth]{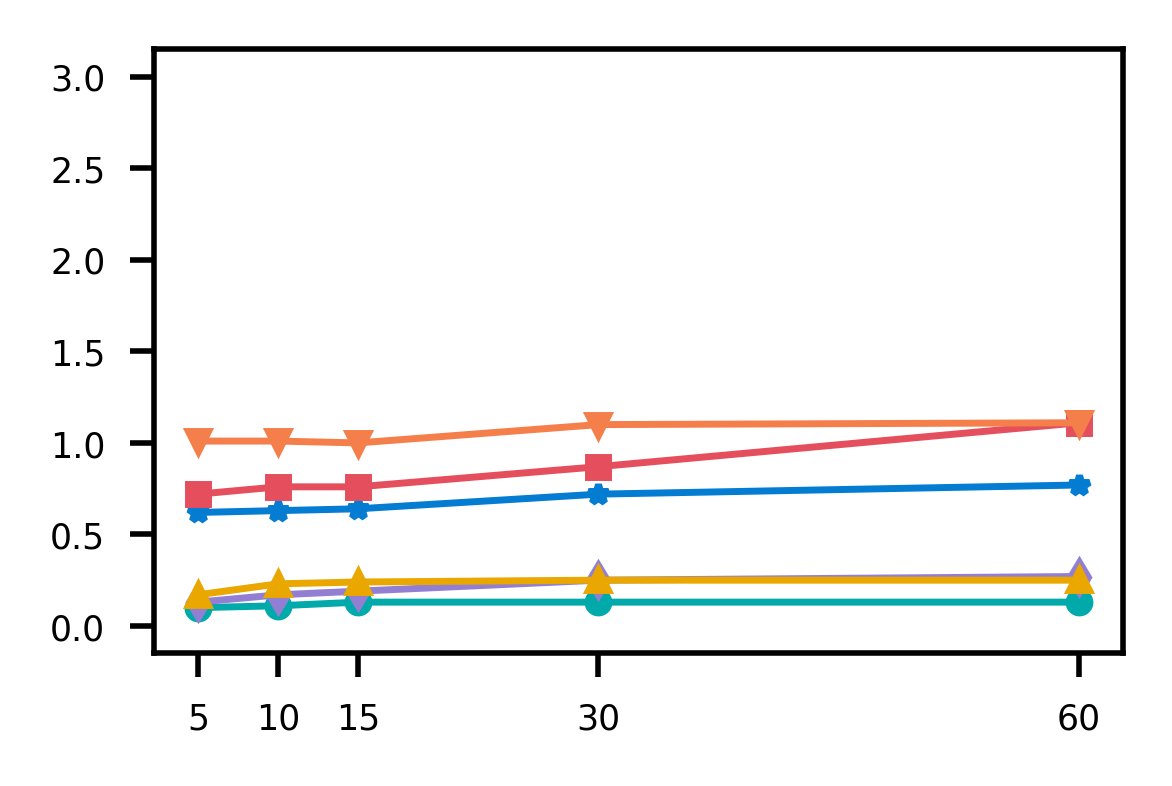}
		\end{minipage}
	}%
	\centering
	\caption{$90^{th}$ percentile CPU utilization
		on all DARPA datasets with varying hyperparameter values.}
	\label{fig:evapre-90th-cpu-all}
  \vspace{-10pt}
\end{figure*}

\begin{figure*}[t]
	\centering
	\subfigure[%
	$|\Phi|$]{
		\begin{minipage}[t]{0.163\paperwidth}
			\centering
			\label{fig:evaluation:node_feature_execution-time-all}
			\includegraphics[width=0.163\paperwidth]{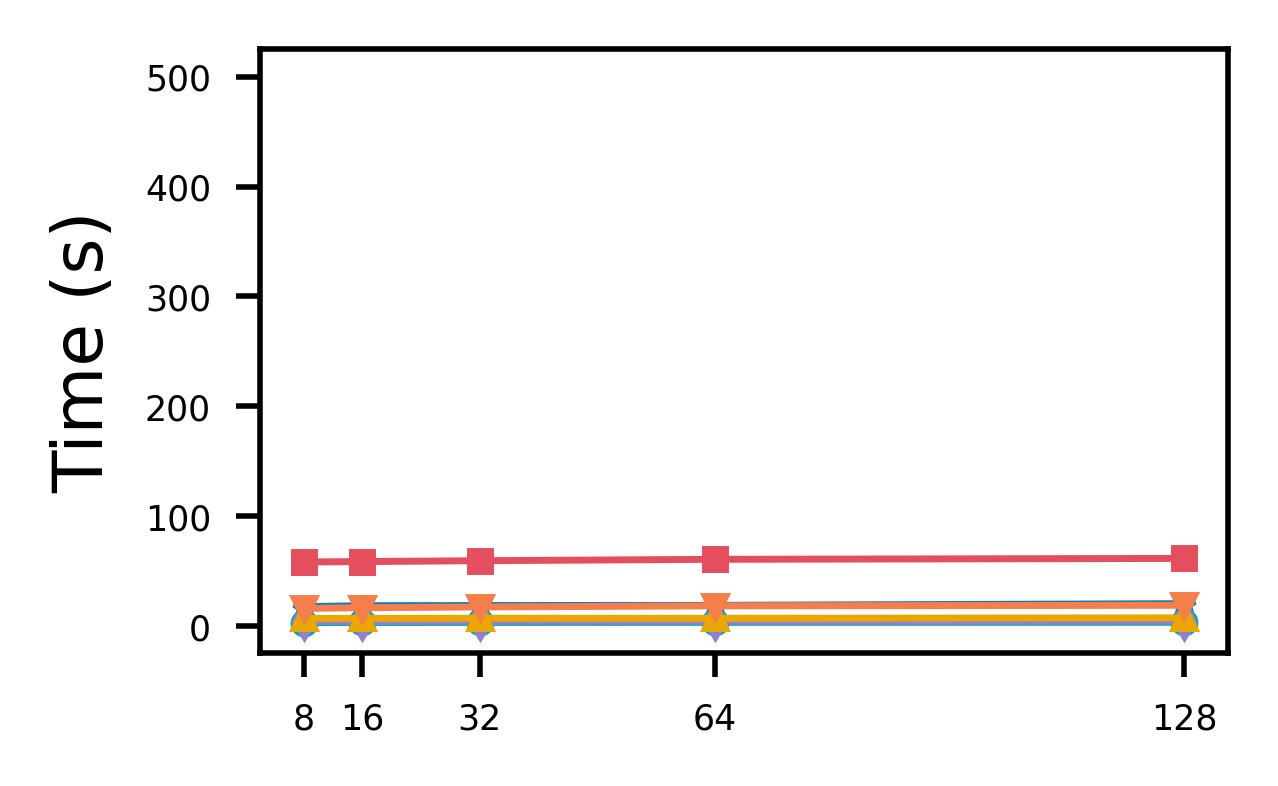}
		\end{minipage}
	}%
	\subfigure[%
	$|\mathbf{s}(v)|$]{
		\begin{minipage}[t]{0.15\paperwidth}
			\centering
			\label{fig:evaluation:node_state_execution-time-all}
			\includegraphics[width=0.15\paperwidth]{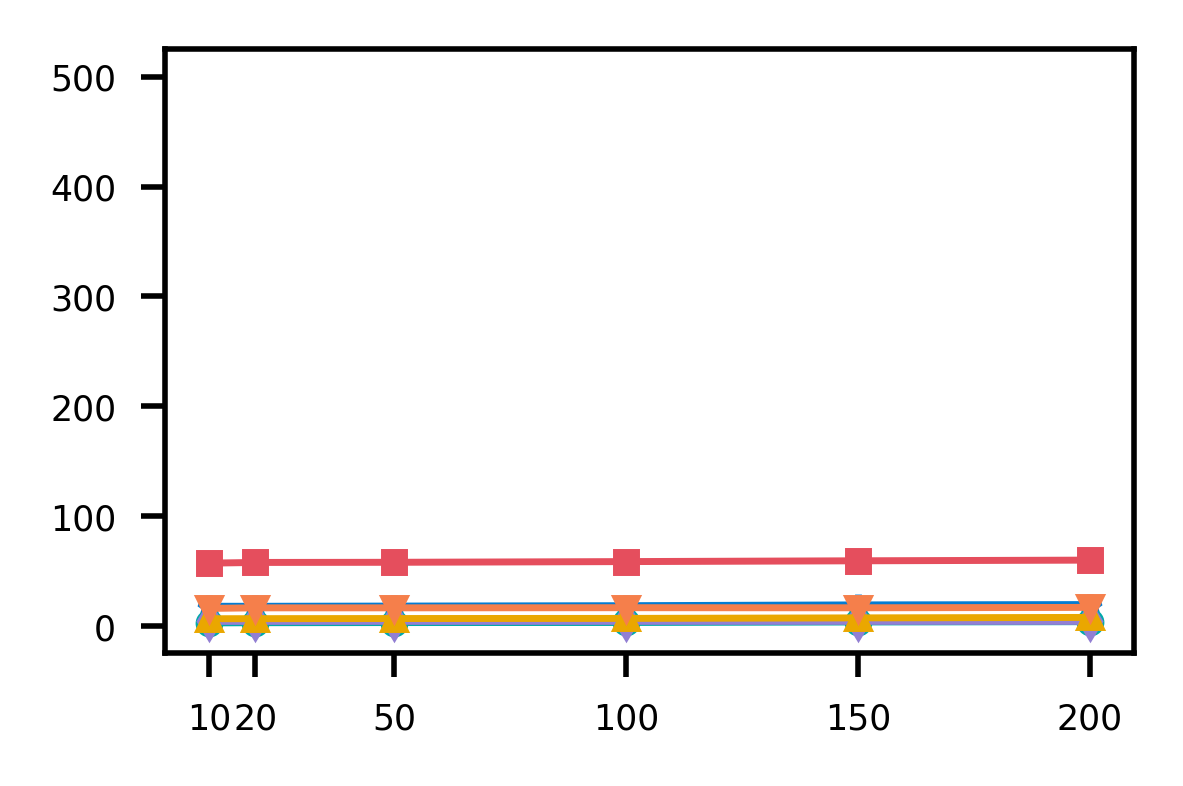}
		\end{minipage}%
	}%
	\subfigure[%
	$|\mathcal{N}|$]{
		\begin{minipage}[t]{0.15\paperwidth}
			\centering
			\label{fig:evaluation:neighbor_size_execution-time-all}
			\includegraphics[width=0.15\paperwidth]{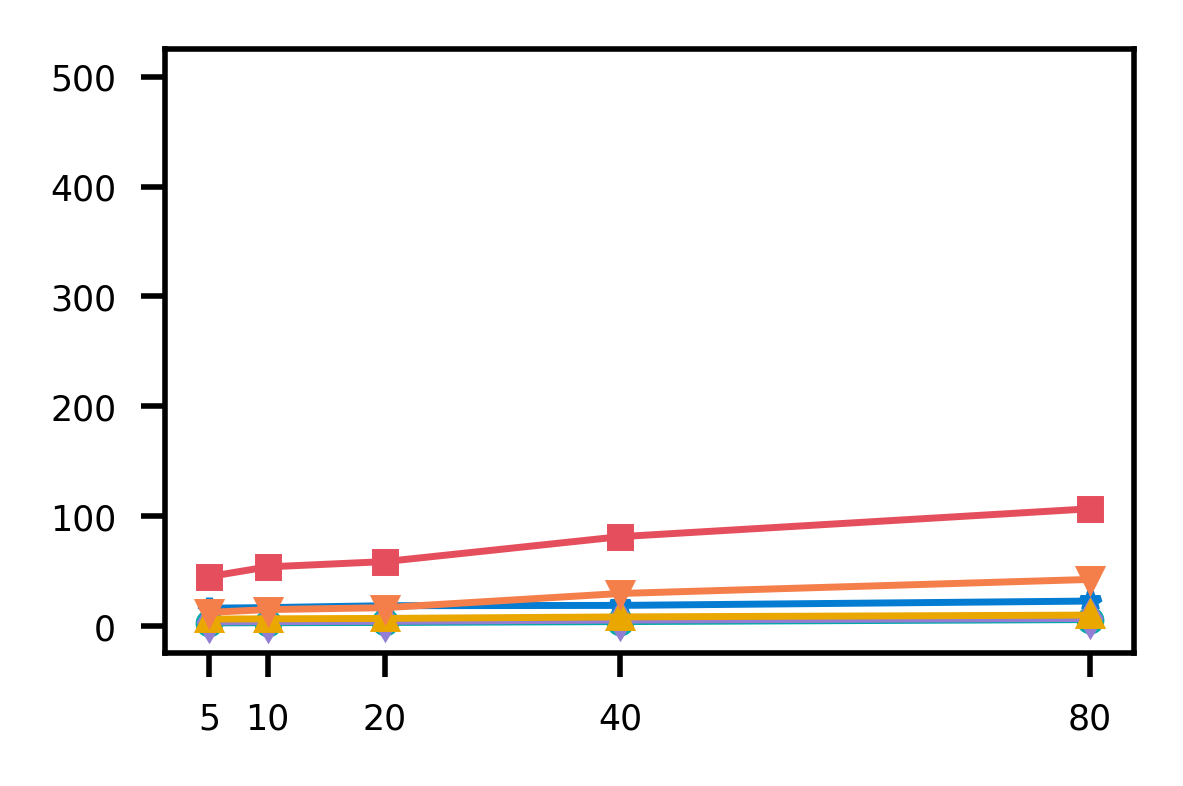}
		\end{minipage}%
	}%
	\subfigure[%
	$|\mathbf{z}|$]{
		\begin{minipage}[t]{0.15\paperwidth}
			\centering
			\label{fig:evaluation:emb_dim_execution-time-all}
			\includegraphics[width=0.15\paperwidth]{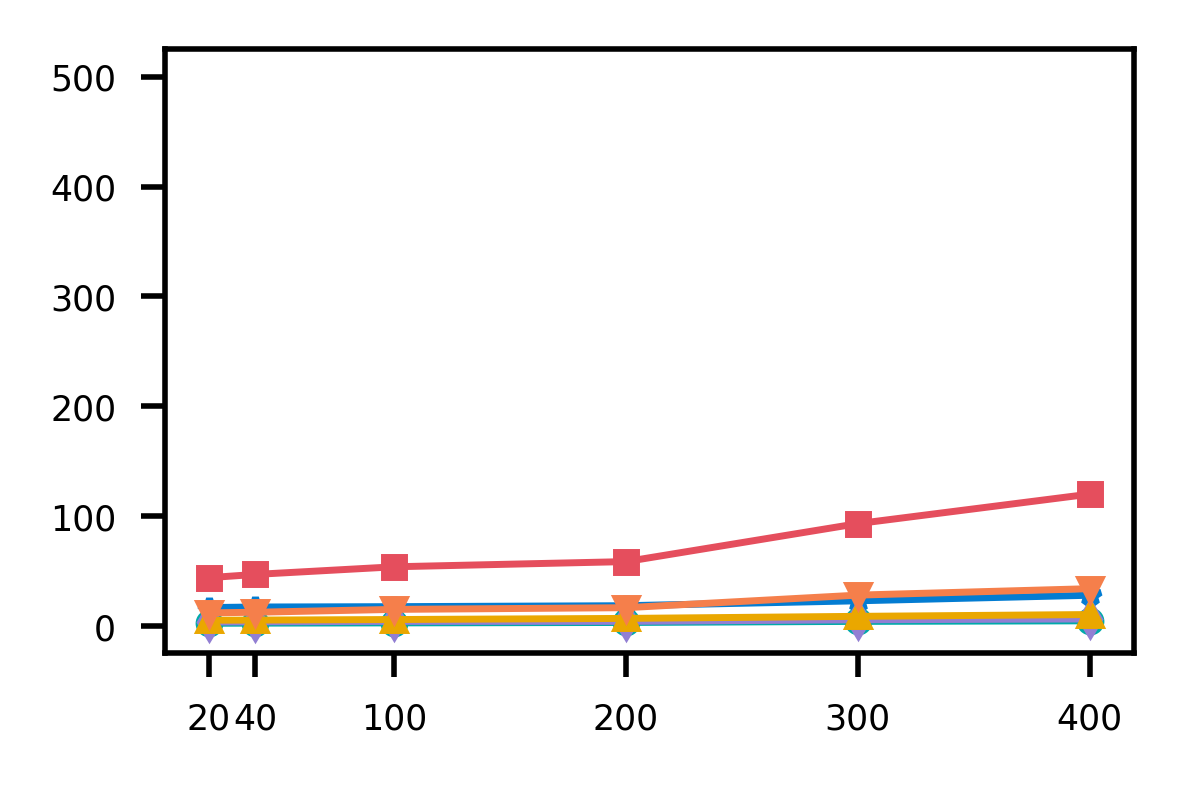}
		\end{minipage}
	}%
	\subfigure[%
	$|\mathbf{tw}|$]{
		\begin{minipage}[t]{0.15\paperwidth}
			\centering
			\label{fig:evaluation:time_window_execution-time-all}
			\includegraphics[width=0.15\paperwidth]{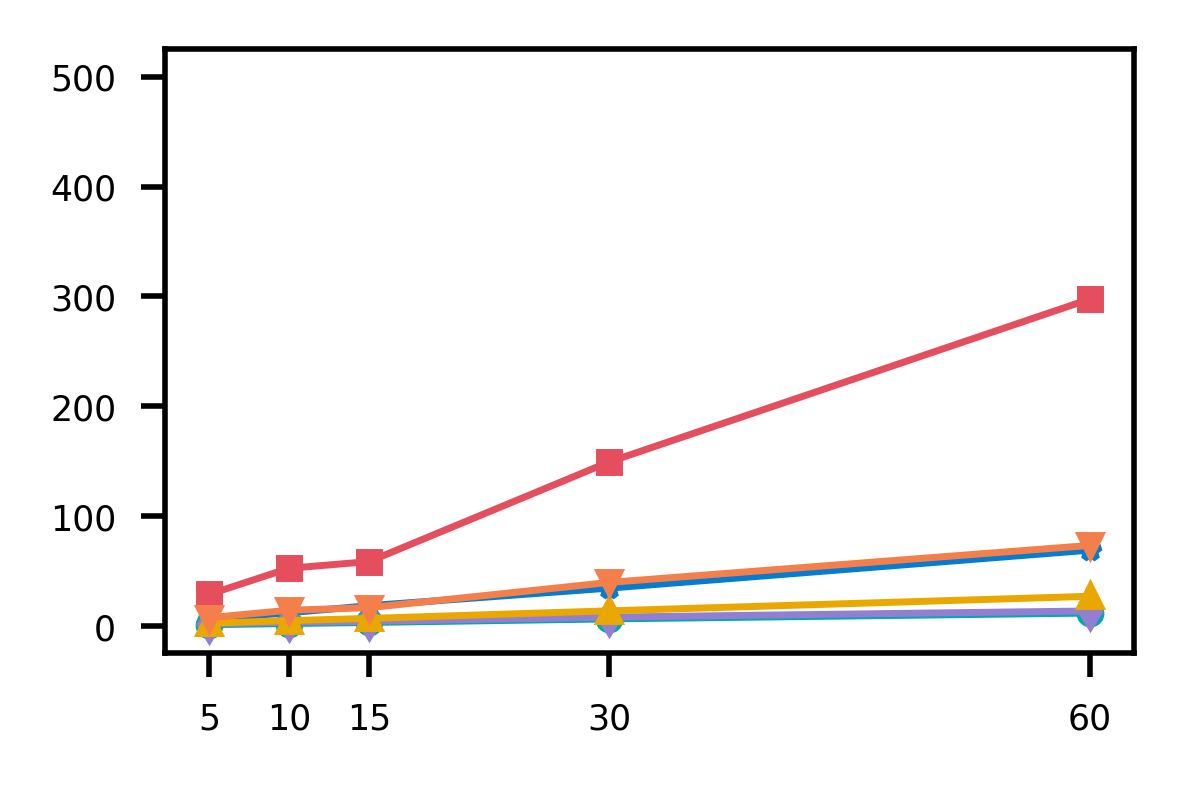}
		\end{minipage}
	}%
	\centering
	\caption{Average execution time on all DARPA datasets with varying hyperparameter values.}
	\label{fig:evapre-time-all}
  \vspace{-15pt}
\end{figure*}

\autoref{fig:evapre-auc-all}
shows AUC results for all DARPA datasets
with varying hyperparameter values.
\autoref{fig:evapre-memory-all}
and \autoref{fig:evapre-90th-cpu-all}
show the corresponding
memory and computational overhead.
\autoref{fig:evapre-time-all} shows
the average time window execution time.

%% file: graph_examples.tex
\vspace{-7pt}
\section{Attack Reconstruction Examples}
\label{sec:appendix:attack_examples}

Due to space constraints,
we provide a subset of candidate graph examples 
from DARPA datasets
in our experiment.
We refer interested readers
to the supplementary material~\cite{kairos-supp}
for full experimental results.
Similarly, we include only
benign summary graph examples of
the corresponding datasets
in~\autoref{sec:appendix:benign_examples}.

\begin{figure*}[!t]
	\centering
	\includegraphics[width=\textwidth]{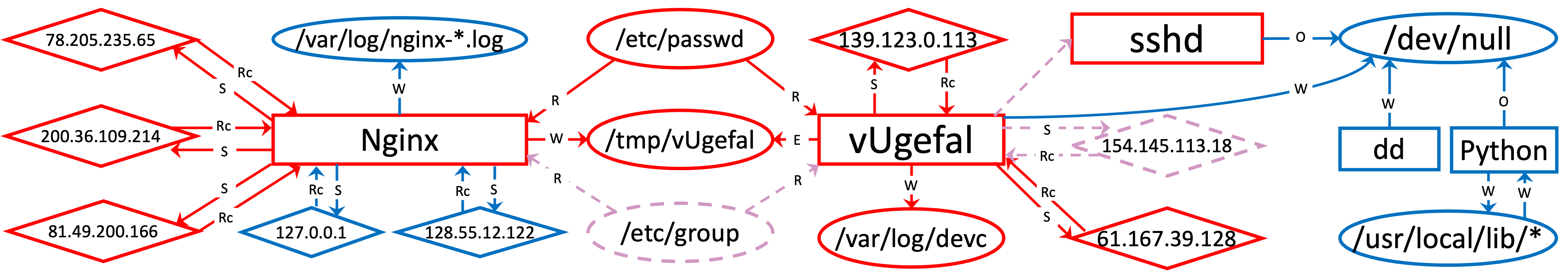}
	\caption{A summary graph that describes attack activity in DARPA's E3-CADETS dataset, as automatically generated by \system.
		\remove{Colors and dashed elements are added to ease comparison with the ground truth.}}
	\label{fig:evaluation:casestudy_CADETS}
 \vspace{-10pt}
\end{figure*}

\noindgras{E3-CADETS (\autoref{fig:evaluation:casestudy_CADETS}).}
The attacker (\texttt{81.49.200.166})
connects to
a vulnerable \texttt{Nginx} server
and obtains a shell.
Through the shell,
the attacker successfully downloads a
malicious payload to \texttt{/tmp/vUgefal}
and executes the payload with root privileges.
The elevated process \texttt{vUgefal}
attempts to move laterally to \texttt{154.145.113.18}
and \texttt{61.167.39.128}.
However,
only the attempt at infecting \texttt{61.167.39.128} is successful.
\texttt{vUgefal} further plans to
inject malicious payload to the \texttt{sshd} process.
To do so,
the attacker downloads the payload to \texttt{/var/log/devc},
but the attempted process injection fails.

\begin{figure*}[t]
	\centering
	\includegraphics[width=\textwidth]{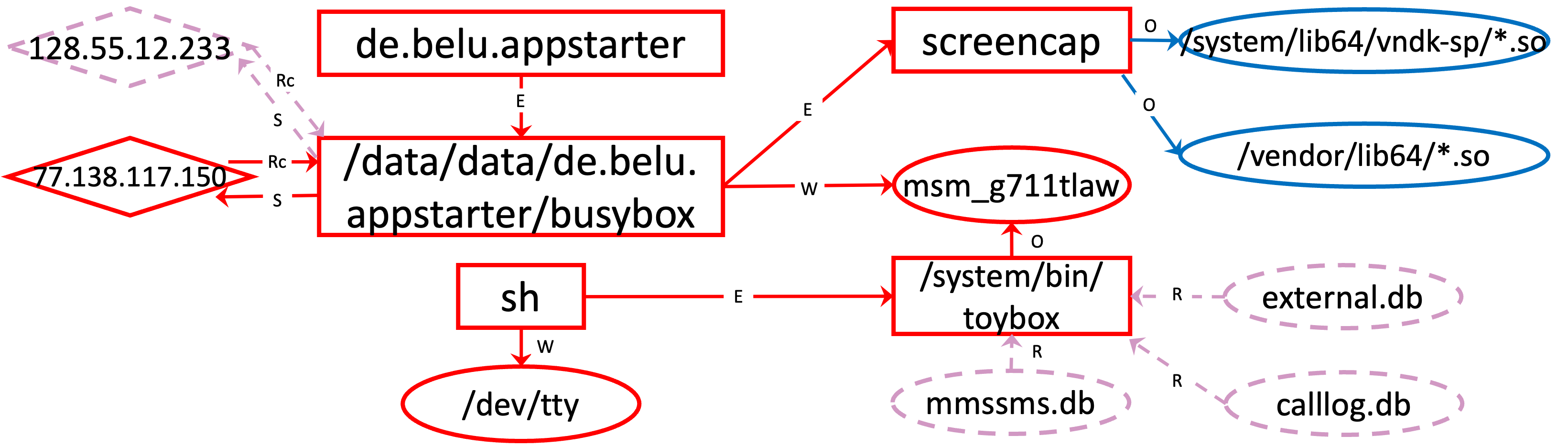}
	\caption{A summary graph that describes attack activity in DARPA's E5-ClearScope dataset, as automatically generated by \system.
		\remove{Colors and dashed elements are added to ease comparison with the ground truth.}}
	\label{fig:evaluation:casestudy_clearscope5}
 \vspace{-10pt}
\end{figure*}

\noindgras{E5-ClearScope (\autoref{fig:evaluation:casestudy_clearscope5}).}
A user accidentally installs
a malicious \texttt{appstarter} APK \texttt{de.belu.appstarter},
which loads an attack module
called \texttt{busybox}.
\remove{during the installation process.}
This module gives the attacker control from \texttt{77.138.117.150}. \remove{to the victim host.}
The attacker then installs the driver \texttt{msm\_g711tlaw}
into the victim host for privilege escalation.
The attack exfiltrates
\texttt{calllog.db},
\texttt{calendar.db},
and \texttt{mmssms.db}
\remove{with elevated privileges }%
and takes a screenshot.
Two days later,
the attacker exploits \texttt{appstarter} again
to try to
connect to the C\&C server (\texttt{128.55.12.233}) but failed.
\remove{Note that t}\change{T}he ground truth also
describes some malicious activity
of attack payloads called \texttt{lockwatch}
and \texttt{mozilla}.
Upon close inspection,
we discover that the provenance data
related to the malicious activity
is corrupted.
\remove{(\eg a unique node identifier does not refer to}%
\remove{any system entity).}%
We remove the corrupted data
and omit the malicious activity in~\autoref{fig:evaluation:casestudy_clearscope5}.

\begin{figure*}[t]
	\centering
	\includegraphics[width=\textwidth]{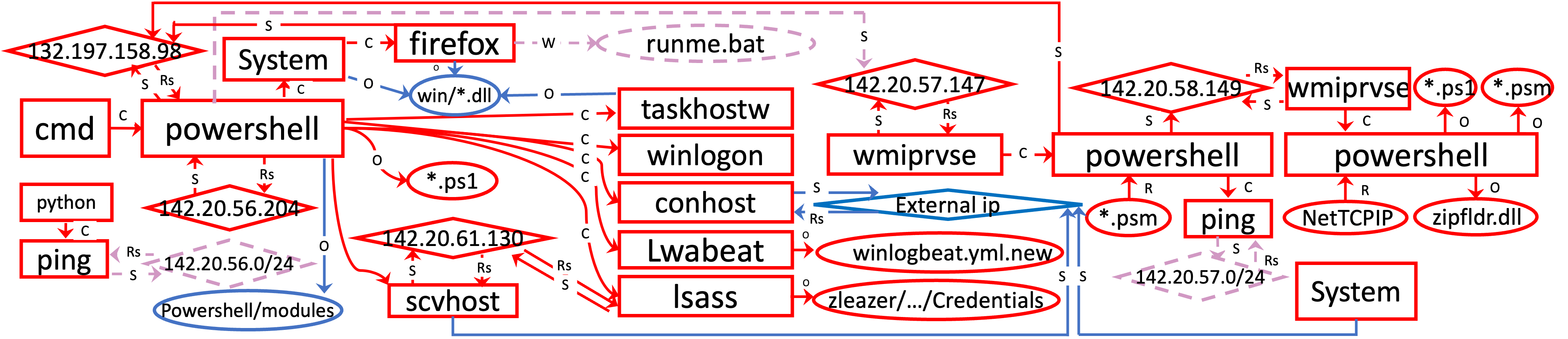}
	\caption{A summary graph that describes attack activity in DARPA's OpTC dataset in day 1, as automatically generated by \system.
		\remove{Colors and dashed elements are added to ease comparison with the ground truth.}}
	\label{fig:evaluation:casestudy_optcday1}
 \vspace{-10pt}
\end{figure*}

\noindgras{OpTC Day 1 (\autoref{fig:evaluation:casestudy_optcday1}).}
The attacker uses a C\&C server (\texttt{132.197.158.98})
to connect to the victim host
and executes a \texttt{powershell} script \texttt{runme.bat}.
The attacker then injects the process \texttt{lsass}
to collect the victim's credential and host information.
The attacker also scans the network
(\eg using \texttt{ping} and \texttt{smb})
and uses \texttt{wmiprvse} to move laterally
to a host at \texttt{142.20.57.147}.
Eventually,
the attacker moves to a host at \texttt{142.20.58.149}
and runs more \texttt{powershell} scripts \texttt{*.ps1}
to collect information. %

\section{Benign Summary Graph Examples}
\label{sec:appendix:benign_examples}

\begin{figure}[!h]
	\centering
	\includegraphics[width=\columnwidth]{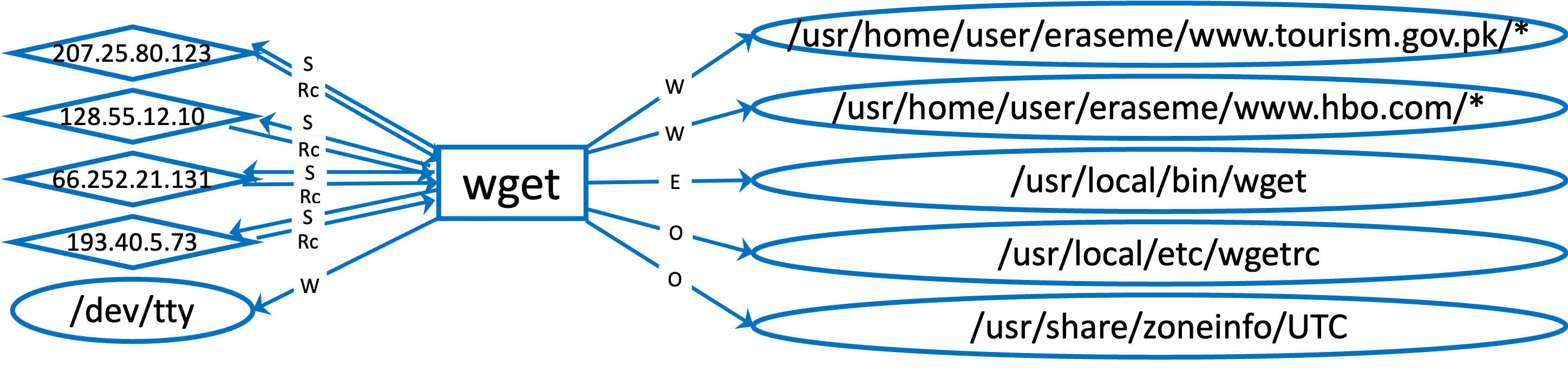}
	\caption{A benign summary graph in DARPA's E3-CADETS dataset.}
	\label{fig:evaluation:casestudy_benign_cadets3}
\end{figure}

\noindgras{E3-CADETS (\autoref{fig:evaluation:casestudy_benign_cadets3}).}
\texttt{wget} is a Linux utility
used to download files from the Internet.
It might connect to any external IP or URL.
To determine
whether \texttt{wget}'s behavior
is related to attack activity,
sysadmins might either
check whether any connected IP
is in a blocklist
or confirm with the user
the identities of the files
they download.
Any file not recognized by the user
might be downloaded
by the attacker
through a C\&C server.

\begin{figure}[!h]
	\centering
	\includegraphics[width=\columnwidth]{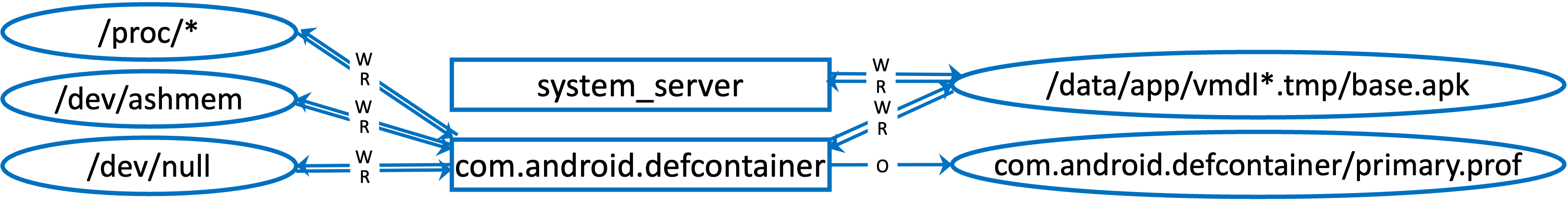}
	\caption{A benign summary graph in DARPA's E5-ClearScope.}
	\label{fig:evaluation:casestudy_benign_clearscope5}
\end{figure}

\noindgras{E5-ClearScope (\autoref{fig:evaluation:casestudy_benign_clearscope5}).}
\texttt{defcontainer} is a system process
associated with APK file installation.
Sysadmins might confirm with the user
the identities of the APK files
they install.
Sysadmins should further inspect
the installed APK files
to ensure that they are from legitimate vendors.

\begin{figure}[!h]
	\centering
	\includegraphics[width=\columnwidth]{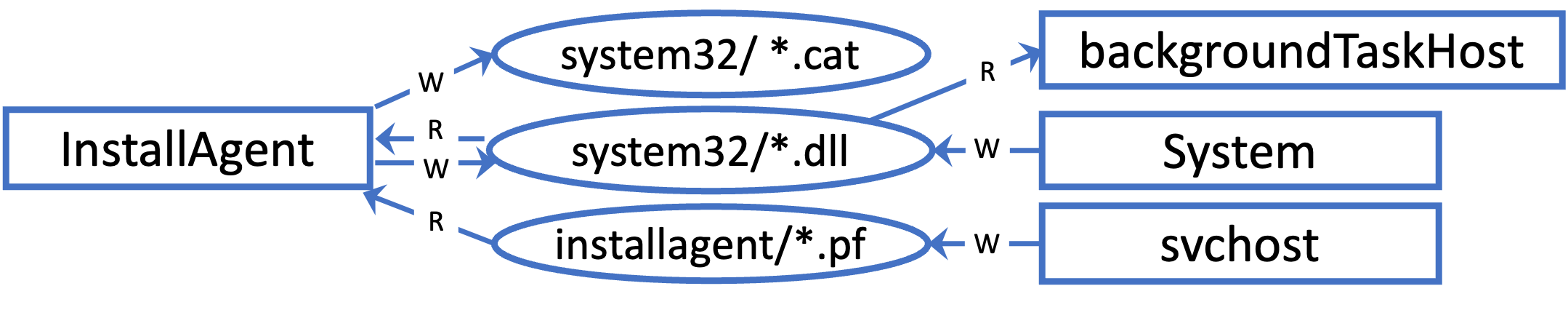}
	\caption{A benign summary graph in DARPA's OpTC.}
	\label{fig:evaluation:casestudy_benign_optc}
\end{figure}

\noindgras{OpTC (\autoref{fig:evaluation:casestudy_benign_optc}).}
\texttt{Installagent} is
Microsoft Windows Store's update agent,
which uses
the system services
\texttt{System}, \texttt{backgroundTaskHost}, and \texttt{svchost}.
Sysadmins need to investigate \texttt{Installagent}
only when suspicious files (\eg files not in the system path)
appear in its activity.